\documentclass[11pt, twoside]{amsart}

\usepackage[utf8]{inputenc}
\usepackage[T1]{fontenc}
\usepackage{amssymb,amsmath,amstext}
\usepackage{hyperref, graphicx}
\usepackage{comment}
\usepackage{mathtools}

\newtheorem{theor}{Theorem}[section]
\newtheorem{lem}[theor]{Lemma}
\newtheorem{defin}[theor]{Definition}

\newtheorem{prop}[theor]{Proposition} 

\newtheorem{notation}[theor]{Notation}
\newtheorem{exam}[theor]{Example}

\newtheorem{cor}[theor]{Corollary}
\newtheorem{rem}[theor]{Remark}

\newtheorem{assump}[theor]{Assumption}

\numberwithin{equation}{section}

\newcommand{\cl}{\mathrm{cl}}

\newcommand{\mr}{\mathrm}

\newcommand{\es}{\emptyset}

\newcommand{\rank}{\mathrm{rank}}

\newcommand{\nts}{\negthickspace}
\newcommand{\uhrc}{\nts \upharpoonright \nts}

\newcommand{\mcA}{\mathcal{A}}
\newcommand{\mcB}{\mathcal{B}}
\newcommand{\mcC}{\mathcal{C}}

\newcommand{\mcG}{\mathcal{G}}

\newcommand{\mcT}{\mathcal{T}}

\newcommand{\mbB}{\mathbf{B}}

\newcommand{\mbE}{\mathbf{E}}

\newcommand{\mbT}{\mathbf{T}}

\newcommand{\mbW}{\mathbf{W}}
\newcommand{\mbX}{\mathbf{X}}
\newcommand{\mbY}{\mathbf{Y}}

\newcommand{\mbbG}{\mathbb{G}}
\newcommand{\mbbP}{\mathbb{P}}
\newcommand{\mbbN}{\mathbb{N}}
\newcommand{\mbbR}{\mathbb{R}}

\newcommand{\msfC}{\mathsf{C}}
\newcommand{\msfN}{\mathsf{N}}

\newcommand{\rng}{\mathrm{rng}}

\title[Random expansions of trees]{Random expansions of trees with bounded height}

\author{Vera Koponen and Yasmin Tousinejad}

\address{Vera Koponen, Department of Mathematics, Uppsala University, Sweden.}
\email{vera.koponen@math.uu.se}

\address{Yasmin Tousinejad, Department of Mathematics, Uppsala University, Sweden.}
\email{yasmin.tousinejad@math.uu.se}

\date{29 March, 2025}

\begin{document}

\maketitle

\begin{abstract}
We consider a sequence $\mbT = (\mcT_n : n \in \mbbN^+)$ of trees $\mcT_n$ where,
for some $\Delta \in \mbbN^+$ every $\mcT_n$ has height at most $\Delta$ and as $n \to \infty$
the minimal number of children of a nonleaf tends to infinity.
We can view every tree as a (first-order) $\tau$-structure where $\tau$ is a signature with one binary relation symbol.
For a fixed (arbitrary) finite and relational signature $\sigma \supseteq \tau$ we consider the set $\mbW_n$ of
expansions of $\mcT_n$ to $\sigma$ and a probability distribution $\mbbP_n$ on $\mbW_n$ which is determined
by a (parametrized/lifted) Probabilistic Graphical Model (PGM) $\mbbG$ which can use the information given
by $\mcT_n$.

The kind of PGM that we consider uses formulas of a many-valued logic that we call $PLA^*$ 
with truth values in the unit interval $[0, 1]$. We also use $PLA^*$ to express queries, or events, on $\mbW_n$.
With this setup we prove that, under some assumptions on $\mbT$, $\mbbG$, and a 
(possibly quite complex) formula $\varphi(x_1, \ldots, x_k)$
of $PLA^*$, as $n \to \infty$, if $a_1, \ldots, a_k$ are vertices of the tree $\mcT_n$ then 
the value of $\varphi(a_1, \ldots, a_k)$ will, with high probability,
be almost the same as the value of $\psi(a_1, \ldots, a_k)$, where $\psi(x_1, \ldots, x_k)$ is a ``simple''
formula the value of which can always be computed quickly (without reference to $n$), and $\psi$ itself can be found by using only
the information that defines $\mbT$, $\mbbG$ and $\varphi$.
A corollary of this, subject to the same conditions, is a probabilistic convergence law for $PLA^*$-formulas.
\end{abstract}

\section{Introduction}

\noindent
Logical convergence laws have been proved or disproved in various contexts since the pioneering work of
Glebskii et al \cite{GKLT} and, independently, Fagin \cite{Fag}.
Since finite structures in the sense of first-order logic can represent, for example, relational databases, such convergence laws
may have impact outside of mathematics itself.
When a convergence law applies to a sentence of a formal logic one can use random sampling of structures on 
a sufficiently large domain to get an estimate
of the probability of the sentence 
on {\em all} large enough domains and the estimate will, with high probability, be as close to the actual probability as we like
(according to some predetermined error marginal).
Moreover, sometimes a logical convergence law can be proved (as in \cite{GKLT}) by 
showing that quantifiers/aggregations can be removed from formulas step by step to produce a new simpler
(e.g. quantifier-free)
formula which is ``asymptotically equivalent'' to the original one, and the simpler formula can be evaluated 
in time which is independent of the domain size. In theory this gives a deterministic algorithm for estminating the 
probability of a sentence (or formula with some parameters/constants) which is independent of the domain size.
In practice, properties of the (generalized) quantifiers or aggregation functions that are involved in the elimination will of course
influence the computational complexity.

Most studies of logical convergence laws have studied contexts in which all relations are modelled probabilistically;
intuitively speaking they are ``uncertain''.
When building a model for inference we may not only want to take into account
properties and relations which are uncertain,
but also ``background information'' that is certain. 
Background information can have many different forms, but arguably some forms of background information are
more common, or more important, than other forms.
Information that categorizes objects into classes, subclasses and so on is prevalent both in theory and in the physical world.
A related form of information is that of a hierarchical structure among objects.
Both forms of information can be represented by trees: the root represents the class of all objects (or the top of the hierarchy),
the children of the root the classes of the first subdivision into subclasses (or the next level of the hierarchy), and so on.
This is a motivation to consider logical convergence laws for random structures that expand a tree, that is, the tree is fixed but in
addition we have random relations and the probability that such a relation, say $R(a, b)$, holds may depend on the positions of $a$ and $b$
in the tree.

As an example, consider a tree in which the children of the 
root represent some communities, where the children of a community represent subcommunities, and
the children of a subcommunity represent individuals.
On top of the background information, represented by the tree, we may consider information
that is given by (conditional) probabilities and is formally
represented by (parametrized) 0/1-valued (i.e. false/true-valued) random variables. 
So we may have random variables (also viewed as atomic logical formulas) $P_1(x)$, $P_2(x)$, and $P_3(x)$,
representing some property that a community, subcommunity, respectively, individual, may have.
The probability that a subcommunity has the property $P_2$ may (for example) depend on whether the community which it is part of has
property $P_1$. Or alternatively, the probability that a community has $P_1$ may depend on the proportion of its
subcommunities that have $P_2$ (and/or the proportion of individuals in the community that have $P_3$).
We may also have a random variable $R(x, y)$, representing some relationship between objects, for example
the perception (good/bad) that $x$ has of $y$. Now, for an individual $x$, the probability that $R(x, y)$ holds may
depend on properties of the (sub)community that $x$ belongs to and on properties of $y$ and/or its (sub)community; 
it may also depend on whether $x$ and $y$ belong to the same (sub)community.

Such uncertain, or probabilistic, information as exemplified above can be represented by a so-called (para\-met\-rized/lif\-ted) 
{\em Probabilistic Graphical Model (PGM)} based on the background knowledge represented by the tree.
(For introductions or surveys about PGMs, a tool in machine learning and statistical relational artifical intelligence, see for example
\cite{BKNP, DKNP, KMG, Koller}.)
A (parametrized/lifted) PGM consists partly of a directed acyclic graph (DAG), the vertices of which are (parametrized/lifted)
random variables. 
The arcs/arrows of the DAG describes the (in)dependencies between the random variables, by stipulating that
a random variable (a vertex of the DAG), say $R$, is independent from all other vertices of the DAG, {\em conditioned} 
on knowing the states (true/false) of the parents of $R$ in the DAG.
In the setup of this article a parametrized random variable will be identified with a relation symbol in the sense of formal logic.
So in the example above the vertex set of the DAG would be $\{P_1, P_2, P_3, R\}$, 
where $P_1, P_2, P_3, R$ are relation symbols, and its arcs would
describe the (conditional) (in)dependencies between the vertices.

So far, nothing has been said about how to express (conditional) probabilities in a PGM and, indeed, there are
different ways of doing it. 
It turns out that a general way of doing it is to use some sort of ``probability logic'',
as was first done (as far as we know) in the contex of PGMs by Jaeger in \cite{Jae98a}.
Here we will use a logic which we call $PLA^*$, 
for {\em probability logic with aggregation functions}  
(Definition~\ref{syntax of PLA*})
to describe (conditional) probabilities associated to the relation symbols of the DAG of a PGM. (The `$*$' in $PLA^*$ just
indicates that it is a more general variant of earlier versions of $PLA$ considered 
by Koponen and Weitkämper in \cite{KW1, KW2}.)
The resulting type of PGM will be called a {\em $PLA^*$-network}
(Definition~\ref{definition of PLA-network}).

$PLA^*$ is a logic such that the (truth) values of its formulas can be any number in the unit interval $[0, 1]$,
so it is suitable for expressing probabilities.
$PLA^*$ achieves this partly by considering every $c \in [0, 1]$ as an atomic formula, and partly 
by using {\em aggregation functions} which take finite sequences of reals in $[0, 1]$ as input and outputs a 
real number in $[0, 1]$.
An example of an aggregation function is the
arithmetic mean, or average, which returns the average of a finite sequence of reals in the unit interval.
So for example, if $R$ is a unary relation symbol then there is a  $PLA^*$-formula (without free variables) such that 
its value is the proportion of elements in the domain which satisfy $R(x)$.
By using the aggregation functions maximum and minimum one can show that every query on finite structures
which can be expressed by a first-order formula can be expressed by a $PLA^*$-formula
(Lemma~\ref{FO is expressible in PLA*}).
Hence, $PLA^*$ subsumes first-order logic.
For a formula $\varphi(x_1, \ldots, x_k)$ of $PLA^*$, finite structure $\mcA$ (with matching signature/vocabulary),
and $a_1, \ldots, a_k \in A$,
we let $\mcA(\varphi(a_1, \ldots, a_k))$ denote the value of $\varphi(a_1, \ldots, a_k)$ in $\mcA$.

Since we are interested in large trees we let 
$\mbT = (\mcT_n : n \in \mbbN^+)$ be a sequence of trees where the number of vertices of $\mcT_n$ tends to infinity as $n \to \infty$.
Each $\mcT_n$ can be represented by a first-order $\tau$-structure where $\tau = \{E\}$ and $E$ is a binary
relation symbol of arity 2.
Let $\sigma$ be a finite relational signature that includes $\tau$.
In the example above we have $\sigma = \{E, P_1, P_2, P_3, R\}$.
Then let $\mbW_n$ be the set of all expansions of $\mcT_n$ to $\sigma$,
where we think of each such expansion $\mcA$ as a ``possible world'', and let $\mbbP_n$ be the probability distribution
on $\mbW_n$ induced by a $PLA^*(\sigma)$-network $\mbbG$ as defined in 
Definition~\ref{the probability distribution induced by an PLA-network}.

Now we have a precise mathematical setting in which we can speak about the probability of
a query (event) on $\mbW_n$ that can be expressed by a $PLA^*$-formula.
The probability of a query depends (in general) on both the hierarchical information given by $\mcT_n$ and by $\mbbG$.
If we use the brute force way of computing the probability of a query, namely that 
we compute the probability of every structure in $\mbW_n$ in which the query is true
and then add all such probabilities, then the time needed for the computation will be 
exponential in the number of vertices of the
tree $\mcT_n$.
In other words, the brute force method of computing the probability of a query does not scale well to large trees $\mcT_n$.
This motivates the search for other methods of computing, or estimating, the probability of a query in the described context.
We do not expect to get a general convergence (or nonconvergence) result that covers every kind of sequence $\mbT$, so
additional assumptions on $\mbT$ will be imposed. 
Koponen's results in \cite{Kop24} apply to sequences $\mbT$ where for some $\Delta$ and all $n$ each vertex in $\mcT_n$ has at
most $\Delta$ children. In this work we instead assume that for some $\Delta$ and all $n$ the height of $\mcT_n$ is at most $\Delta$.
It will also be assumed that for every $k$ and all large enough $n$, every nonleaf in $\mcT_n$ has at least $k$ children.
Such trees can (for large $n$) describe hierarchies (categorizations) in with few levels compared to the number of objects of each type
in the hierarchy (or few partitions into subcategories compared to the number of objects in each category).
In the above example the height of the tree, is 3 (or 4, depending on our conventions),
but the number of individuals in even the smallest subcommunity that we consider may be much larger than 3.

In this article we identify
\begin{enumerate}
\item certain additional conditions on the sequence of trees  $\mbT = (\mcT_n : n \in \mbbN^+)$
(the stronger Assumption~\ref{properties of the trees} and the milder Assumption~\ref{less properties of the trees}),
\item certain conditions on the $PLA^*(\sigma)$-network  $\mbbG$ used for defining a  probability distribution on $\mbW_n $, 
roughly meaning that all aggregation functions used in the formulas associated to $\mbbG$ satisfy a continuity property, 
\end{enumerate}
such that if 
\begin{itemize}
\item[(3)] $\varphi(x_1, \ldots, x_k)$ (with free variables $x_1, \ldots, x_k$) is a $PLA^*(\sigma)$-formula 
in which all aggregation functions satisfy the continuity property,
\end{itemize}
then there is a simpler, ``closure-basic'',
formula, say $\psi(x_1, \ldots, x_k)$, such that,
for all $n$ and $a_1, \ldots, a_k \in T_n$, 
\begin{itemize}
\item[(a)] the truth value of $\psi(a_1, \ldots, a_k)$ in every structure $\mcA \in \mbW_n$
can be evaluated by using only (the fixed information defining) $\mbT$, $\mbbG$, $\psi$, and
the (bounded number of) ancestors of $a_1, \ldots, a_k$ in $\mcT_n$, and
\item[(b)] for every $\varepsilon > 0$, if $n$ is large enough, then, with probability at least $1 - \varepsilon$
(with the distribution induced by $\mbbG$),
the values of $\varphi(a_1, \ldots, a_k)$ and $\psi(a_1, \ldots, a_k)$ differ by at most $\varepsilon$.
(We will say that $\varphi$ and $\psi$ are {\em asymptotically equivalent}.)
\end{itemize}
This intuitively means that, for large enough $n$, with high probability the value of $\varphi(a_1, \ldots, a_k)$
is approximated (with as good accuracy as we like) by the value of \\
$\psi(a_1, \ldots, a_k)$ for all vertices $a_1, \ldots, a_k$ in the tree $\mcT_n$.
Moreover, under the same conditions (1)--(3),
one can find such $\psi$ by using only $\varphi$, $\mbT$ and $\mbbG$,
so the time needed to compute the approximation is independent of  $n$, that is, of the size of the tree $\mcT_n$.
A corollary of the above result is that (under the same conditions) 
we get a {\em convergence law} for $PLA^*$-formulas, which of course has to be
suitably formulated in the context of many-valued formulas and
trees as ``background'', or ``base'', structures.
All three conditions (1)--(3) are necessary for the conclusions described to hold, as we will show.

Although condition~(1) in the form of Assumption~\ref{properties of the trees}
has a rather technical statement, the general intuition is just that each tree $\mcT_n$ is ``sufficiently homogeneous''
in the sense that for every nonroot vertex $a$ and subtree $\mcT'$ of $\mcT_n$ 
rooted in $a$ there are ``sufficiently many'' siblings $b$ of $a$ such that
the number of subtrees of $\mcT_n$ that are rooted in $b$ and isomorphic to $\mcT'$ is rougly the same
as the number of subtrees of $\mcT_n$ that are rooted in $a$ and isomorphic to $\mcT'$.
For example,
suppose that, in a clinical test say, a group of $n$ persons, where $n$ is large, is first divided into ``many'' groups of roughly the same 
(large) size, then each group is subdivided into roughly equally many subclasses of roughly the same (still large) size, and so on
for, say 5, subdivisions. 
(The divisions may be based on putting people with similar values with respect to various measurements
into the same group.)
Given that $n$, ``many'', and ``large'' are large enough,  the tree $\mcT_n$ 
representing the subdivisions will be of the kind considered in~(1).
More precisely
(according to Assumption~\ref{properties of the trees}),
it suffices that ``many'' and ``large'' means ``growing faster'' than $c \ln n$ for every positive contant $c$.

The main results, stated in Section~\ref{main result}, have the general form described above.
We first prove Theorem~\ref{main result}.
Then we make some observations which allow us to easily obtain some corollaries which are variations of Theorem~\ref{main result}.
The essence of these corollaries is (i) that if we consider stronger assumptions on the $PLA^*(\sigma)$-network $\mbbG$
(i.e. less general probability distributions), then we get results that apply to more queries, 
and (ii) if we loosen the assumptions on the sequence of trees $\mbT$ then we get results that apply to fewer, 
but still interesting, queries.

\medskip

\noindent
{\bf Related work.} 
Instead of bounding the height of the trees, as we do in this article, one could consider trees with a bound on the 
number of children that vertices may have. 
Such sequences of trees are covered by the (more general) context considered by Koponen in \cite{Kop24},
where probability distributions were defined by $PLA^*(\sigma)$-networks and queries by $PLA^*$-formulas.
Other work that we are aware of on logical convergence laws, and related issues, in the setting of 
expansions of ``base structures'' uses first-order logic 
(or $L_{\infty, \omega}^\omega$) as the query language.
Baldwin \cite{Bal} and Shelah \cite{She} considered a base sequence where each $\mcB_n$ is a directed path of length $n$.
In both cases $\mbW_n$ consists of all expansions of $\mcB_n$ to a new binary relation symbol $R$ interpreted as an irreflexive
and symmetric relation.
In \cite{Bal}  the probability distribution on $\mbW_n$ is defined by letting an $R(x, y)$ be true with probability
$n^{-\alpha}$ for some irrational $\alpha \in (0, 1)$, independently of whether it is true for other pairs. 
In \cite{She} the probability of $R(x, y)$ is $d^{-\alpha}$ where $d$ is the distance in $\mcB_n$ between $x$ and $y$
and $\alpha \in (0, 1)$ is irrational.
Lynch \cite[Corollary~2.16]{Lyn} and later
Abu Zaid, Dawar, Grädel and Pakusa \cite{ADGP} and Dawar, Grädel and Hoelzel \cite{DGH} 
considered the context where $\mcB_n$ is a product of finite cyclic groups and the probability distribution considered is the uniform one.
Ahlman and Koponen \cite{AK17} considered base structures $\mcB_n$ with a (nicely behaved)
pregeometry and a probability distribution defined by a kind of stochastic block model where the ``blocks'' correspond to subspaces.
Lynch \cite{Lyn} formulated a condition ($k$-extendibility) that makes sense for base sequences $\mbB$ in general
and which guarantees that a
convergence law holds for first-order logic and the uniform probability distribution.

Besides this article and \cite{Kop24},
other studies that use a PGM for generating  a probability distribution and that prove some result about asymptotic probabilities
do not consider background information in the form of a sequence of nontrivial base structures.
But in the context of no background information, when we let $\sigma$ be a finite and relational signature,
$\mbW_n$ the set of all $\sigma$-structures with domain $\{1, \ldots, n\}$,
$\mbbP_n$ a probability distribution induced by a PGM, there are some studies.
The most closely related to this one are 
\cite{KW1, KW2, KW3} by Koponen and Weitkämper and \cite{Jae98a} by Jaeger.
But we also have \cite{Kop20} by Koponen, \cite{WeiPLP, WeiFLBN} by Weitkämper, and \cite{CM} 
by Cozman and Maua.
The first result about ``asymptotic equivalence'' between complex first-order formulas and simpler formulas was probably
given by Glebskii et. al. \cite{GKLT} and was used to prove the classical zero-one law for first-order logic,
also proved independently by somewhat different methods by Fagin \cite{Fag}.
Since the pioneering work in \cite{GKLT} and \cite{Fag} a large number of studies on logical convergence laws and related
issues have been conducted in the fields of finite model theory and probabilistic combinatorics, but those studies consider
less flexible ways of generating probability distributions than PGMs (most often the uniform distribution).

The present work and \cite{KW1, KW2, Kop24} consider similar formalisms for inducing probability and for defining queries/events,
so we point out some (unavoidable) differences.
In \cite{KW1, KW2} all relations are uncertain (probabilistically modelled) and every formula to which the main results(s) 
applies is asymptotically equivalent to a formula without aggregation functions and without quantifiers.
In the present work and in \cite{Kop24} which also considers deterministic background information, 
represented by ``base structures'',
it is only shown that every formula to which some of the main results applies is asymptotically equivalent to a 
formula which may use (only)  ``{\em local}'' aggregations/quantifications where the notion of ``locality'' is 
adapted to the kind of base structures considered, which in the present work are trees.
It is not possible to remove the use of local aggregations/quantifications in this work (or in \cite{Kop24}) 
because aggregation/quantifier-free formulas cannot describe the structure of trees (or the base structures considered in \cite{Kop24}).
Besides this difference compared with \cite{KW1, KW2}, another difference is that in the present work (and \cite{Kop24})
we only asymptotically eliminate aggregations that range over elements, or tuples of elements, 
that satisfy some constraints with respect to the underlying base structure. 
It seems that one can not in general remove this assumption
because probabilities may depend on the base structure.

\medskip

\noindent
{\bf Organization.}
Section~\ref{Preliminaries} specifies some notation and terminology that will be used, as well as 
a couple of probability theoretic results.
Section~\ref{Probability logic with aggregation functions}
defines the formal logic $PLA^*$, and some related notions, that will be used in PGMs and for defining queries (events).
Section~\ref{A general method} describes a general method (introduced in \cite{KW3})
for ``asymptotically eliminating aggregation functions''.
Section~\ref{Section about trees} 
gives the necessary background about directed acyclic graphs and trees,
and some other notions related to trees, that will be used in the rest of the article.
Section~\ref{The base sequence of trees} 
specifies the assumptions on the sequence of trees $\mbT = (\mcT_n : n \in \mbbN^+)$ that we make
and also explains why we make these assumptions. 
It then defines the notion of a $PLA^*$-network, which is the kind of PGM that we use, and how it induces a 
probability distribution on the set $\mbW_n$ of all expansions to $\sigma$ of $\mcT_n$.
In section~\ref{Convergence and balance} we prove technical results that imply that the assumptions of
Theorem~\ref{general asymptotic elimination}
in Section~\ref{A general method} are satisfied.
Therefore, we can, in the last subsection of Section~\ref{Convergence and balance},
use Theorem~\ref{general asymptotic elimination} 
and the results from earlier subsections of Section~\ref{Convergence and balance}
to prove a result about ``asymptotic elimination of aggregation functions''.
In Section~\ref{The main results} we use induction on the height of the $PLA^*$-network 
and the results from Section~\ref{Convergence and balance} to prove our main results.

\section{Preliminaries}\label{Preliminaries}

\noindent
Structures in the sense of first-order logic are denoted by calligraphic letters $\mcA, \mcB, \mcC, \ldots$ and their
domains (universes) by the corresponding noncalligraphic letter $A, B, C, \ldots$.
Finite sequences (tuples) of objects are denoted by $\bar{a}, \bar{b}, \ldots, \bar{x}, \bar{y}, \ldots$.
We usually denote logical variables by $x, y, z, u, v, w$. 
Unless stated otherwise, when $\bar{x}$ is a sequence of variables we assume that $\bar{x}$ does not repeat a variable.
But if $\bar{a}$ denotes a sequence of elements from the domain of a structure then repetitions may occur.

We let $\mbbN$ and $\mbbN^+$ denote the set of nonnegative integers and the set of positive integers, respectively.
For a set $S$, $|S|$ denotes its cardinality, and for a finite sequence $\bar{s}$, $|\bar{s}|$ denotes its length
and $\rng(\bar{s})$ denotes the set of elements in $\bar{s}$.
For a set $S$, $S^{<\omega}$ denotes the set of finite nonempty sequences (where repetitions are allowed) of elements from $S$,
so $S^{<\omega} = \bigcup_{n\in \mbbN^+}S^n$.
In particular, $[0, 1]^{<\omega}$ denotes the set of all finite nonempty sequences of reals from the
unit interval $[0, 1]$.

A signature (vocabulary) is called {\em finite (and) relational} if it is finite and contains only relation symbols.
Let $\sigma$ be a signature and let $\mcA$ be a $\sigma$-structure.
If $\tau \subseteq \sigma$ then $\mcA \uhrc \tau$ denotes the {\em reduct} of $\mcA$ to $\tau$.
If $B \subseteq A$ then $\mcA \uhrc B$ denotes the {\em substructure} of $\mcA$ generated by $B$.
If $R \in \sigma$ is a relation symbol then $R^\mcA$ denotes the interpretation of $R$ in $\mcA$.

A random variable will be called {\em binary} if it can only take the value $0$ or $1$.
The following is a direct consequence of \cite[Corollary~A.1.14]{AlonSpencer} which in turn follows from the
Chernoff bound \cite{Chernoff}:

\begin{lem}\label{independent bernoulli trials}
Let $Z$ be the sum of $n$ independent binary random variables, each one with probability $p$ of having the value 1,
where $p > 0$.
For every $\varepsilon > 0$ there is $c_\varepsilon > 0$, depending only on $\varepsilon$, such that the probability that
$|Z - pn| > \varepsilon p n$ is less than $2 e^{-c_\varepsilon p n}$.
(If $p = 0$ then the same statement holds if `$2 e^{-c_\varepsilon p n}$' is replaced by (for example) `$e^{-n}$'.)
\end{lem}

\noindent
The following is a straightforward corollary (proved in \cite{KW2}):

\begin{cor}\label{independent bernoulli trials, second version}
Let $p \in [0, 1]$ and let $\varepsilon > 0$.
Let $Z$ be the sum of $n$ independent binary random variables $Z_1, \ldots, Z_n$, where for each $i = 1, \ldots, n$ the
probability that $Z_i$ equals 1 belongs to the interval $[p-\varepsilon, p+\varepsilon]$.
Then there is $c > 0$, depending only on $p$ and $\varepsilon$, such that the probability that
$Z > (1 + \varepsilon)(p + \varepsilon) n$ or 
$Z < (1 - \varepsilon)(p - \varepsilon) n$ 
is less than $2 e^{-c n}$.
\end{cor}

\noindent
The following lemma follows easily from the definition of conditional probability.

\begin{lem}\label{basic fact about conditional probabilities}
Suppose that $\mbbP$ is a probability measure on a set $\Omega$.
Let $X \subseteq \Omega$ and $Y \subseteq \Omega$ be measurable.
Also suppose that $Y = Y_1 \cup \ldots \cup Y_k$, $Y_i \cap Y_j = \es$ if $i \neq j$, and that each $Y_i$ is measurable.
If $\alpha \in [0, 1]$, $\varepsilon > 0$, and
$\mbbP(X \ | \ Y_i) \in [\alpha - \varepsilon, \alpha + \varepsilon]$ for all $i = 1, \ldots, k$,
then $\mbbP(X \ | \ Y) \in [\alpha - \varepsilon, \alpha + \varepsilon]$.
\end{lem}

\section{Probability logic with aggregation functions}\label{Probability logic with aggregation functions}

\noindent
We consider a logic that we call {\em probability logic with aggregation functions}, or $PLA^*$, where
$PLA^*$ is a more general version of $PLA$ and $PLA^+$
which were considered in \cite{KW1} and
\cite{KW2}, respectively, and of the {\em probability logic} considered by Jaeger in \cite{Jae98a, Jae98b}.
$PLA^*$ is a logic with (truth) values in the unit interval $[0, 1]$.
With $PLA^*$ we can, for example, express the proportion of all elements in a domain that
have some property, and the proportion need not be 0 or 1.
As a nontrivial example of what $PLA^*$ can express,
Example~\ref{example of page rank} shows that the PageRank can be expressed with $PLA^*$.
On finite structures, all queries that can be expressed by first-order logic can also be expressed by $PLA^*$
(as stated by Lemma~\ref{FO is expressible in PLA*}).

Recall that $[0, 1]^{<\omega}$ denotes the set of all finite nonempty sequences of reals in 
the unit interval $[0, 1]$.

\begin{defin}\label{definition of connective and aggregation function} {\rm
Let $k \in \mbbN^+$.\\
(i) A function $\msfC : [0, 1]^k \to [0, 1]$ will also be called a {\bf \em ($k$-ary) connective}.\\
(ii) A function $F : \big([0, 1]^{<\omega}\big)^k \to [0, 1]$ which is symmetric in the following sense
will be called a {\bf \em ($k$-ary) aggregation function}:
if $\bar{p}_1, \ldots, \bar{p}_k \in [0, 1]^{<\omega}$ and, for $i = 1, \ldots, k$,
$\bar{q}_i$ is a reordering of the entries in $\bar{p}_i$,
then $F(\bar{q}_1, \ldots, \bar{q}_k) = F(\bar{p}_1, \ldots, \bar{p}_k)$.
}\end{defin}

\noindent
In the definition below we extend the usual 0/1-valued connectives according to the semantics of 
Lukasiewicz logic 
(see for example \cite[Section~11.2]{Ber}, or \cite{LT})
to get connectives which are continous as functions from $[0, 1]$, or from $[0, 1] \times [0, 1]$, to $[0, 1]$.

\begin{defin}\label{special connectives} {\rm
Let
\begin{enumerate}
\item $\neg : [0, 1] \to [0, 1]$ be defined by $\neg(x) = 1 - x$,
\item $\wedge : [0, 1]^2 \to [0, 1]$ be defined by $\wedge(x, y) = \min(x, y)$,
\item $\vee : [0, 1]^2 \to [0, 1]$ be defined by $\vee(x, y) = \max(x, y)$, and
\item $\rightarrow : [0, 1]^2 \to [0, 1]$ be defined by $\rightarrow(x, y) = \min(1, \ 1 - x + y)$.
\end{enumerate}
}\end{defin}

\noindent
The following aggregation functions are useful:

\begin{defin}\label{examples of aggregation functions} {\rm
For $\bar{p} = (p_1, \ldots, p_n) \in [0, 1]^{<\omega}$, define
\begin{enumerate}
\item $\max(\bar{p})$ to be the {\em maximum} of $p_1, \ldots, p_n$,
\item $\min(\bar{p})$ to be the {\em minimum} of $p_1, \ldots, p_n$,
\item $\mr{am}(\bar{p}) = (p_1 + \ldots + p_n)/n$, so `am' is the {\em arithmetic mean}, or {\em average}, 
\item $\mr{gm}(\bar{p}) = \big(\prod_{i=1}^n p_i\big)^{(1/n)}$, so `gm' is the {\em geometric mean}, and
\item for every $\beta \in (0, 1]$, $\mr{length}^{-\beta}(\bar{p}) = |\bar{p}|^{-\beta}$.
\end{enumerate}
}\end{defin}

\noindent
The aggregation functions above take only one sequence from $[0, 1]^{<\omega}$ as input.
But there are useful aggregation functions of higher arities, i.e. taking two or more sequences as input,
as shown in Examples~5.5 -- 5.7 in \cite{KW1} and in Example~6.4 in \cite{KW1}.

\medskip

\noindent
{\bf \em For the rest of this section we fix a finite and relational signature $\sigma$.}

\begin{defin}\label{syntax of PLA*}{\bf (Syntax of $PLA^*$)} {\rm 
We define the formulas of $PLA^*(\sigma)$, as well as the set of free variables  of a formula $\varphi$, 
denoted $Fv(\varphi)$, as follows:
\begin{enumerate}
\item  For each $c \in [0, 1]$, $c \in PLA^*(\sigma)$ (i.e. $c$ is a formula) and $Fv(c) = \es$. 
We also let $\bot$ and $\top$
denote $0$ and $1$, respectively.

\item For all variables $x$ and $y$, `$x = y$' belongs to $PLA^*(\sigma)$ and $Fv(x = y) = \{x, y\}$.

\item For every $R \in \sigma$, say of arity $r$, and any choice of variables $x_1, \ldots, x_r$, $R(x_1, \ldots, x_r)$ belongs to 
$PLA^*(\sigma)$ and  $Fv(R(x_1, \ldots, x_r)) = \{x_1, \ldots, x_r\}$.

\item If $k \in \mbbN^+$, $\varphi_1, \ldots, \varphi_k \in PLA^*(\sigma)$ and
$\msfC : [0, 1]^k \to [0, 1]$ is a continuous connective, then 
$\msfC(\varphi_1, \ldots, \varphi_k)$ is a formula of $PLA^*(\sigma)$ and
its set of free variables is $Fv(\varphi_1) \cup \ldots \cup Fv(\varphi_k)$.

\item Suppose that $\varphi_1, \ldots, \varphi_k \in PLA^*(\sigma)$,
$\chi_1, \ldots, \chi_k \in PLA^*(\sigma)$,
$\bar{y}$ is a sequence of distinct variables,
and that $F : \big( [0, 1]^{<\omega} \big)^k \to [0, 1]$ is an aggregation function.
Then 
\[
F(\varphi_1, \ldots, \varphi_k : \bar{y} : 
\chi_1, \ldots, \chi_k)
\]
is a formula of $PLA^*(\sigma)$ and its set of free variables is
\[
\bigg( \bigcup_{i=1}^k Fv(\varphi_i)\bigg) \setminus \rng(\bar{y}),
\] 
so this construction binds the variables in $\bar{y}$.
The construction $F(\varphi_1, \ldots, \varphi_k : \bar{y} : \chi_1, \ldots, \chi_k)$ will be called an
{\bf \em aggregation (over $\bar{y}$)} and the formulas $\chi_1, \ldots, \chi_k$ are called the {\bf \em conditioning formulas} of this aggregation.
\end{enumerate}
}\end{defin}

\begin{defin}\label{definition of subformula} {\rm
(i) A formula in $PLA^*(\sigma)$ without free variables is called a {\bf \em sentence}.\\
(ii) In part~(4) of Definition~\ref{syntax of PLA*} the formulas $\varphi_1, \ldots, \varphi_k$
are called {\bf \em subformulas} of $\msfC(\varphi_1, \ldots, \varphi_k)$.\\
(iii) In part~(5) of Definition~\ref{syntax of PLA*} the formulas $\varphi_1, \ldots, \varphi_k$ and
$\chi_1, \ldots, \chi_k$ are called {\bf \em subformulas} of
$F(\varphi_1, \ldots, \varphi_k : \bar{y} : \chi_1, \ldots, \chi_k)$.
We also call $\chi_1, \ldots, \chi_k$ {\bf \em conditioning subformulas} of the formula
$F(\varphi_1, \ldots, \varphi_k : \bar{y} : \chi_1, \ldots, \chi_k)$, and 
we say that (this instance of the aggregation) $F$ {\bf \em is conditioned on $\chi_1, \ldots, \chi_k$}.\\
(iv) We stipulate the following {\em transitivity properties}:
If $\psi_1$ is a subformula of $\psi_2$ and $\psi_2$ is a subformula of $\psi_3$,
then $\psi_1$ is a subformula of $\psi_3$. 
If $\psi_1$ is a conditioning subformula of $\psi_2$ and $\psi_2$ is a subformula of $\psi_3$, then $\psi_1$
is a conditioning subformula of $\psi_3$.
}\end{defin}

\begin{notation}\label{abbreviation when using aggregation functions}{\rm
When denoting a formula in $PLA^*(\sigma)$ by for example $\varphi(\bar{x})$ then we assume that $\bar{x}$ is a sequence
of different variables and that every {\em free} variable in the formula denoted by $\varphi(\bar{x})$ belongs to $\rng(\bar{x})$
(but we do not require that every variable in $\rng(\bar{x})$ actually occurs as a free variable in the formula).
If a formula is denoted by $\varphi(\bar{x}, \bar{y})$ then we assume that $\rng(\bar{x}) \cap \rng(\bar{y}) = \es$.
}\end{notation}

\begin{defin}\label{definition of literal} {\rm
The $PLA^*(\sigma)$-formulas described in parts~(2) and~(3) of 
Definition~\ref{syntax of PLA*}
are called {\bf \em first-order atomic $\sigma$-formulas}.
A $PLA^*(\sigma)$-formula is called a {\bf \em $\sigma$-literal} if it has the form $\varphi(\bar{x})$
or $\neg\varphi(\bar{x})$, where $\varphi(\bar{x})$ is a first-order atomic formula and
$\neg$ is like in Definition~\ref{special connectives}
(so it corresponds to negation when truth values are restricted to 0 and 1).
}\end{defin}

\begin{defin}\label{semantics of PLA*}{\bf (Semantics of $PLA^*$)} {\rm
For every $\varphi \in PLA^*(\sigma)$ and every sequence of distinct variables $\bar{x}$ such that 
$Fv(\varphi) \subseteq \rng(\bar{x})$ we associate a mapping from pairs $(\mcA, \bar{a})$,
where $\mcA$ is a finite $\sigma$-structure and $\bar{a} \in A^{|\bar{x}|}$, to $[0, 1]$.
The number in $[0, 1]$ to which $(\mcA,\bar{a})$ is mapped is denoted $\mcA(\varphi(\bar{a}))$
and is defined by induction on the complexity of formulas, as follows:
\begin{enumerate}
\item If $\varphi(\bar{x})$ is a constant $c$ from $[0, 1]$, then $\mcA(\varphi(\bar{a})) = c$.

\item If $\varphi(\bar{x})$ has the form $x_i = x_j$, then $\mcA(\varphi(\bar{a})) = 1$ if $a_i = a_j$,
and otherwise $\mcA(\varphi(\bar{a})) = 0$.

\item For every $R \in \sigma$, of arity $r$ say, if $\varphi(\bar{x})$ has the form $R(x_{i_1}, \ldots, x_{i_r})$,
then $\mcA(\varphi(\bar{a})) = 1$ if $\mcA \models R(a_{i_1}, \ldots, a_{i_r})$
(where `$\models$' has the usual meaning
of first-order logic), and otherwise $\mcA(\varphi(\bar{a})) = 0$.

\item If $\varphi(\bar{x})$ has the form $\msfC(\varphi_1(\bar{x}), \ldots, \varphi_k(\bar{x}))$,
where $\msfC : [0, 1]^k \to [0, 1]$ is a continuous connective, then
\[
\mcA\big(\varphi(\bar{a})\big) \ = \ 
\msfC\big(\mcA(\varphi_1(\bar{a})), \ldots, \mcA(\varphi_k(\bar{a}))\big).
\]

\item Suppose that $\varphi(\bar{x})$ has the form 
\[
F(\varphi_1(\bar{x}, \bar{y}), \ldots, \varphi_k(\bar{x}, \bar{y}) : \bar{y} : 
\chi_1(\bar{x}, \bar{y}), \ldots, \chi_k(\bar{x}, \bar{y}))
\] 
where $\bar{x}$ and $\bar{y}$ are sequences of distinct variables, $\rng(\bar{x}) \cap \rng(\bar{y}) = \es$, and
$F : \big( [0, 1]^{<\omega} \big)^k \to [0, 1]$ is an aggregation function.
If, for every $i = 1, \ldots, k$, the set 
$\{\bar{b} \in A^{|\bar{y}|} : \mcA(\chi_i(\bar{a}, \bar{b})) = 1\}$ is nonempty, then
let 
\[
\bar{p}_i = 
\big(\mcA\big(\varphi_i(\bar{a}, \bar{b})\big) : \bar{b} \in A^{|\bar{y}|} \text{ and } 
\mcA\big(\chi_i(\bar{a}, \bar{b})\big) = 1\big)
\]
and 
\[
\mcA\big(\varphi(\bar{a})\big) = F(\bar{p}_1, \ldots, \bar{p}_k).
\]
Otherwise let $\mcA\big(\varphi(\bar{a})\big) = 0$.
\end{enumerate}
}\end{defin}

\begin{defin}\label{definition of equivalence} {\rm
Let $\varphi(\bar{x}), \psi(\bar{x}) \in PLA^*(\sigma)$. We say that $\varphi$ and $\psi$
are {\bf \em equivalent} if for every finite $\sigma$-structure $\mcA$ and every $\bar{a} \in A^{|\bar{x}|}$,
$\mcA(\varphi(\bar{a})) = \mcA(\psi(\bar{a}))$.
}\end{defin}

\begin{notation}{\rm
(i) For any formula $\varphi(\bar{x}, \bar{y}) \in PLA^*(\sigma)$, finite $\sigma$-structure $\mcA$ and $\bar{a} \in A^{|\bar{x}|}$, let
\[
\varphi(\bar{a}, \mcA) = \{\bar{b} \in A^{|\bar{y}|} : \mcA(\varphi(\bar{a}, \bar{b})) = 1 \}.
\]
(ii) Let $\varphi(\bar{x}) \in PLA^*(\sigma)$, let $\mcA$ be a finite $\sigma$-structure, and let $\bar{a} \in A^{|\bar{x}|}$.
Then `$\mcA \models \varphi(\bar{a})$' means the same as `$\mcA(\varphi(\bar{a})) = 1$'.\\
(iii) Let $\varphi(\bar{x}), \psi(\bar{x}) \in PLA^*(\sigma)$. 
When writing `$\varphi(\bar{x}) \models \psi(\bar{x})$'
we mean that for every finite $\sigma$-structure $\mcA$
and all $\bar{a} \in A^{|\bar{x}|}$, if $\mcA \models  \varphi(\bar{a})$ then $\mcA \models \psi(\bar{a})$.
}\end{notation}

\begin{defin}\label{L-basic subformula} {\rm 
(i) A formula in $PLA^*(\sigma)$ such that no aggregation function occurs in it is called {\bf \em aggregation-free}.\\
(ii) If $\varphi(\bar{x}) \in PLA^*(\sigma)$ and 
for every finite $\sigma$-structure $\mcA$,
 and every $\bar{a} \in A^{|\bar{x}|}$,
$\mcA(\varphi(\bar{a}))$ is 0 or 1, then we call $\varphi(\bar{x})$ {\bf \em $0/1$-valued}.\\
(iii) If $L \subseteq PLA^*(\sigma)$ and every formula in $L$ is 0/1-valued, then 
we say that $L$ is {\bf \em $0/1$-valued}. \\
(iv) Let $L \subseteq PLA^*(\sigma)$. 
A formula of $PLA^*(\sigma)$ is called an {\bf \em $L$-basic formula} if it has the form
$\bigwedge_{i=1}^k \big(\varphi_i(\bar{x}) \to c_i\big)$ where $\varphi_i(\bar{x}) \in L$ 
and $c_i \in [0, 1]$ for all $i = 1, \ldots, k$.
(We will only use this notion when $L$ is $0/1$-valued.)
}\end{defin}

\begin{exam}\label{example of page rank} {\rm
We exemplify what can be expressed with $PLA^*(\sigma)$, provided that it
contains a binary relation symbol, with the notion of PageRank \cite{BP}.
The PageRank of an internet site can be approximated in ``stages'' as follows
(if we supress the ``damping factor'' for simplicity), where $IN_x$ is the set of sites that link to $x$,
and $OUT_y$ is the set of sites that $y$ links to:
\begin{align*}
&PR_0(x) = 1/N \text{ where $N$ is the number of sites,} \\
&PR_{k+1}(x) = \sum_{y \in IN_x} \frac{PR_k(y)}{|OUT_y|}.
\end{align*}
It is not difficult to prove, by induction on $k$, that for every $k$ the sum of all $PR_k(x)$ as $x$ ranges over all sites
is 1. Hence the sum in the definition of $PR_{k+1}$ is less or equal to 1 (and this will matter later).
Let $E \in \sigma$ be a binary relation symbol representing a link.
Then $PR_0(x)$ is expressed by the $PLA^*(\sigma)$-formula $\mr{length}^{-1}(x = x : y : \top)$.

Suppose that $PR_k(x)$ is expressed by $\varphi_k(x) \in PLA^*(\sigma)$.
Note that multiplication is a continuous connective from $[0, 1]^2$ to $[0, 1]$ so it can be used in $PLA^*(\sigma)$-formulas.
Then observe that the quantity $|OUT_y|^{-1}$ is expressed by the $PLA^*(\sigma)$-formula
\[
\mr{length}^{-1}\big(y=y : z : E(y, z)\big)
\]
which we denote by $\psi(y)$.
Let $\mr{tsum} : [0, 1]^{<\omega} \to [0, 1]$ be the ``truncated sum'' defined by letting
$\mr{tsum}(\bar{p})$ be the sum of all entries in $\bar{p}$ if the sum is at most 1, and otherwise $\mr{tsum}(\bar{p}) = 1$.
Then $PR_{k+1}(x)$ is expressed by the $PLA^*(\sigma)$-formula
\[
\mr{tsum}\big(x = x \wedge (\varphi_k(y) \cdot \psi(y)) : y : E(y, x)\big).
\]
With $PLA^*(\sigma)$ we can also define all stages of the SimRank \cite{JW} in a simpler way than done
in \cite{KW1} with the sublogic $PLA(\sigma) \subseteq PLA^*(\sigma)$.
}\end{exam}

\begin{lem}\label{FO is expressible in PLA*}
Suppose that $\varphi(\bar{x})$ is a first-order formula over $\sigma$.
Then there is a 0/1-valued $\psi(\bar{x}) \in PLA^*(\sigma)$ such that for every finite $\sigma$-structure $\mcA$
and every $\bar{a} \in A^{|\bar{x}|}$, $\mcA \models \varphi(\bar{a})$ if and only if $\mcA(\psi(\bar{a})) = 1$.
\end{lem}

\noindent
{\bf Proof.}
We argue by induction on the complexity of $\varphi$.
If it is atomic then the conclusion follows from parts~(2) and~(3) of the syntax and semantics of $PLA^*$.
The inductive step for the connectives of first-order logic follows since in $PLA^*$ we can use their extensions 
to the unit interval as defined
in Definition~\ref{special connectives}.

Now suppose that $\varphi(\bar{x})$ has the form $\exists y \varphi(\bar{x}, y)$.
By the induction hypothesis, there is a 0/1-valued $\psi(\bar{x}, y) \in PLA^*(\sigma)$
such that for all finite $\sigma$-structures $\mcA$, all $\bar{a} \in A^{|\bar{x}|}$, and $b \in A$,
$\mcA \models \varphi(\bar{a}, b)$ if and only if $\mcA(\psi(\bar{a}, b)) = 1$. 
Let $\mcA$ be a finite $\sigma$-structure and $\bar{a} \in A^{|\bar{x}|}$. Then
\[
\mcA \models \exists y \varphi(\bar{a}, y) \text{ if and only if } 
\mcA\big(\max(\psi(\bar{a}, y) : y : \top)\big) = 1
\]
where $\max(\psi(\bar{a}, y) : y : \top)$ is 0/1-valued.
In a similar way, the aggregation function min can play the role of $\forall$.
\hfill $\square$

\begin{notation}\label{using first-order notation} {\bf (Using $\exists$ and $\forall$ as abbreviations)} {\rm
Due to Lemma~\ref{FO is expressible in PLA*}, if $\varphi(\bar{x}, \bar{y}) \in PLA^*(\sigma)$ 
is a 0/1-valued formula
then we will often
write
`$\exists \bar{y} \varphi(\bar{x}, \bar{y})$' to mean the same as
`$\max(\varphi(\bar{x}, \bar{y}) : \bar{y} : \top)$', and
`$\forall \bar{y} \varphi(\bar{x}, \bar{y})$' to mean the same as
`$\min(\varphi(\bar{x}, \bar{y}) : \bar{y} : \top)$'.
}\end{notation}

\noindent
The next basic lemma has an analog in first-order logic and is proved straightforwardly by induction on the complexity of 
$PLA^*$-formulas.

\begin{lem}\label{truth values only depend on the relation symbols used}
Suppose that $\sigma' \subseteq \sigma$,
$\varphi(\bar{x}) \in PLA^*(\sigma')$, $\mcA$ is a finite $\sigma$-structure, $\mcA' = \mcA \uhrc \sigma'$,
and $\bar{a} \in A^{|\bar{x}|}$.
Then $\mcA(\varphi(\bar{a})) = \mcA'(\varphi(\bar{a}))$.
\end{lem}

\section{A general method for asymptotic elimination of aggregation functions}\label{A general method}

\noindent
{\bf \em Let $\sigma$ be a finite and relational signature.
 In all of this section let $D_n$, $n \in \mbbN^+$, be finite sets such that $\lim_{n\to\infty}|D_n| = \infty$, and
let $\mbW_n$ be a set of $\sigma$-structures with domain $D_n$.}
We will describe a method for ``asymptotically eliminating aggregation functions'' 
which is studied in more detail by Koponen and Weitkämper in \cite{KW3}.
The definitions below and 
Theorem~\ref{general asymptotic elimination}
come from \cite{KW3} where
Theorem~\ref{general asymptotic elimination} will be used later 
(in Section~\ref{Asymptotic elimination of aggregation functions}) to prove our main results.

\begin{defin}\label{definition of asymptotic probability distribution}{\rm
By a {\bf \em sequence of probability distributions (for $(\mbW_n : n \in \mbbN^+)$)} we mean a sequence
$\mbbP = (\mbbP_n : n \in \mbbN^+)$ such that for every $n$, $\mbbP_n$ is a probability distribution on $\mbW_n$.
}\end{defin}

\begin{defin}\label{definition of asymptotically equivalent formulas} {\rm
Let $\varphi(\bar{x}), \psi(\bar{x}) \in  PLA^*(\sigma)$ and 
let $\mbbP = (\mbbP_n : n \in \mbbN^+)$ be a sequence of probability distributions.
We say that $\varphi(\bar{x})$ and $\psi(\bar{x})$ are {\bf \em asymptotically equivalent (with respect to $\mbbP$)} 
if for all $\varepsilon > 0$
\[
\lim_{n\to\infty} \mbbP_n\Big(\big\{\mcA \in \mbW_n : \text{ for all $\bar{a} \in (D_n)^{|\bar{x}|}$, 
$|\mcA(\varphi(\bar{a})) - \mcA(\psi(\bar{a}))| \leq \varepsilon$}\big\} \Big) \ = \ 1.
\]
}\end{defin}

\noindent
{\bf \em For the rest of this section we fix a sequence $\mbbP = (\mbbP_n : n \in \mbbN^+)$ of probability distributions.}

\medskip

\noindent
To define the notions of continuity that we will use we need the notion of
{\em convergence testing sequence} which generalizes a similar
notion used by Jaeger in \cite{Jae98a}.

\begin{defin}\label{definition of convergence testing} {\rm 
(i) A sequence $\bar{p}_n \in [0, 1]^{<\omega}$, $n \in \mbbN$,  is called {\bf \em convergence testing for parameters} 
$c_1, \ldots, c_k \in [0,1]$ and $\alpha_1, \ldots \alpha_k  \in  [0,1]$ if the following hold, 
where $p_{n,i}$ denotes the $i$th entry of $\bar{p}_n$:
\begin{enumerate}
\item $|\bar{p}_n| < |\bar{p}_{n+1}|$ for all $n \in \mbbN$.
\item For every disjoint family of open (with respect to the induced topology on $[0, 1]$)
intervals $I_1, \ldots I_k \subseteq [0,1]$ 
 such that $c_i \in I_i$ for each $i$, 
there is an $ N \in \mbbN$ such that $\mathrm{rng}(\bar{p}_n) \subseteq \bigcup\limits_{j=1}^{k} I_j$ for all $n \geq N$, 
and for every $j \in \{1, \ldots, k \}$, 
\[
\lim\limits_{n \rightarrow \infty} \frac{\left| \{ i  : p_{n,i} \in I_j \} \right| }{|\bar{p}_n|} = \alpha_j.
\]
\end{enumerate}   
(ii) More generally, a sequence of $m$-tuples 
$(\bar{p}_{1, n}, \ldots, \bar{p}_{m, n}) \in  \big([0, 1]^{<\omega}\big)^m$, $n \in \mbbN$,  
is called {\bf \em convergence testing for parameters} $c_{i,j} \in [0,1]$ and $\alpha_{i,j} \in [0,1]$, 
where $i \in \{1, \ldots, m\}$, $j \in \{ 1, \ldots, k_i \}$ and $k_1, \ldots k_m \in \mbbN^+$, if  for every fixed 
$i \in \{1, \ldots, m \}$
the sequence $\bar{p}_{i, n}$, $n \in \mbbN$, is convergence testing for $c_{i,1}, \ldots, c_{i, k_i}$, and 
$\alpha_{i,1}, \ldots, \alpha_{i, k_i}$.
}\end{defin}

\begin{defin} \label{definition of ct-continuous} {\rm 
An aggregation function \\ 
$F : \big([0, 1]^{<\omega}\big)^m \to [0, 1]$ is called 
{\bf \em ct-continuous (convergence test continuous)} with respect to the sequence of parameters
$c_{i,j}, \alpha_{i,j} \in [0, 1]$, $i = 1, \ldots, m$, $j = 1, \ldots, k_i$, 
if the following condition holds:
\begin{enumerate}
\item[] For all convergence testing sequences of $m$-tuples
$(\bar{p}_{1, n}, \ldots, \bar{p}_{m, n}) \in  \big([0, 1]^{<\omega}\big)^m$, $n \in \mbbN$,
and $(\bar{q}_{1, n}, \ldots, \bar{q}_{m, n}) \in  \big([0, 1]^{<\omega}\big)^m$, $n \in \mbbN$,
with the same parameters $c_{i,j}, \alpha_{i,j} \in [0, 1]$, 
$\underset{n \rightarrow \infty}{\lim}  
|F(\bar{p}_{1, n}, \ldots, \bar{p}_{m, n}) - F(\bar{q}_{1, n}, \ldots, \bar{q}_{m, n})| = 0$.
\end{enumerate}
}\end{defin}

\begin{defin}\label{definition of strong admissibility} {\rm
Let $F : \big([0, 1]^{<\omega}\big)^m \to [0, 1]$.\\
(i) We call $F$ {\bf \em continuous} (or {\bf \em strongly admissible}) if $F$ is ct-continuous with respect to {\em every} choice
of parameters $c_{i, j}, \alpha_{i, j} \in [0, 1]$, $i = 1, \ldots, m$ and $j = 1, \ldots, k_i$ (for arbitrary $m$ and $k_i$).\\
(ii) We call $F$ {\bf \em admissible} if $F$ is ct-continuous with respect to every choice of parameters
$c_{i, j}, \alpha_{i, j} \in [0, 1]$, $i = 1, \ldots, m$ and $j = 1, \ldots, k_i$ (for arbitrary $k_i$)
{\em such that $\alpha_{i, j} > 0$ for all $i$ and $j$}.
}\end{defin}

\begin{exam}\label{example of continuous aggregation functions} {\rm
The aggregation functions am, gm and $\mr{length}^{-\beta}$ are continuous,
which is proved in \cite{KW1} in the case of am and gm.
In the case of $\mr{length}^{-\beta}$ the claim is easy to prove.
The aggregation functions max and min are admissible (which is proved in \cite{KW1}) but not continuous.
To see that max is not continuous, consider, for $n \in \mbbN$, $\bar{p}_n$ and $\bar{q}_n$, both of length $n+1$,
where all entries of $\bar{p}_n$ are 0, the first entry of $\bar{q}_n$ is 1 and the rest of the entries are 0.
Let $c_1 = 0$, $c_2 = 1$, $\alpha_1 = 1$, and $\alpha_2 = 0$.
It is straightforward to verify that both $(\bar{p}_n : n \in \mbbN)$ and $(\bar{q}_n : n \in \mbbN)$ are convergence
testing with parameters $c_1, c_2$ and $\alpha_1, \alpha_2$.
But clearly $\max(\bar{p}_n) = 0$ and $\max(\bar{q}_n) = 1$ for all $n$, so 
$|\max(\bar{p}_n) - \max(\bar{q}_n)|$ does not tend to 0 as $n \to \infty$.

The aggregation function $\text{noisy-or}((p_1, \ldots, p_n)) = 1 - \prod_{i=1}^n(1 - p_n)$ is not even admissible
(which is not hard to prove).
For more examples of admissible, or even continuous, aggregation functions (of higher arity) see
Example~6.4 and Proposition~6.5 in \cite{KW1}.
}\end{exam}

\noindent
The method that we consider for asymptotically eliminating continuous or admissible
aggregation functions can be applied if one can find
sets $L_0, L_1 \subseteq PLA^*(\sigma)$ of 0/1-valued formulas that satisfy the conditions of 
Assumption~\ref{assumptions on the basic logic} below (which comes from \cite{KW3}).
The intuition behind the technical part~(2) of the assumption is that for the sets $L_0, L_1 \subseteq PLA^*(\sigma)$ 
and every $\varphi(\bar{x}, \bar{y}) \in L_0$ there is a 
set $L_{\varphi(\bar{x}, \bar{y})} \subseteq L_1$ of formulas defining some ``allowed'' conditions (with respect to $\varphi(\bar{x}, \bar{y})$)
and there are some 
$\varphi'_1(\bar{x}), \ldots, \varphi'_s(\bar{x}) \in L_0$ such that if $\mcA \models \varphi'_i(\bar{a})$
and $\chi(\bar{x}, \bar{y}) \in L_{\varphi(\bar{x}, \bar{y})}$, 
then the fraction $|\varphi(\bar{a}, \mcA) \cap \chi(\bar{a}, \mcA)| / |\chi(\bar{a}, \mcA)|$ is with high probability close to a 
number $\alpha_i$ that depends only on 
$\varphi(\bar{x}, \bar{y}), \chi(\bar{x}, \bar{y}), \varphi_i(\bar{x})$
and the sequence of probability distributions $\mbbP$.
As we allow aggregation functions with arity $m > 1$, part~(2) needs to simultaneously
speak of a sequence  $\varphi_1(\bar{x}, \bar{y}), \ldots, \varphi_m(\bar{x}, \bar{y})\in L_0$.

\begin{assump}\label{assumptions on the basic logic}{\rm
Suppose that $L_0, L_1 \subseteq PLA^*(\sigma)$
are $0/1$-valued and that the following conditions hold:
\begin{enumerate}
\item For every aggregation-free $\varphi(\bar{x}) \in PLA^*(\sigma)$ there is an $L_0$-basic formula $\varphi'(\bar{x})$ 
which is asymptotically equivalent to $\varphi(\bar{x})$ with respect to $\mbbP$.

\item For every $m \in \mbbN^+$ and all $\varphi_1(\bar{x}, \bar{y}), \ldots, \varphi_m(\bar{x}, \bar{y})\in L_0$, 
there are $L_{\varphi_1(\bar{x}, \bar{y})}, \ldots, L_{\varphi_m(\bar{x}, \bar{y})} \subseteq L_1$
such that if $\chi_j(\bar{x}, \bar{y}) \in L_{\varphi_j(\bar{x}, \bar{y})}$
for $j = 1, \ldots, m$, then 
there are $s, t \in \mbbN^+$, $\varphi'_i(\bar{x}) \in L_0$, $\alpha_{i, j} \in [0, 1]$,
for $i = 1, \ldots, s$, $j = 1, \ldots, m$,
and $\chi'_i(\bar{x}) \in L_0$, for $i = 1, \ldots, t$,
such that for every $\varepsilon > 0$ and $n$ there is $\mbY^\varepsilon_n \subseteq \mbW_n$ such that
$\lim_{n\to\infty}\mbbP_n(\mbY^\varepsilon_n) = 1$ and 
for every $\mcA \in \mbY^\varepsilon_n$ the following hold:
\begin{align*}
&(a) \ \mcA \models \forall \bar{x} \bigvee_{i=1}^s \varphi'_i(\bar{x}), \\
&(b) \ \text{if $i \neq j$ then } 
\mcA \models \forall \bar{x} \neg(\varphi'_i(\bar{x}) \wedge \varphi'_j(\bar{x})), \\
&(c) \ \mcA \models \forall \bar{x}
\Big(\Big(\bigvee_{i=1}^m \neg\exists\bar{y}\chi_i(\bar{x}, \bar{y})\Big) \leftrightarrow
\Big(\bigvee_{i=1}^t \chi'_i(\bar{x})\Big)\Big), \text{ and}\\
&(d) \ \text{for all $i = 1, \ldots, s$ and $j = 1, \ldots, m$, if $\bar{a} \in (D_n)^{|\bar{x}|}$, and
$\mcA \models \varphi'_i(\bar{a})$,} \\
&\text{ then } (\alpha_{i, j} - \varepsilon)|\chi_j(\bar{a}, \mcA)| \ \leq \
|\varphi_j(\bar{a}, \mcA) \cap \chi_j(\bar{a}, \mcA)|
\ \leq \ (\alpha_{i, j} + \varepsilon)|\chi_j(\bar{a}, \mcA)|.
\end{align*}
\end{enumerate}
}\end{assump}

\noindent
In Section~\ref{Asymptotic elimination of aggregation functions}
we will use the following result, which is a less detailed version of Theorem~5.9 in \cite{KW3}.

\begin{theor}\label{general asymptotic elimination} {\rm \cite{KW3}} 
Suppose that $L_0, L_1 \subseteq PLA^*(\sigma)$ are 0/1-valued and that 
Assumption~\ref{assumptions on the basic logic} holds.
Let $F : \big([0, 1]^{<\omega}\big)^m \to [0, 1]$, 
let $\psi_i(\bar{x}, \bar{y}) \in PLA^*(\sigma)$, for $i = 1, \ldots, m$, and suppose that each 
$\psi_i(\bar{x}, \bar{y})$
is asymptotically equivalent to an $L_0$-basic formula 
\[
\bigwedge_{k=1}^{s_i} (\psi_{i, k}(\bar{x}, \bar{y}) \to c_{i, k}) \quad
\text{(so $\psi_{i, k} \in L_0$ for all $i$ and $k$).}
\]
Suppose that for $i = 1, \ldots, m$, 
$\chi_i(\bar{x}, \bar{y}) \in \bigcap_{k=1}^{s_i} L_{\psi_{i, k}(\bar{x}, \bar{y})}$.
Let $\varphi(\bar{x})$ denote the $PLA^*(\sigma)$-formula
\[
F\big(\psi_1(\bar{x}, \bar{y}), \ldots, \psi_m(\bar{x}, \bar{y}) : \bar{y} : 
\chi_1(\bar{x}, \bar{y}), \ldots, \chi_m(\bar{x}, \bar{y})\big).
\]
(i) If $F$ is continuous then $\varphi(\bar{x})$ is asymptotically equivalent to an $L_0$-basic formula with respect to $\mbbP$.\\
(ii) Suppose, in addition, that the following holds 
if $\varphi_j(\bar{x}, \bar{y}) \in L_0$, $\chi_j(\bar{x}, \bar{y}) \in L_1$, $\varphi'_i(\bar{x}) \in L_0$,
$\mbY^\varepsilon_n$ and $\alpha_{i, j}$ are  like in 
part~2 of Asumption~\ref{assumptions on the basic logic}:
If $\alpha_{i, j} = 0$
then, for all sufficiently large $n$, all $\bar{a} \in (D_n)^{|\bar{x}|}$ and all $\mcA \in \mbY^\varepsilon_n$,
if $\mcA \models \varphi'_i(\bar{a})$
then $\varphi_j(\bar{a}, \mcA) \cap \chi_j(\bar{a}, \mcA) = \es$.
Then it follows that if $F$ is admissible then $\varphi(\bar{x})$ is asymptotically equivalent to an $L_0$-basic formula with respect to $\mbbP$.
\end{theor}

\section{Directed acyclic graphs, trees and closure types}\label{Section about trees}

\noindent
{\bf \em For the rest of this article we let $\tau = \{E\}$ where $E$ is a binary relation symbol.
Moreover, $\sigma$ will always denote a finite relational signature such that $\tau \subseteq \sigma$.}

\begin{defin}\label{definition of DAG}{\rm
(i) By a {\bf \em directed acyclic graph (DAG)} we mean a {\em finite} $\tau$-structure  $\mcG$ such that
\begin{enumerate}
\item $\mcG \models \forall x, y \big(E(x, y) \rightarrow (x \neq y \wedge \neg E(y, x)) \big)$, and

\item there do {\em not} exist $k \in \mbbN^+$ and distinct $a_0, \ldots, a_k \in G$ such that $\mcG \models E(a_i, a_{i+1})$
for all $i = 0, \ldots, k-1$ and $\mcG \models E(a_k, a_0)$.
\end{enumerate}
}\end{defin}

\noindent
{\bf Note} that we allow the domain of a DAG $\mcG$ to be empty.

\begin{defin}\label{notions in a DAG}{\rm
Suppose that $\mcG$ is a DAG.\\
(i) If $a, b \in G$ and $\mcG \models E(b, a)$ then we call $b$ a {\bf \em parent} of $a$ and $a$ a {\bf \em child} of $b$,
and the set of parents of $a$ is denoted $\mr{par}(a)$.\\
(ii) If $\mcG$ is a DAG and $a \in G$ then $a$ is called a {\bf \em root} if $\mcG \models \neg \exists x E(x, a)$.\\
(iii) Let $a, b \in G$. A {\bf \em directed path (of length $l$)} from $a$ to $b$
is a sequence $c_0, c_1, \ldots, c_l \in T$ such that $l \geq 1$, 
$\mcG \models E(c_i, c_{i+1})$ for all $i = 0, \ldots, l-1$, $a = c_0$ and $b = c_l$. 
If there is a directed path from $a$ to $b$ then $a$ is called an {\bf \em ancestor} of $b$ and $b$ is called
a {\bf \em successor} of $a$.\\
(iv) The {\bf \em level $0$} of $\mcG$ is the set of all roots of $\mcG$.
An element $a \in G$ belongs to {\bf \em level $l+1$} (of $\mcG$) if $a$ has a parent in level $l$.\\
(v) If the domain of $\mcG$ is nonempty, then the 
{\bf \em height} of $\mcG$ is the largest $l$ such that level $l$ of $\mcG$ is nonempty.
We also adopt the convention that if the domain of $\mcG$ is empty then its {\bf \em height} is $-1$.\\
}\end{defin}

\begin{defin}\label{definition of tree}{\rm
By a {\bf \em tree} we mean a {\em finite} $\tau$-structure $\mcT$ such that 
\begin{enumerate}
\item $\mcT \models \forall x, y \big(E(x, y) \rightarrow (x \neq y \wedge \neg E(y, x)) \big)$,

\item there is a unique element $a \in T$, called {\bf \em the root}, such that $\mcT \models \neg\exists x E(x, a)$, 

\item for all $a \in T$, if $a$ is not the root then there is a unique $b \in T$, 
called the {\bf \em parent} of $a$, such that $\mcT \models E(b, a)$,
and in this case $a$ is called a {\bf \em child} of $b$, and

\item there do {\em not} exist $n \in \mbbN^+$ and distinct $a_0, \ldots, a_n \in T$ such that $\mcT \models E(a_n, a_0)$
and $\mcT \models E(a_k, a_{k+1})$ for all $k = 0, \ldots, n-1$.
\end{enumerate}
}\end{defin}

\noindent
Note that every tree is a DAG.

\begin{defin}\label{definition of subtree}{\rm
Let $\mcT$ be a tree. \\
(i) A tree $\mcT'$ is a {\bf \em subtree} of $\mcT$ if it is a substructure of $\mcT$ in the model theoretic sense. \\
(ii) A subtree $\mcT'$ of $\mcT$ is {\bf \em rooted} in $a \in T$ if $a$ is the root of $\mcT'$.\\
(iii) Let $\mcT'$ be a tree and $a \in T$. Then $\msfN_\mcT(a, \mcT')$ denotes the number of subtrees of $\mcT$ that
are rooted in $a$ and isomorphic to $\mcT'$. If $\mcT$ is clear from the context we may just write $\msfN(a, \mcT')$.
}\end{defin}

\begin{defin}\label{definition of closure}{\rm
Let $\mcT$ be a tree.\\
(i) Two elements of the tree with a common parent are called {\bf \em siblings}.\\
(ii) If $B \subseteq T$ then the {\bf \em closure} of $B$ (in $\mcT$), denoted $\cl_\mcT(B)$ or just $\cl(B)$,
is defined by 
\[
\cl_\mcT(B) = \big\{a \in T : \text{ $a \in B$, or $a$ is the root, or $a$ is an ancestor of some element in $B$}\big\}.
\]
If $\bar{a}$ is a sequence of elements from $T$ then we define $\cl_\mcT(\bar{a}) = \cl_\mcT(\rng(\bar{a}))$. \\
(iii) A set $B \subseteq T$ (sequence $\bar{b}$) is {\bf \em closed (in $\mcT$)} if $\cl_\mcT(B) = B$
($\cl_\mcT(\bar{b}) = \rng(\bar{b})$).\\
(iv) Let $\mcA$ be a $\sigma$-structure such that $\mcT = \mcA \uhrc \tau$ is a tree.
If $B \subseteq A$ then the {\bf \em closure} of $B$ (in $\mcA$), denoted $\cl_\mcA(B)$ or just $\cl(B)$,
is defined by $\cl_\mcA(B) = cl_\mcT(B)$, and similarly for sequences of elements from $A$.
We say that $B \subseteq A$ is {\bf \em closed in $\mcA$} if it is closed in $\mcT$.\\
(v) If $\mcA$ is as in the previous part then we may use notions such as root, child, level, etcetera, with reference
to the underlying tree $\mcA \uhrc \tau$.
}\end{defin}

\begin{rem}\label{remarks on closure}{\rm
Observe that for every tree $\mcT$ and every $B \subseteq T$, $\cl_\mcT(\cl_\mcT(B)) = \cl_\mcT(B)$.
Note also that the statement ``$\{x_1, \ldots, x_k\}$ is closed'' is expressed, in every tree, by the formula
\[
\forall z \bigg( \Big( \bigvee_{i=1}^k E(z, x_i) \Big) \rightarrow \Big( \bigvee_{i=1}^k z = x_i \Big)\bigg).
\]
Suppose that $\mcT$ is a tree, $a \in T$, $\{b_1, \ldots, b_k\} \subseteq T$ and $a \in \cl_\mcT(b_1, \ldots, b_k)$.
Then either $a = b_i$ for some $i$, or for some $l \in \mbbN^+$ and $i$ there is a directed path from $a$ to some $b_i$ of length $l$.
The statement ``there is a directed path from $x$ to $y$ of length at most $l$'' can be expressed by a first-order formula,
say $\xi_l(x, y)$. It follows that in all trees with height at most $l$ the property `$x \in \cl(y)$' is expressed by the formula $\xi_l(x, y)$.

Finally, observe that if $\mcT$ is a tree and $B \subseteq T$ is closed, then the 
substructure $\mcT \uhrc B$ is also a tree.
}\end{rem}

\noindent
{\bf \em For the rest of this section we fix some $\Delta \in \mbbN^+$ and we assume that all trees mentioned have height at most $\Delta$,
so ``tree'' will mean ``tree of height at most $\Delta$''.}
With this assumption the property  `$x \in \cl(y)$' is expressed by the formula $\xi_\Delta(x, y)$ 
from Remark~\ref{remarks on closure} and we will use the more intuitive expression   `$x \in \cl(y)$' to denote that formula.
More generally, the expression `$x \in \cl(y_1, \ldots, y_k)$' will mean `$\bigvee_{i=1}^k x \in \cl(y_i)$'.

\begin{defin}\label{definition of atomic sigma-type}{\rm
Let $\tau \subseteq \sigma$ and let $\mcT$ be a tree.
A formula $\varphi(x_1, \ldots, x_k)$ is called an {\bf \em atomic type over $\sigma$ with respect to $\mcT$} if 
\begin{enumerate}
\item there is a $\sigma$-structure $\mcA$ that expands $\mcT$ and $\mcA \models \exists x_1, \ldots, x_k \varphi(x_1, \ldots, x_k)$,
\item $\varphi(x_1, \ldots, x_k)$ is a conjunction of $\sigma$-literals,
\item for all $i, j \in \{1, \ldots, k\}$, either $E(x_i, x_j)$ or $\neg E(x_i, x_j)$ is a conjunct of $\varphi(x_1, \ldots, x_k)$, and
\item for all different $i$ and $j$ in $\{1, \ldots, k\}$, $x_i \neq x_j$ is a conjunct of $\varphi(x_1, \ldots, x_k)$.
\end{enumerate}
If, moreover, for every $R \in \sigma$, of arity $r$ say, and all $i_1, \ldots, i_r \in \{1, \ldots, k\}$, 
either $R(x_{i_1}, \ldots, x_{i_r})$ or $\neg R(x_{i_1}, \ldots, x_{i_r})$ is a conjunct of $\varphi(x_1, \ldots, x_k)$,
then we call $\varphi(x_1, \ldots, x_k)$ a {\bf \em complete atomic type over $\sigma$ with respect to $\mcA$}.
If the particular tree $\mcT$ is not important in the context we may omit the phrase `with respect to $\mcT$'.
}\end{defin}

\noindent
Note that every atomic type over $\tau$ (with respect to some tree $\mcT$) is a {\em complete} atomic type over $\tau$,
but this implication does not (in general) hold if $\tau$ is replaced by a proper expansion $\sigma \supset \tau$.

\begin{defin}\label{definition of closure type}{\rm
Let $\tau \subseteq \sigma$ and let $\mcT$ be a tree.
A formula $\psi(x_1, \ldots, x_k)$ is called a {\bf \em closure type over $\sigma$ with respect to $\mcT$} if it is equivalent to a 
formula of the form 
\begin{align*}
\exists y_1, \ldots, y_m \Big(\varphi(x_1, \ldots, x_k, y_1, \ldots, y_m) \ \wedge \ 
\text{ `$\{x_1, \ldots, x_k, y_1, \ldots, y_m\}$ is closed'} \Big).
\end{align*}
where $\varphi(x_1, \ldots, x_k, y_1, \ldots, y_m)$ is an atomic type over $\sigma$ with respect to $\mcT$
and 
\[
\varphi(x_1, \ldots, x_k, y_1, \ldots, y_m) \models \bigwedge_{i=1}^m y_i \in \cl(x_1, \ldots, x_k).
\]
If, in addition, the formula $\varphi$ is a {\em complete} atomic type over $\sigma$, then we call 
$\psi$ a {\bf \em complete closure type over $\sigma$ with respect to $\mcT$}.
If the particular tree $\mcT$ is not important in the context we may omit the phrase `with respect to $\mcT$'.

We allow the (important special) case when the quantifier prefix `$\exists y_1, \ldots, y_m$' is empty and the variables
`$y_1, \ldots, y_m$' do not occur.
We also allow the case when the sequence $x_1, \ldots, x_k$ is empty, and in this case $\psi$
is a sentence, $m=1$, $y_1$ denotes the root, and $\psi$ expresses which relations the root satisfies.
}\end{defin}

\noindent
Observe that every closure type over $\tau$ (with respect to some tree $\mcT$) is a {\em complete} closure type over $\tau$,
but this implication does not (in general) hold if $\tau$ is replaced by a proper expansion $\sigma \supset \tau$.
It is easy to see that if $\varphi(\bar{x})$ is a closure-type over $\sigma \supseteq \tau$ with respect to a tree $\mcT$, then 
there are infinitely many trees $\mcT'$ such that $\mcT'$ has the same height as $\mcT$,
$\mcT$ is a subtree (that is, substructure) of $\mcT'$, and $\varphi(\bar{x})$ is a
closure-type over $\sigma$ with respect to $\mcT'$.

\begin{defin}\label{definitions of restrictions of types}{\rm
Let $\tau \subseteq \sigma' \subseteq \sigma$ and let $p(\bar{x})$ be a (complete) closure type over $\sigma$
with respect to a tree $\mcT$.
Also let $\bar{y}$ be a subsequence of $\bar{x}$.\\
(i) The {\bf \em restriction of $p(\bar{x})$ to $\sigma'$}, denoted $p \uhrc \sigma'$,
is a closure-type $p'(\bar{x})$ over $\sigma'$ (with respect to $\mcT$) such that $p(\bar{x}) \models p'(\bar{x})$, and for every
closure-type $p^*(\bar{x})$ over $\sigma'$, if $p(\bar{x}) \models p^*(\bar{x})$ then $p'(\bar{x}) \models p^*(\bar{x})$.
(The restriction is not syntactically unique, but it is unique up to logical equivalence, so technically we just choose one of the
formulas that satisfy the condition of being a restriction.)\\
(ii) The {\bf \em restriction of $p(\bar{x})$ to $\bar{y}$}, denoted $p \uhrc \bar{y}$, is a 
closure type over $\sigma$ (with respect to $\mcT$) $p'(\bar{y})$ such that $p(\bar{x}) \models p'(\bar{y})$, and for every
closure-type $p^*(\bar{y})$ over $\sigma$, if $p(\bar{x}) \models p^*(\bar{y})$ then $p'(\bar{y}) \models p^*(\bar{y})$.
(Again, the restriction is unique up to logical equivalence.)
}\end{defin}

\noindent
It is not hard to see that, under the same assumptions as in the above definition, if $p(\bar{x})$ is a closure type over $\sigma$
with respect to the tree $\mcT$, then $p \uhrc \sigma'$ is a closure-type over $\sigma'$ with respect to $\mcT$,
and $p \uhrc \bar{y}$ is a closure type over $\sigma$ with respect to $\mcT$.
It will be convenient to use `$\models_{tree}$' to denote consequence restricted to expansions of trees, as defined below.

\begin{defin}\label{definition of consequence in trees}{\rm
Suppose that $\tau \subseteq \sigma$ and let $\varphi(\bar{x}), \psi(\bar{x}) \in PLA^*(\sigma)$ be $0/1$-valued formulas.
By the expression $\varphi(\bar{x}) \models_{tree} \psi(\bar{x})$ we mean that if $\mcA$ is a $\sigma$-structure
that expands a tree (of height at most $\Delta$, $\bar{a} \in (A)^{|\bar{x}|}$, and $\mcA \models \varphi(\bar{a})$, 
then $\mcA \models \psi(\bar{a})$.
The expression $\models_{tree} \psi(\bar{x})$ means the same as $\top \models_{tree} \psi(\bar{x})$, or equivalently,
that $\forall \bar{x} \psi(\bar{x})$ is true in all $\sigma$-structures that expand a tree.
}\end{defin}

\begin{defin}\label{definition of self-contained}{\rm
Let $\tau \subseteq \sigma$ and let 
$p(\bar{x})$ be a closure type over $\sigma$ (with respect to some tree $\mcT$) where $\bar{x} = (x_1, \ldots, x_k)$.
We call $p(\bar{x})$ {\bf \em self-contained} if $p(\bar{x})$ implies that $\bar{x}$ is closed, or more formally, if
\[
p(\bar{x}) \ \models \ \forall z \bigg( \Big( \bigvee_{i=1}^k E(z, x_i) \Big) \rightarrow \Big( \bigvee_{i=1}^k z = x_i \Big)\bigg).
\]
}\end{defin}

\noindent
Note that if $p(\bar{x})$ is a self-contained closure type over $\sigma \supseteq \tau$,
then $p(\bar{x})$ is equivalent to a formula of the form $p^*(\bar{x}) \wedge \text{ `$\bar{x}$ is closed'}$,
where $p^*(\bar{x})$ is an atomic type over $\sigma$.

\begin{lem}\label{bijection to self-contained types}
Let $\tau \subseteq \sigma$ and
let $p(\bar{x})$ be a closure type over $\sigma$.
Then there is a self-contained closure type over $\sigma$ $p^*(\bar{x}, \bar{y})$ such that if $\mcA$ is a
$\sigma$-structure that expands a tree $\mcT$, then
\[
\models_{tree} \forall \bar{x}\big(p(\bar{x}) \leftrightarrow \exists ! \bar{y} p^*(\bar{x}, \bar{y})\big)
\]
where `$\exists ! \bar{y}$' means `there exists unique $\bar{y}$'.
It follows that $|p(\mcA)| = |p^*(\mcA)|$ and if $\bar{x} = \bar{u}\bar{v}$ and $\bar{a} \in A^{|\bar{u}|}$,
then $|p(\bar{a}, \mcA)| = |p^*(\bar{a}, \mcA)|$. 
\end{lem}

\noindent
{\bf Proof.}
Let $p(\bar{x})$ be a closure type over $\sigma$. 
Then $p(\bar{x})$ is equivalent to a formula of the form 
\[
\exists \bar{y} \big( \varphi(\bar{x}, \bar{y}) \ \wedge \ `\bar{x}\bar{y} \text{ is closed'} \big)
\]
where $\varphi(\bar{x}, \bar{y})$ is an atomic type over $\sigma$, $\bar{y} = (y_1, \ldots, y_m)$, and
\[
\varphi(\bar{x}, \bar{y}) \models_{tree} \bigwedge_{i=1}^m y_i \in \cl(\bar{x}).
\]
Let $p^*(\bar{x}, \bar{y})$ be the formula 
\[
\varphi(\bar{x}, \bar{y}) \ \wedge \ `\bar{x}\bar{y} \text{ is closed'}.
\]
Then $p^*(\bar{x}, \bar{y})$ is a closure type over $\sigma$ with empty sequence of quantifiers in the beginning.
Since $p^*(\bar{x}, \bar{y})$ implies that `$\bar{x}\bar{y}$ is closed' it follows that $p^*$ is self-contained.
It is evident that if $\mcA$ is a $\sigma$-structure then
$\mcA \models \forall \bar{x}\big(p(\bar{x}) \leftrightarrow \exists \bar{y} p^*(\bar{x}, \bar{y})\big)$.
Suppose, moreover, that $\mcA$ expands a tree and $\mcA \models p^*(\bar{a}, \bar{b})$.
Then, for every $b \in \rng(\bar{b})$, either $b \in \rng(\bar{a})$
or there is a directed path of a particular length, say $l_b$, from $b$ to
some $a_b \in \rng(\bar{a})$.
In the latter case, since $\mcA \uhrc \tau$ is a tree, it follows that $b$ is the unique element such that there is a directed path from
$b$ to $a_b$ of length $l_b$.
Therefore $\bar{b}$ is the unique tuple such that $\mcA \models p^*(\bar{a}, \bar{b})$.
\hfill $\square$
\\

\noindent
The following lemma follows straightforwardly from the definition of closure-type, so we omit the proof.

\begin{lem}\label{some properties of closure-types}
Suppose that $\tau \subseteq \sigma$ and $p(\bar{x})$ is a closure-type over $\sigma$
where $\bar{x} = (x_1, \ldots, x_k)$.\\
(i) For all distinct $i, j \in \{1, \ldots, k\}$ either 
$p(\bar{x}) \models_{tree}$ ``$x_j$ is an ancestor of $x_i$'', or
$p(\bar{x}) \models_{tree}$ ``$x_j$ is {\em not} an ancestor of $x_i$''.\\
(ii) For every subsequence $\bar{y}$ of $\bar{x}$ and every $x_i \in \rng(\bar{x}) \setminus \rng(\bar{y})$,
either $p(\bar{x}) \models_{tree} x_i \in \cl(\bar{y})$, or
$p(\bar{x}) \models_{tree} x_i \notin \cl(\bar{y})$.
\end{lem}

\noindent
With the above lemma the following definition makes sense.

\begin{defin}\label{definition of y-independent}{\rm
Let $\tau \subseteq \sigma$ and let 
$p(\bar{x}, \bar{y})$ be a closure type over $\sigma$,
where $\bar{y} = (y_1, \ldots, y_k)$.
We call $p(\bar{x}, \bar{y})$ {\bf \em $\bar{y}$-independent} if, for all $i = 1, \ldots, k$,
$p(\bar{x}, \bar{y}) \models_{tree} y_i \notin  \cl(\bar{x})$.
}\end{defin}

\noindent
In the next lemma, and later, if $\bar{y} = (y_1, \ldots, y_k)$ the expression `$\rng(\bar{y}) \subseteq \cl(\bar{x})$' denotes the formula
`$\bigwedge_{i=1}^k y_i \in \cl(\bar{x})$'.

\begin{lem}\label{getting a y-independent type}
Suppose that $\tau \subseteq \sigma$ and that $p(\bar{x}, \bar{y}, \bar{z})$ is a closure type over $\sigma$
such that $p(\bar{x}, \bar{y}, \bar{z}) \models \rng(\bar{y}) \subseteq \cl(\bar{x})$ and $\bar{z}$ is nonempty.
Suppose that $\mcA$ is a $\sigma$-structure that expands a tree
and $\mcA \models p(\bar{a}, \bar{b}, \bar{c}) \wedge p(\bar{a}, \bar{b}', \bar{c}')$ where
$\bar{a} \in A^{|\bar{x}|}$, $\bar{b}, \bar{b}' \in A^{|\bar{y}|}$, and $\bar{c}, \bar{c}' \in A^{|\bar{z}|}$.
Then $\bar{b} = \bar{b}'$ and hence $|p(\bar{a}, \mcA)| = |p(\bar{a}, \bar{b}, \mcA)|$.
\end{lem}

\noindent
{\bf Proof.}
Suppose that $\mcA \models p(\bar{a}, \bar{b}, \bar{c}) \wedge p(\bar{a}, \bar{b}', \bar{c}')$,
where $\bar{b} = (b_1, \ldots, b_k)$ and $\bar{b}' = (b'_1, \ldots, b'_k)$,
and $p(\bar{x}, \bar{y}, \bar{z}) \models \rng(\bar{y}) \subseteq \cl(\bar{x})$.
The last assumption implies that for each $i = 1, \ldots, k$ there are $j_i$ and $l_i$ such that 
there is a directed path from $b_i$ to $a_{j_i}$ of length $l_i$ and a directed path from $b'_i$ to $a_{j_i}$
of length $l_i$. As $\mcA \uhrc \tau$ is a tree it follows that $b_i = b'_i$.
\hfill $\square$

\begin{rem}\label{remark on differences of closures}{\rm
Suppose that $p(\bar{x}, \bar{y})$ is a closure-type over $\tau$.
By Lemma~\ref{bijection to self-contained types},
there is a self-contained closure type $p^*(\bar{x}, \bar{y}, \bar{z})$ over $\tau$
such that 
\[
\models_{tree} \forall \bar{x}, \bar{y}
 \big(p(\bar{x}, \bar{y}) \leftrightarrow \exists ! \bar{z} p^*(\bar{x}, \bar{y}, \bar{z})\big).
\]
Recall that by the definition of closure-type $p^* \models u \neq v$ whenever $u$ and $v$ are different entries
from the sequence of variables $\bar{x}\bar{y}\bar{z}$.
Let $\bar{y} = (y_1, \ldots, y_k)$ and $\bar{z} = (z_1, \ldots, z_l)$.
Lemma~\ref{some properties of closure-types}
implies that,
by reordering $\bar{y}$ and $\bar{z}$ if necessary we may (without loss of generality) assume that, 
for some $1 \leq m_0 \leq k$, $1 \leq m_1 \leq l$ and $m_1 \leq m_2 \leq l$,
$p^*(\bar{x}, \bar{y}, \bar{z}) \models_{tree} y_i \notin \cl(\bar{x})$ if and only if $i > m_0$, 
$p^*(\bar{x}, \bar{y}, \bar{z}) \models_{tree} z_i \in \cl(\bar{y})$ if and only if $i > m_1$, and
for $i > m_1$,
$p^*(\bar{x}, \bar{y}, \bar{z}) \models_{tree} z_i \notin \cl(\bar{x})$ if and only if $i > m_2$.

It follows that for every tree $\mcT$ and all $\bar{a} \in T^{|\bar{x}|}$ and $\bar{b} \in T^{|\bar{y}|}$,
if $\mcT \models p(\bar{a}, \bar{b})$ (so $\mcT \models p^*(\bar{a}, \bar{b}, \bar{c})$ for some unique $\bar{c} \in T^{|\bar{z}|}$),
then
$|\cl_\mcT(\bar{b}) \setminus \cl_\mcT(\bar{a})| = (k - m_0) + (l - m_2)$.
Thus the number $|\cl_\mcT(\bar{b}) \setminus \cl_\mcT(\bar{a})|$ depends only on $p(\bar{x}, \bar{y})$.
}\end{rem}

\begin{defin}\label{definition of rank}{\rm
Let $\tau \subseteq \sigma$, let $p(\bar{x}, \bar{y})$ be a closure type over $\sigma$, and
let $p_\tau(\bar{x}, \bar{y})$ be the restriction of $p$ to $\tau$.
We define the {\bf \em $\bar{y}$-rank of p}, denoted $\rank_{\bar{y}}(p)$, to be the number
\[
\rank_{\bar{y}}(p) = |\cl_\mcT(\bar{b}) \setminus \cl_\mcT(\bar{a})|
\]
where $\mcT$ is any tree and $\bar{a} \in T^{|\bar{x}|}$ and $\bar{b} \in T^{|\bar{y}|}$ are any tuples (of the specified lengths)
such that $\mcT \models p_\tau(\bar{a}, \bar{b})$.
By Remark~\ref{remark on differences of closures}, this definition depends only on $p_\tau(\bar{x}, \bar{y})$.
}\end{defin}

\begin{rem}\label{characterization of rank 0}{\rm
It follows from Remark~\ref{remark on differences of closures} and
Definition~\ref{definition of rank} 
that if $p(\bar{x}, \bar{y})$ is a closure type over $\sigma$ (where $\tau \subseteq \sigma$) with respect to 
a tree $\mcT$,
then $\rank_{\bar{y}}(p) = 0$ if and only if $p(\bar{x}, \bar{y}) \models_{tree} \rng(\bar{y}) \subseteq \cl(\bar{x})$.
}\end{rem}

\begin{rem}\label{remark on getting a y-independent type}
{\bf (The relevance of self-contained closure types)} {\rm
Suppose that $\tau \subseteq \sigma$ and that $p(\bar{x}, \bar{y})$ is a closure type over $\sigma$
such that $\rank_{\bar{y}}(p) > 0$.
By Lemma~\ref{bijection to self-contained types} there is a sequence of variables $\bar{z}$ which extends $\bar{y}$ and a
{\em self-contained} closure type over $\sigma$ $p^*(\bar{x}, \bar{z})$ 
such that $|p(\bar{a}, \mcA)| = |p^*(\bar{a}, \mcA)|$ for all $\sigma$-structures $\mcA$ such that
$\mcA \uhrc \tau$ is a tree and all $\bar{a} \in A^{|\bar{x}|}$.
By Lemma~\ref{getting a y-independent type} we can write $\bar{z} = \bar{u}\bar{v}$ (if $\bar{z}$ is reordered if necessary)
so that $p^*$ is {\em $\bar{v}$-independent} and, for all $\bar{a} \in A^{|\bar{x}|}$ and $\bar{b} \in A^{|\bar{u}|}$
such that $p^*(\bar{a}, \bar{b}, \mcA) \neq \es$,
$|p^*(\bar{a}, \mcA)| = |p^*(\bar{a}, \bar{b}, \mcA)|$ and hence $|p(\bar{a}, \mcA)| = |p^*(\bar{a}, \bar{b}, \mcA)|$.
(That $\bar{v}$ is nonempty follows from the assumption that $\rank_{\bar{y}}(p) > 0$.)
Note that the root of $\mcA \uhrc \tau$ belongs to $\rng(\bar{a}\bar{b})$, so $\bar{x}\bar{z}$ is a nonempty sequence
(even if $\bar{x}$ is empty).
Also observe that if $p(\bar{x}, \bar{y})$ is a {\em complete} closure type over $\sigma$
then $p^*(\bar{x}, \bar{z})$ will be a {\em complete} closure type over $\sigma$.
{\em This justifies that we will, in some technical proofs that will follow, work only with
closure types over $\sigma$ $p(\bar{x}, \bar{y})$ that are self-contained and $\bar{y}$-independent and where
$\bar{x}$ is nonempty.}
}\end{rem}

\begin{rem}\label{decomposing a type of higher rank} {\bf (Decomposing a closure type over $\sigma$ of rank $\geq 2$)} {\rm
Let $\tau \subseteq \sigma$ and suppose that $p(\bar{x}, \bar{y})$ is a self-contained and 
$\bar{y}$-independent closure type over $\sigma$
such that $\bar{x}$ is nonempty and $\rank_ {\bar{y}}(p) = r+1$ where $r \geq 1$. 
Then $|\bar{y}| = r+1$, 
$p(\bar{x}, \bar{y}) \models_{tree}$ ``$\bar{x}$ contains the root'', 
and there is $u \in \rng(\bar{y})$ such that
\[
p(\bar{x}, \bar{y}) \models_{tree} \text{ ``$u$ is the child of a member of $\bar{x}$''.}
\]
Let $\bar{v}$ be the sequence obtained from $\bar{y}$ by removing $u$.
By reordering $\bar{y}$ if necessary we may assume that $\bar{y} = u\bar{v}$.
Let $q(\bar{x}, u) = p \uhrc \bar{x}u$.
By the choice of $u$ if follows that $q(\bar{x}, u)$ is self-contained and $u$-independent, and that
$q(\bar{x}, u) \models u \notin \cl(\bar{x})$.
Hence $\rank_u(q) = 1$.
By the choice of $u$, $p(\bar{x}, u, \bar{v}) \models$ ``$\cl(u) \cap \rng(\bar{v}) = \es$''.
Since $p$ is $\bar{y}$-independent and $\bar{v}$ is a subsequence of $\bar{y}$ it follows that $p$ is $\bar{v}$-independent.

Suppose that $\mcT$ is a tree and $\mcT \models p(\bar{a}, b, \bar{c})$ where $\bar{a} \in T^{|\bar{x}|}$,
$b \in T$ and $\bar{c} \in T^{|\bar{v}|}$. Then, 
by the choice of $u$ and since $p$ is self-contained and $\bar{y}$-independent,
$\cl_\mcT(\bar{c}) \setminus \cl_\mcT(\bar{a}b) = \rng(\bar{c})$
where $|\rng(\bar{c})| = |\bar{c}|$. Hence $\rank_{\bar{v}}(p) = |\bar{c}| = |\bar{v}| =  r$.
}\end{rem}

\noindent
Our main results will show that every $PLA^*$-formula that satisfies certain conditions is asymptotically equivalent to 
a closure-basic formula as defined below.

\begin{defin}\label{definition of closure-basic formula}
Let $\tau \subseteq \sigma$.
A {\bf \em closure-basic formula over $\sigma$} is a formula of the form 
$\bigwedge_{i=1}^{k}(\varphi_i(\bar{x})\rightarrow c_i)$ where for each $i = 1, \ldots, k$, 
$\varphi_i(\bar{x})$ is a complete closure type over $\sigma$ and $c_i\in [0,1]$.
\end{defin}

\noindent
At one point of the proof of the main results we will use induction on the complexity on formulas,
and the base case of the induction uses the following result.

\begin{lem}\label{connectives and basic formulas}
(i) Suppose that $\varphi_1(\bar{x}), \ldots, \varphi_k(\bar{x})$ are closure-basic formulas over $\sigma$ and that
$\msfC : [0, 1]^k \to [0, 1]$. Then the formula $\msfC(\varphi_1(\bar{x}), \ldots, \varphi_k(\bar{x}))$
is equivalent to a closure-basic formula over $\sigma$.\\
(ii) If $\varphi(\bar{x}) \in PLA^*(\sigma)$ is aggregation-free then it is equivalent 
to a closure-basic formula.
\end{lem}

\noindent
{\bf Proof.}
(i) Suppose that $\varphi_i(\bar{x})$, $i = 1, \ldots, k$, is a closure-basic formula over $\sigma$.
Let $q_1(\bar{x}), \ldots, q_m(\bar{x})$ enumerate, up to logical equivalence, all complete closure types over $\sigma$
in the free variables $\bar{x}$.
Suppose that  $\mcA$ is a finite $\sigma$-structure and $\bar{a} \in A^{|\bar{x}|}$.
Observe that for each $i$ the value $\mcA(\varphi_i(\bar{a}))$ depends only on which $q_j(\bar{x})$ the sequence $\bar{a}$ satisfies.
So let $c_{i, j} = \mcA(\varphi_i(\bar{a}))$ if $\mcA \models q_j(\bar{a})$.
Then let $d_j = \msfC(c_{1, j}, \ldots, c_{k, j})$ for $j = 1, \ldots, m$.
Now $\varphi(\bar{x})$ is equivalent  to 
the closure-basic formula $\bigwedge_{j = 1}^m (q_j(\bar{x}) \to d_j)$.

(ii) Let $\varphi(\bar{x}) \in PLA^*(\sigma)$ be aggregation-free.
The proof proceeds by induction on the number of connectives in $\varphi$.
If the number of connectives is 0 then $\varphi(\bar{x})$ can be a constant from $[0, 1]$, or
it can have the form $R(\bar{x}')$ for some $R \in \sigma$ and subsequence $\bar{x}'$ of $\bar{x}$, or it can have the 
form $u = v$ for some $u, v \in \rng(\bar{x})$. 
It is easy to verify that in each case $\varphi(\bar{x})$ is equivalent 
to a closure-basic formula.
The inductive step follows from part~(i) of this lemma.
\hfill $\square$

\section{The base sequence of trees, expansions of them, and probabilities}\label{The base sequence of trees}

\noindent
{\bf \em We adopt the assumptions from the previous section, so $\sigma$ is a finite relational signature, $\tau \subseteq \sigma$,
and $\tau = \{E\}$ where $E$ is a binary relation symbol.}
We consider a {\em base sequence  $\mbT = (\mcT_n : n \in \mbbN^+)$} where each $\mcT_n$ is a tree,
so in particular $\mcT_n$ is a $\tau$-structure, 
and $|T_n| \to \infty$ as $n\to\infty$. 
Then $\mbW_n$ will be the set of all expansions of $\mcT_n$ to $\sigma$,
and a probability distribution $\mbbP_n$ will be defined on $\mbW_n$ via a $PLA^*(\sigma)$-network.
Our goal is to identify some constraints on $\mbT$ and on the $PLA^*(\sigma)$-network 
such that under these constraints we can prove
results about asymptotic equivalence of complex formulas to simpler (closure basic) formulas, and 
derive convergence results.

For example, we could fix an integer $\Delta \geq 1$ and let $\mcT_n$ be a tree in which every element 
(that is, vertex) has
at most $\Delta$ children. However, this case is covered by the context studied by Koponen in \cite{Kop24}.
So we will allow, in fact require, that there is a function, say $g_1$, such that $\lim_{n\to\infty}g_1(n) = \infty$
and every nonleaf of $\mcT_n$ has at least $g_1(n)$ children.
This condition alone does not imply the kind of results that we are looking for.
So, in addition, we will assume that there is an integer $\Delta \geq 1$ such that, for all $n$, the height of $\mcT_n$ is at most $\Delta$.

The perhaps simplest sequence $\mbT = (\mcT_n : n \in \mbbN^+)$ for which the main results of this article hold is obtained 
by fixing some integer $\Delta \geq 1$ and letting $\mcT_n$ be a tree of height $\Delta$ such that every leaf is on level $\Delta$
and all nonleaves have exactly $n$ children.
But in order to increase the applicability, we prove results for more general sequences $\mbT$.
For example, we want to allow different elements (that is, vertices) of $\mcT_n$ to have different numbers of children.
However, if we put no constraints on the relative number of children that different elements of a tree can have,
then we do not get the results that we aim for.
The next couple of examples illustrate this and also motivate the constraints that we will impose on the sequence $\mbT$ of trees.

\begin{exam}\label{example with leaves on different levels}{\rm
Let $\sigma = \tau \cup \{R\}$ where $R$ has arity 1.
For even $n \in \mbbN^+$ let $\mcT_n$ be a tree of height $2$ such that the root has $n$ children,
half of the children of the root have $n$ children, and the rest of the children of the root have no child at all.
For odd $n \in \mbbN^+$ let $\mcT$ be a tree of height $2$ such that the root has $n$ children,
$\lfloor n/3 \rfloor$ children of the root have $n$ children, and the rest of the children of  the root have no child at all.
Let $\mbW_n$ the set of all $\sigma$-structures that expand $\mcT_n$ and let $\mbbP_n$ 
be the uniform probability distribution on $\mbW_n$, or equivalently, let $\mbbP_n$ be such that for every $a \in T_n$,
the probability that $\mcA \models R(a)$ for a random $\mcA \in \mbW_n$ is $1/2$, 
independently of whether $\mcA \models R(b)$ for $b \neq a$.

Let $q(x)$ be a formula which expresses that  ``$x$ is a child of the root'', 
let $p(x, y)$ be a formula which expresses that ``$q(x)$ and $y$ is a child of $x$'', and
let $\varphi(x)$ be the formula
\[
\mr{am}\big( p(x, y) \wedge R(x) \wedge R(y) : y : p(x, y) \big).
\]
Suppose that $n$ is a large and even. 
By Lemma~\ref{independent bernoulli trials} and since $|T_n|$ is bounded by a polynomial in $n$ we can argue 
(somewhat informally) like this.
Let $a \in T_n$ be a child of the root such that $a$ is a child of the root with a child, hence $n$ children.

With high probability (approaching 1 as $n\to\infty$), if $\mcA \in \mbW_n$ is chosen at random
and if we condition on $\mcA \models R(a)$, then 
$\mcA(\varphi(a)) \approx 1/2$.
If we condition on $\mcA \not\models R(a)$ then $\mcA(\varphi(a)) = 0$.
If $a$ is a child of the root without any child, then (by the semantics of $PLA^*$) $\mcA(\varphi(a)) = 0$
(no matter whether $\mcA \models R(a)$ or not).
The probability that $\mcA \models R(a)$ is $1/2$.
If we put this together we get
$\mcA\big(\mr{am}( \varphi(x) : x : q(x)) \big) \approx (\frac{n}{2} \cdot \frac{1}{2} + \frac{n}{2} \cdot 0)/n = 1/4$.
So if $\chi = \mr{am}( \varphi(x) : x : q(x))$ then, with high likelihood, $\mcA(\chi) \approx 1/4$.

By a similar argument, for large odd $n$, if $\mcA \in \mbW_n$ is chosen at random, then, with high likelihood,
$\mcA(\chi) \approx 1/6$. 
It follows that, under the given assumptions, we do not have a convergence result as in part~(ii) of
Theorem~\ref{main result}.
It is not too difficult to show that under the same assumptions part~(i) of Theorem~\ref{main result} does not hold either.
In this example we can blame the failure on the fact that we allow that
different leaves in $\mcT_n$ are on different levels.
So in the next example we stipulate that all leaves are on the same level (which turns out to be insufficient for convergence).
}\end{exam}

\begin{exam}\label{example with few vertices with many more children}{\rm
Let $\sigma$ be as in Example~\ref{example with leaves on different levels}.
For odd $n \in \mbbN^+$ let $\mcT_n$ be a tree of height $2$ such that the root has $n$ children,
one child of the root has $2n^3$ children, and all other children of the root have $n$ children.
For even $n \in \mbbN^+$ let $\mcT_n$ be a tree of height $2$ such that the root has $n$ children,
two children of the root have $n^3$ children, and all other children of the root have $n$ children.
Let $\mbW_n$ be the set of all expansions of $\mcT_n$ to $\sigma$ and let $\mbbP_n$ be the uniform 
probability distribution on $\mbW_n$.

Suppose that $n$ is odd and large.
Let $b_n \in T_n$ be the unique child of the root that has $2n^3$ children.
Note that $\mcT_n$ has $(n-1)n + 2n^3$ leaves and that $2n^3$ of these leaves are children of $b_n$.
As $n$ is large, almost all leaves are children of $b_n$.
Let $q(\bar{x})$ and $p(\bar{x}, \bar{y})$ be as in
Example~\ref{example with leaves on different levels}.
Hence $\exists x p(x, y)$ expresses, in every $\mcT_n$, that ``$y$ is a leaf''.
Let $\varphi$ be
\[
\mr{am}\big( \exists x (p(x, y) \wedge R(x) \wedge R(y)) : y : \exists x p(x, y) \big).
\]
For a random $\mcA \in \mbW_n$ the probability that  $\mcA \models R(b_n)$ is $1/2$,
and conditioned on $\mcA \models R(b_n)$ then, with high probability, $\mcA(\varphi) \approx 1/2$, and 
conditioned on $\mcA \not\models R(b_n)$,
$\mcA(\varphi) \approx 0$.
So if $n$ is odd and large then, with probability roughly $1/2$ we have $\mcA(\varphi) \approx 1/2$, and with probability roughly $1/2$
we have $\mcA(\varphi) \approx 0$.

By reasoning in a similar way it also follows that if $n$ is even and large then,
with probability roughly $1/4$ we have $\mcA(\varphi) \approx 1/2$, with probability roughly $1/2$ we have
$\mcA(\varphi) \approx 1/4$, and with probability roughly $1/4$ we have $\mcA(\varphi) \approx 0$.
It follows that we do not have a convergence result as in part~(ii) of
Theorem~\ref{main result}.
One can also show that neither does part~(i) of the same theorem hold.
}\end{exam}

\noindent
In Example~\ref{example with few vertices with many more children}
an obstacle to convergence is that we have a varying and bounded number (either one or two) of children of 
the root with many more vertices than the other children of the root.
If we have trees $\mcT_n$ such that all leaves are on level 2,
the root has $n$ children, and, for every child $a$ of the root of $\mcT_n$, there is
$k_n$, where $\lim_{n\to\infty}k_n = \infty$ but not too slow, and there are at least $k_n$ children $a'$ of the root such that the
number of children of $a'$ is not too different from the number of children of $a$, then we do not encounter the same problem.

However, in general it is not sufficient to just have some (not too small) lower bound on the number of siblings in $\mcT_n$ with a similar 
number of children. But we actually need, for every fixed tree $\mcT'$, to put a (not too large) lower bound on
the number of siblings $a'$ of (any vertex) $a$ such that $\msfN_{\mcT_n}(a', \mcT')$ is not too different from 
$\msfN_{\mcT_n}(a, \mcT')$,
where we recall that $\msfN_{\mcT_n}(a, \mcT')$ is the number of subtrees of $\mcT_n$ that are rooted in $a$ and
isomorphic to $\mcT'$.
(It is not hard to modify Example~\ref{example with few vertices with many more children}, by considering trees
in which all leaves are on level 3, so that it shows that this stronger constraint is necessary.)
Also, we will assume that there is a polynomial upper bound on the number of children that any member of $\mcT_n$ can have.
This will allow us to use Corollary~\ref{independent bernoulli trials, second version} in such a way that the main results follow.

The above considerations motivate the following assumption:

\begin{assump}\label{properties of the trees} {\bf (The base sequence of trees)} {\rm
$\text{   }$
\begin{enumerate}
\item $\Delta \in \mbbN^+$,
\item $g_1, g_2, g_3, g_4$ are functions from  $\mbbN$ to the positive reals,
\begin{enumerate}
\item for $i = 1, 2, 3, 4$, $\lim_{n\to\infty} g_i(n) = \infty$, 
\item for every $\alpha \in \mbbR$, $\lim_{n\to\infty}(g_3(n) - \alpha\ln(n)) = \lim_{n\to\infty}(g_1(n) - g_3(n))= \infty$,
\item $\lim_{n\to\infty}\frac{g_2(n)}{g_1(n)} = 0$, and 
\item $g_4$ is a polynomial,
\end{enumerate}
\item $\mbT = (\mcT_n : n \in \mbbN^+)$ and each $\mcT_n$ is a tree such that
\begin{enumerate}
\item the height of $\mcT_n$ is $\Delta$ and all leaves of $\mcT$ are on level $\Delta$,
\item every nonleaf has at least $g_1(n)$ children and at most $g_4(n)$ children, and
\item for every tree $\mcT'$ of height at least 1, if $n$ is sufficently large then the following holds:
if $a \in T_n$ is not the root and $\msfN_{\mcT_n}(a, \mcT') > 0$, 
then $a$ has at least $g_3(n)$ siblings $b$ such that 
$\msfN_{\mcT_n}(a, \mcT') - g_2(n) < \msfN_{\mcT_n}(b, \mcT') < \msfN_{\mcT_n}(a, \mcT') + g_2(n)$.
\end{enumerate}
\end{enumerate}
}\end{assump}

\noindent
Although the above assumption may look complicated, the requirement on $g_1, g_2$ and $g_3$ is just that they grow faster than
every logarithm and slower that some polynomial, and that $g_1$ grows faster than both $g_2$ and $g_3$.
So there are many possible choices of such functions, for example 
$g_1(n) = (\ln n)^3$, $g_2(n) = \ln n$, and $g_3(n) = (\ln n)^2$ for all $n > 1$.

{\bf \em From now until and including Section~\ref{Convergence and balance}
let $\mbT = (\mcT_n : n \in \mbbN^+)$ be a sequence of trees that satisfies 
Assumption~\ref{properties of the trees}.}

\begin{rem}\label{remark on the properties of the trees}{\rm 
(i) Note that (by  Assumption~\ref{properties of the trees})
$|T_n| \leq g_4(n)^\Delta$ for all $n$, where $g_4(x)^\Delta$ is a polynomial function.\\
(ii) Suppose that $\mcT'$ is a tree of height $\geq 1$. Let $a \in T_n$.
For all sufficiently large $n$ we have that if $\msfN_{\mcT_n}(a, \mcT') > 0$, then
$\msfN_{\mcT_n}(a, \mcT') \geq g_1(n)$ as we now demonstrate.
Suppose that $\msfN_{\mcT_n}(a, \mcT') > 0$.
Then there is a subtree $\mcT''$ of $\mcT_n$ that is isomorphic to $\mcT'$ and rooted in $a$.
Let $b$ be a nonleaf vertex of $\mcT''$ such that all children of $b$ are leaves.
Let $b_1, \ldots, b_m$ be all children of $b$ in $\mcT_n$, so $m \geq g_1(n)$,  and let $b'_1, \ldots, b'_k$ be all
children of $b$ in $\mcT''$. Assuming that $n$ is large enough, $\{b'_1, \ldots, b'_k\}$ is a proper subset of $\{b_1, \ldots, b_m\}$.
For every choice of $B \subseteq \{b_1, \ldots, b_m\}$ such that $|B| = k$ we get a subtree $\mcT_B$ of $\mcT_n$ that is rooted in $a$
and isomorphic to $\mcT''$ (hence to $\mcT'$) by replacing $b'_1, \ldots, b'_k$ in $\mcT''$ by the vertices in $B$.
The subtree $\mcT_B$ can also be described as the subtree of $\mcT_n$ generated by $(T'' \setminus \{b'_1, \ldots, b'_k\}) \cup B$.
There are $\binom{m}{k} \geq \binom{g_1(n)}{k} \geq g_1(n)$ choices of $B$ and therefore 
$\msfN_{\mcT_n}(a, \mcT') \geq g_1(n)$.

Moreover, if $n$ is large enough, then whether $\msfN_{\mcT_n}(a, \mcT') > 0$ or not depends only on which level $a$ belongs to,
because all leaves of $\mcT_n$ are on level $\Delta$ and all nonleaves have at least $g_1(n)$ children where $g_1(n) \to \infty$.
}\end{rem}

\noindent
One part of part~2(b) of Assumption~\ref{properties of the trees} will be used in the form of the following lemma.

\begin{lem}\label{consequences of the first assumption about the trees} 
Let $k \in \mbbN^+$.
For every $\alpha > 0$, if $n$ is sufficiently large then
$\frac{n^k}{e^{\alpha g_3(n)}} \leq e^{-\frac{1}{2}\alpha g_3(n)}$.
\end{lem}

\noindent
{\bf Proof.}
Let $k \in \mbbN^+$.
We first show that for every $\alpha > 0$, $\lim_{n\to\infty} \frac{n^k}{e^{\alpha g_3(n)}} = 0$.
Since $\ln(x)$ is bounded on the interval $[a, b]$ for all $a, b \in \mbbR^+$ such that $a < b$,
and $\lim_{x\to\infty}\ln(x) = \infty$ it suffices to show that, for all $k \in \mbbN^+$ and $\alpha > 0$,
$\lim_{n\to\infty} \ln \frac{e^{\alpha g_3(n)}}{n^k} = \infty$.
We have 
$
\lim_{n\to\infty} \ln \frac{e^{\alpha g_3(n)}}{n^k} = \lim_{n\to\infty} (\ln e^{\alpha g_3(n)} - \ln(n^k)) = 
\lim_{n\to\infty} (\alpha g_3(n) - k\ln(n)) = \lim_{n\to\infty} \alpha(g_3(n) - \frac{k}{\alpha}\ln(n)) = \infty,
$
where the last identity follows from Assumption~\ref{properties of the trees}.
For all $k \in \mbbN^+$ and $\alpha > 0$ we now get
$\frac{n^k}{e^{\alpha g_3(n)}} \leq \frac{n^{2k}}{e^{\alpha g_3(n)}} = 
\frac{n^{2k}}{e^{\frac{1}{2}\alpha g_3(n)}} \cdot \frac{1}{e^{\frac{1}{2}\alpha g_3(n)}}
\leq e^{-\frac{1}{2}\alpha g_3(n)}$
if $n$ is large enough, by what we just proved.
\hfill $\square$

\begin{defin}\label{definition of W}{\rm
For all $n \in \mbbN^+$ let $\mbW_n$ be the set of all $\sigma$-structures $\mcA$ such that $\mcA \uhrc \tau = \mcT_n$.
}\end{defin}

\noindent
Note that if $\sigma = \tau$ then $\mbW_n = \{\mcT_n\}$.

\begin{defin}\label{definition of cofinally satisfiable}
Let $\varphi(\bar{x}) \in PLA^*(\sigma)$.
We call $\varphi(\bar{x})$ {\bf \em cofinally satisfiable} if there are infinitely many $n \in \mbbN^+$ 
such that there is $\bar{a} \in (T_n)^{|\bar{x}|}$ and $\mcA \in \mbW_n$ such that $\mcA(\varphi(\bar{a})) = 1$.
\end{defin}

\noindent
Suppose that $p(\bar{x})$ is a closure-type over $\sigma$ and let $p_\tau(\bar{x}) = p \uhrc \tau$.
If $p_\tau(\bar{x})$ is satisfied in $\mcT_n$ for some $n$, then (because of the assumptions on $\mbT$) it is cofinally satisfiable.
Since $\mbW_n$ contains all expansions to $\sigma$ of $\mcT_n$ it follows that if $p_\tau(\bar{x})$ 
is satisified in $\mcT_n$ for some $n$, then $p(\bar{x})$ is cofinally satisfiable.

\begin{defin}\label{definition of PLA-network} {\rm
(i) A {\bf \em $PLA^*(\sigma)$-network based on $\tau$} is specified by the following two parts:
\begin{enumerate}
\item A DAG $\mbbG$ with vertex set (or domain) $\sigma  \setminus  \tau$.

\item To each relation symbol $R \in \sigma \setminus \tau$ a formula 
$\theta_R(\bar{x}) \in PLA^*(\mr{par}(R) \cup \tau)$ 
is associated where $|\bar{x}|$ equals the arity of $R$ and
$\mr{par}(R)$ is the set of parents of $R$ in the DAG $\mbbG$.
We call $\theta_R$ the {\bf \em formula associated to $R$} by the $PLA^*(\sigma)$-network.
\end{enumerate}
We will denote a $PLA^*(\sigma)$-network by the same symbol (usually $\mbbG$, possibly with a sub or superscript)
as its underlying DAG.\\
(ii) Let $\mbbG$ denote a $PLA^*(\sigma)$-network based on $\tau$, 
let $\tau \subseteq \sigma' \subseteq \sigma$, and suppose that for
every $R \in \sigma'$, $\mr{par}(R) \subseteq \sigma'$. 
Then the $PLA^*(\sigma')$-network specified by the induced subgraph of $\mbbG$
with vertex set $\sigma' \setminus \tau$ 
and the associated formulas $\theta_R$ for all $R \in \sigma' \setminus \tau$ will be called the
{\bf \em $PLA^*(\sigma')$-subnetwork of $\mbbG$ induced by $\sigma'$}.
}\end{defin}

\noindent
Observe that if $\sigma = \tau$ then the DAG of a $PLA^*(\sigma)$-network has an empty vertex set
and hence there is no probability formula associated to it. 
Example~\ref{example of PLA-network} gives an example of a $PLA^*(\sigma)$-network.

\medskip
\noindent
{\bf  \em From now on let $\mbbG$ be a $PLA^*(\sigma)$-network based on $\tau$.}

\begin{defin}\label{the probability distribution induced by an PLA-network} {\rm 
(i) If $\sigma = \tau$  then $\mbbP_n$, the {\bf \em probability distribution on $\mbW_n$ induced by $\mbbG$}, 
is the unique probability distribution on (the singleton set) $\mbW_n$.\\
(ii) Now suppose that $\tau$ is a proper subset of $\sigma$.
For each $R \in \sigma$ let $\nu_R$ denote its arity, and let $\theta_R(\bar{x})$, where $|\bar{x}| = \nu_R$,
be the formula which $\mbbG$ associates to $R$.
Suppose that the underlying DAG of $\mbbG$ has height $\rho$.
For each $0 \leq l \leq \rho$ let $\mbbG_l$ be the subnetwork which is
induced by $\sigma_l = \{R \in \sigma : \text{$R$ is on level $i$ and $i \leq l$}\}$ and note that 
$\sigma_\rho = \sigma$ and $\mbbG_\rho = \mbbG$.
Also let $\sigma_{-1} = \tau$ and let $\mbbP^{-1}_n$ be the unique probability distribution on $\mbW^{-1}_n = \{\mcT_n\}$.
By induction on $r$ we define, for every $l = 0, 1, \ldots, \rho$, a probability distribution $\mbbP^l_n$ on the set 
$\mbW^l_n = \{ \mcA \uhrc \sigma_l : \mcA \in \mbW_n \}$
as follows: \\
For every $\mcA \in \mbW^l_n$, let $\mcA' = \mcA \uhrc \sigma_{l-1}$ and 
\[
\mbbP^l_n(\mcA) \ = \ 
\mbbP^{l-1}_n(\mcA') 
\prod_{R \in \sigma_l \setminus \sigma_{l-1}} \ 
\prod_{\bar{a} \in R^\mcA } 
\mcA' \big(\theta_R(\bar{a})\big) \ 
\prod_{\bar{a} \in \ (T_n)^{^{\nu_R}} \ \setminus \ R^\mcA } 
\big( 1 - \mcA' \big(\theta_R(\bar{a})\big) \big).
\]
Finally we let $\mbbP_n = \mbbP^\rho_n$ and note that $\mbW_n = \mbW^\rho_n$, so 
$\mbbP_n$ is a probability distribution on $\mbW_n$ which we call
{\bf \em the probability distribution on $\mbW_n$ induced by $\mbbG$}.
We also call $\mbbP = (\mbbP_n : n \in \mbbN^+)$
{\bf \em the sequence of probability distributions induced by $\mbbG$}.
}\end{defin}

\noindent
From
Definition~\ref{the probability distribution induced by an PLA-network}
of $\mbbP^l_n$ and $\mbbP_n$ we immediately get the following:

\begin{lem}\label{conditioning on A'}
Let $\rho \in \mbbN$, $l \in \{0, \ldots, \rho\}$, and let $\sigma_l$, $\mbW^l_n$, $\mbbP^l_n$ and $\mbbP_n$ be as in 
Definition~\ref{the probability distribution induced by an PLA-network}.\\
(i) Let $R \in \sigma_l \setminus \sigma_{l-1}$, 
$n \in \mbbN^+$, $\bar{a} \in (T_n)^{\nu_R}$ (where $\nu_R$ is the arity of $R$), and $\mcA' \in \mbW^{l-1}_n$.
Then
\[
\mbbP^l_n\big( \{ \mcA \in \mbW^l_n : \mcA \models R(\bar{a}) \} \ | \ 
\{ \mcA \in \mbW_n : \mcA \uhrc \sigma_{l-1} = \mcA' \}\big) \  = \ 
\mcA'(\theta_R(\bar{a})).
\]
(ii) Let $R_1, \ldots, R_t \in \sigma_l \setminus \sigma_{l-1}$
where we allow that $R_i = R_j$ even if $i \neq j$, 
$n \in \mbbN^+$, $\bar{a}_i \in (T_n)^{\nu_i}$ for $i = 1, \ldots, t$, and $\mcA' \in \mbW^{l-1}_n$.
Suppose that 
for $i = 1, \ldots, t$, $\varphi_i(\bar{x})$ is a literal in which $R_i$ occurs, 
and if $i \neq j$ then $\bar{a}_i \neq \bar{a}_j$ or $R_i \neq R_j$.
Using the probability distribution $\mbbP_n^l$,
the event $\mbE_n^{\varphi_1(\bar{a})} = \{\mcA \in \mbW^l_n : \mcA \models \varphi_1(\bar{a}) \}$ is 
independent of the event $\bigcap_{i = 2}^t \mbE_n^{\varphi_i(\bar{a})}$
(where $\mbE_n^{\varphi_i(\bar{a})}$ is defined similarly), {\bf \em conditioned} on the event 
$\{\mcA \in \mbW_n : \mcA \uhrc \sigma_{l-1} = \mcA'\}$. \\
(iii) Suppose that $\mbX_n \subseteq \mbW_n^{l-1}$ and $\mbY_n = \{\mcA \in \mbW_n^l : \mcA \uhrc \sigma_{l-1} \in \mbX_n \}$.
Then $\mbbP_n^l(\mbY_n) = \mbbP_n^{l-1}(\mbX_n)$.
\end{lem}

\begin{exam}\label{example of PLA-network}{\rm
Recall the very informal example in the introduction where $\sigma = \tau \cup \{P_1, P_2, P_3, R\}$ where 
$P_1, P_2, P_3$ are unary and $R$ binary. A $PLA^*(\sigma)$-network based on $\tau$ can (for example) have the 
following underlying DAG:

\begin{figure}[h!]
\begin{center}
\includegraphics[scale=0.9]{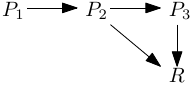}
\end{center}
\end{figure}

\noindent
The $PLA^*(\sigma)$-network must also associate a $PLA^*(\sigma)$-formula to each relation symbol $P_1, P_2, P_3, R$.
Examples of such formulas will be described informally, and for the descriptions to make sense we imagine that
they refer to a (large) tree $\mcT_n$ of height 3 as in 
Assumption~\ref{properties of the trees} (with $\Delta = 3$).
Moreover, we describe how the ``output'' of such formulas (a number in $[0, 1]$) depends on the properties of its input
(a vertex, or pair of vertices).
\begin{enumerate}
\item[] $\theta_{P_1}(x)$ =  ``If $x$ is a child of the root then $1/3$, else 0.''

\item[]$\theta_{P_2}(x)$ = `` If $x$ is not on level 2 then 0, else
\begin{itemize}
\item[] if the parent of $x$ satisfies $P_1$ then $2/3$, else $1/3$.''
\end{itemize}

\item[]$\theta_{P_3}(x)$ = ``If $x$ is not on level 3, then 0, else
\begin{itemize}
\item[] if the parent of $x$ satisfies $P_2$ then $1/3$, else $1/4$.''
\end{itemize}

\item[]$\theta_R(x, y)$ =  ``If $x$ is not on level 3 then 0, else 
\begin{itemize}
\item[] if $y$ is on level 3 then
\begin{itemize}
\item[] if $y$ is a sibling of $x$, then if $P_3(y)$ then $3/4$, else $1/2$, else
\item[] the proportion of $y$ on level 3 such that $P_3(y)$, multiplied by $3/4$, else
\end{itemize}
\item[] if $y$ is on level 2 then
\begin{itemize}
\item[] if $y$ is a parent of $x$, then if $P_2(y)$ then $3/4$, else $1/2$, else
\item[] the proportion of $y$ on level 2 such that $P_2(y)$ among the $y$ that are not a parent of $x$, else
\end{itemize}
\item[] if $y$ is on level 1 then the proportion of children of $y$ such that $P_1(y)$, multiplied by $2/3$.''
\end{itemize}

\end{enumerate}

}\end{exam}

\section{Convergence, balance, and asymptotic elimination of aggregation functions}\label{Convergence and balance}

\noindent
In this section we consider two technical notions, convergent pairs of formulas,
and balanced triples of formulas, which are at the heart of the proofs of the main results.
Intuitively speaking, a pair $(\varphi(\bar{x}), \psi(\bar{x}))$ of 0/1-valued formulas from $PLA^*(\sigma)$
converges to $\alpha$ if the probability that $\varphi(\bar{x})$ holds, conditioned on $\psi(\bar{x})$ being true,
converges to $\alpha$ as $n\to\infty$.
And, also intuitively speaking, a triple $(\varphi(\bar{x}, \bar{y}), \psi(\bar{x}, \bar{y}), \chi(\bar{x}))$ of 0/1-valued formulas 
is $\alpha$-balanced if with probability tending to 1 as $n \to \infty$,
a random $\mcA \in \mbW_n$ such that $\mcA \models \chi(\bar{a})$ satisfies that
$|\varphi(\bar{a}, \mcA)| / |\psi(\bar{a}, \mcA)| \approx \alpha$.

Recall the assumptions made on $\mbT = (\mcT_n : n \in \mbbN^+)$, $\mbW_n$, $\mbbG$,
and $\mbbP = (\mbbP_n : n \in \mbbN^+)$ in the previous two sections, which we keep in this section.

\subsection{Definitions and some immediate consequences}

\begin{defin}\label{definition of E}{\rm
Suppose that $\varphi(\bar{x}) \in PLA^*(\sigma)$ and $\bar{a} \in (T_n)^{|\bar{x}|}$ (for some $n$).
Then 
\[
\mbE_n^{\varphi(\bar{a})} = \big\{ \mcA \in \mbW_n : \mcA(\varphi(\bar{a})) = 1 \big\}.
\]
}\end{defin}

\begin{defin}\label{definition of convergence}{\rm
Let $\varphi(\bar{x}), \psi(\bar{x}) \in PLA^*(\sigma)$.
We say that $\big(\varphi, \psi \big)$ {\bf \em converges (to $\alpha \in [0, 1]$) with respect to $\mbbG$} if, 
for all $\varepsilon > 0$, there exists $n_0$, such that for all $n\geq n_0$ and all $\bar{a} \in T_n^{|\bar{x}|}$, 
\[
\big|\mathbb{P}_n\big(\mathbf{E}_n^{\varphi(\bar{a})} \ \big| \ 
\mathbf{E}_n^{\psi(\bar{a})}\big) - \alpha \big| \leq \varepsilon
\quad \text{if} \quad
\mbbP_n\big(\mbE_n^{\psi(\bar{a})}\big) > 0.
\]
If the stronger condition holds that 
$\mathbb{P}_n\big(\mathbf{E}_n^{\varphi(\bar{a})} \ \big| \ \mathbf{E}_n^{\psi(\bar{a})}\big) = \alpha$
whenever $n$ is large enough and $\mbbP_n\big(\mbE_n^{\psi(\bar{a})}\big) > 0$, then
we say that $(\varphi, \psi)$ is {\bf \em eventually constant (with value $\alpha$) with respect to $\mbbG$}.
}\end{defin}

\begin{rem}\label{remark on trivial convergences} {\rm
Let $\varphi(\bar{x}), \psi(\bar{x}) \in PLA^*(\sigma)$.
It follows from the definition that
if there are only finitely many $n$ such that there is $\bar{a} \in (T_n)^{|\bar{x}|}$ such that 
$\mbbP_n\big(\mbE_n^{\psi(\bar{a})}\big) > 0$ then, for every $\alpha$,
$(\varphi, \psi)$ converges to $\alpha$ with respect to $\mbbG$.
It also follows from the definition that if there are infinitely many 
$n$ such that $\mbbP_n\big(\mbE_n^{\psi(\bar{a})}\big) > 0$ for some $\bar{a}$,
but there are only finitely many $n$ such that there is 
$\bar{a} \in (T_n)^{|\bar{x}|}$ such that 
$\mbbP_n\big(\mbE_n^{\varphi(\bar{a}) \wedge \psi(\bar{a})}\big) > 0$ then 
$(\varphi, \psi)$ converges to 0 with respect to $\mbbG$.

Therefore, when proving that a pair $(\varphi, \psi)$ converges we will assume that 
there are infinitely many $n$ such that there is $\bar{a} \in (T_n)^{|\bar{x}|}$
such that $\mbbP_n\big(\mbE_n^{\varphi(\bar{a}) \wedge \psi(\bar{a})}\big) > 0$, where the positive probability
implies that for some $\mcA \in  \mbW_n$, $\mcA \models \varphi(\bar{a}) \wedge \psi(\bar{a})$.
}\end{rem}

\begin{defin}\label{definition of balance}{\rm
Let $\varphi(\bar{x}, \bar{y}), \psi(\bar{x}, \bar{y}), \chi(\bar{x}) \in PLA^*(\sigma)$.\\
(i) Let $\alpha \in [0, 1]$, $\varepsilon > 0$ and let $\mcA$ be a finite $\sigma$-structure.
The triple $(\varphi, \psi, \chi)$ is called {\bf \em $(\alpha, \varepsilon)$-balanced in $\mcA$} if
whenever $\bar{a} \in A^{|\bar{x}|}$ and $\mcA(\chi(\bar{a})) = 1$, then
\[
(\alpha - \varepsilon)|\psi(\bar{a}, \mcA)| \leq
|\varphi(\bar{a}, \mcA) \cap \psi(\bar{a}, \mcA)| \leq (\alpha + \varepsilon)|\psi(\bar{a}, \mcA)|.
\]
(ii) Let $\alpha \in [0, 1]$. 
The triple $(\varphi, \psi, \chi)$ is {\bf \em $\alpha$-balanced with respect to $\mbbG$}
if for all $\varepsilon > 0$, if 
\[
\mbX_n^\varepsilon = 
\big\{\mcA \in \mbW_n : \text{ $(\varphi, \psi, \chi)$ is $(\alpha, \varepsilon)$-balanced in $\mcA$} \big\}
\]
then $\lim_{n\to\infty} \mbbP_n\big(\mbX_n^\varepsilon\big) = 1$.
The triple $(\varphi, \psi, \chi)$ is {\bf \em balanced with respect to $\mbbG$} if, for some $\alpha \in [0, 1]$, it is
$\alpha$-balanced with respect to $\mbbG$.
If, in addition, $\alpha > 0$ then we call $(\varphi, \psi, \chi)$ {\bf \em positively balanced (with respect to $\mbbG$)}.
}\end{defin}

\begin{rem}\label{balance follows from inconsistency}{\rm
Let $\varphi(\bar{x}, \bar{y}), \psi(\bar{x}, \bar{y}), \chi(\bar{x}) \in PLA^*(\sigma)$ be $0/1$-valued.
Suppose that $\varphi \wedge \psi \wedge \chi$ is not cofinally satisfiable.
We claim that then $(\varphi, \psi, \chi)$ is 0-balanced with respect to $\mbbG$.
It suffices to show that, for all sufficiently large $n$, if $\mcA \in \mbW_n$, $\mcA \models \chi(\bar{a})$,
and $\mcA \models \psi(\bar{a}, \bar{b})$, then $\mcA \not\models \varphi(\bar{a}, \bar{b})$.
But this is immediate from the assumption that  $\varphi \wedge \psi \wedge \chi$ is not cofinally satisfiable.
{\em For this reason we may assume, when proving results about balanced triples, that the conjunction of the involved formulas
is cofinally satisfiable and in particular consistent.}
}\end{rem}

\begin{rem}\label{0-dimensional types are balanced}{\rm
Let $p(\bar{x}, \bar{y})$ be a closure type over $\sigma$, let $q(\bar{x}) = p \uhrc \bar{x}$, 
and let $p_\tau(\bar{x}, \bar{y}) = p \uhrc \tau$ (so $p \models q$ and $p \models p_\tau$).
Suppose that $\rank_{\bar{y}}(p_\tau) = 0$. 
We claim that $(p, p_\tau, q)$ is 1-balanced with respect to $\mbbG$.
Since $\rank_{\bar{y}}(p_\tau) = 0$ we have $p_\tau(\bar{x}, \bar{y}) \models \rng(\bar{y}) \subseteq \cl(\bar{x})$.
(So if $\bar{x}$ is empty then $\bar{y}$ is a single variable, say $y$,  and $p_\tau(y)$ expresses that $y$ is the root.)
It follows that $q(\bar{x}) \models \exists \bar{y} p(\bar{x}, \bar{y})$
(so if $\bar{x}$ is empty then $q$ is a sentence which expresses which relations the root satisfies), and
that if $\bar{a} \in (T_n)^{|\bar{x}|}$ and $\bar{b} \in (T_n)^{|\bar{y}|}$,
$\mcT_n \models p_\tau(\bar{a}, \bar{b})$, then $\bar{b}$ is the unique tuple that satisfies $p_\tau(\bar{a}, \bar{y})$ in $\mcT_n$.
So for every $n$, $\bar{a} \in (T_n)^{|\bar{x}|}$, and $\mcA \in \mbW_n$, 
if  $\mcA \models q(\bar{a})$ then $|p(\bar{a}, \mcA)| = |p_\tau(\bar{a}, \mcA)| =1$.
Hence $(p, p_\tau, q)$ is 1-balanced.
}\end{rem}

\subsection{Proofs of results about convergence and balance}\label{Proofs of results about convergence and balance}

We will use induction on the height of the underlying DAG of $\mbbG$ to prove that some pairs of formulas converge and that some
triples of formulas are balanced. 
The base case will {\em not} be when the height is 0, but it will be when the height is $-1$, where we recall the convention
(from Definition~\ref{notions in a DAG}) that an empty $DAG$ (one without any element/vertex) has height $-1$.
The assumption that $\tau = \sigma$ is equivalent to the assumption that the height of the underlying DAG of $\mbbG$ is $-1$.
So the base case considers the case when $\tau = \sigma$.
The following lemma shows that the induction hypothesis 
(Assumption~\ref{induction hypothesis})
that we will use holds in the base case when $\tau = \sigma$.

\begin{lem}\label{base case}
Suppose that $\tau = \sigma$, so $\mbW_n = \{\mcT_n\}$ for each $n$ and
let $\mbbG$ be the unique $PLA^*(\sigma)$-network over $\tau$.
For each $n$ let $\mbbP_n$ be the probability distribution on $\mbW_n$ which is induced by $\mbbG$ (so $\mbbP_n(\mcT_n) = 1$). Then:
\begin{enumerate}
\item If $p(\bar{x})$ and $q(\bar{x})$ are closure types over $\tau$ then $(p, q)$ converges with respect to $\mbbG$.

\item Suppose that $p(\bar{x}, \bar{y})$, $r(\bar{x}, \bar{y})$ and $q(\bar{x})$ are closure types over $\tau$ and
suppose that $p \wedge r \wedge q$ is cofinally satisfiable.
Then there is $\alpha \in [0, 1]$ such that for all $\varepsilon > 0$ there is $c > 0$ such that for all sufficiently
 large $n$, if $\bar{a} \in (T_n)^{|\bar{x}|}$, $B \subseteq r(\bar{a}, \mcT_n)$, and $|B| \geq g_3(n)$, then
 \begin{align*}
 &\mbbP_n \big( \big\{ \mcA \in \mbW_n : \text{ if $\mcA \models q(\bar{a})$ then }
 (\alpha - \varepsilon)|B| \leq |p(\bar{a}, \mcA) \cap B| \leq (\alpha + \varepsilon)|B| \big\} \big) \ \geq \\
 &1 - e^{-c g_3(n)}.
 \end{align*}
\end{enumerate}
\end{lem}

\noindent
{\bf Proof.}
(1) Suppose that $p(\bar{x})$ and $q(\bar{x})$ are closure types over $\tau$.
By Remark~\ref{remark on trivial convergences}
we may assume that $p(\bar{x}) \wedge q(\bar{x})$ is satisfiable in some $\mcT_n$.
It follows from Definition~\ref{definition of closure type}
of closure type over $\tau$ that $p$ and $q$ are equivalent, so for all $n$ we have 
$\mathbb{P}_n\big(\mathbf{E}_n^{p(\bar{a})} \ \big| \ \mathbf{E}_n^{q(\bar{a})}\big) = 1$
if $\mcT_n \models q(\bar{a})$, or equivalently, if
$\mbbP_n\big(\mbE_n^{q(\bar{a})}\big) > 0$.
Hence $(p, q)$ converges to 1 with respect to $\mbbG$.
In fact $(p, q)$ is eventually constant with value 1.

(2) Suppose that $p(\bar{x}, \bar{y})$, $r(\bar{x}, \bar{y})$ and $q(\bar{x})$ are closure types over $\tau$ and
suppose that $p \wedge r \wedge q$ is cofinally satisfiable.
Then $p$ and $r$ are equivalent and $p$ implies $q$.
Suppose that $B \subseteq r(\bar{a}, \mcT_n)$, and $|B| \geq g_3(n)$.
As $\lim_{n\to\infty}g_3(n) = \infty$ this means that $B \neq \es$ if $n$ is large enough.
Suppose that $\mcA \in \mbW_n$, so $\mcA = \mcT_n$. Also suppose that $\mcT_n \models q(\bar{a})$. 
Let $\bar{b} \in B$, so $\mcT_n \models r(\bar{a}, \bar{b})$ and hence $\mcT_n \models p(\bar{a}, \bar{b})$
(as $p$ and $r$ are equivalent).
Thus $B \subseteq p(\bar{a}, \mcT_n)$ and $|B \cap p(\bar{a}, \mcA)| = |B|$, so for every $\varepsilon > 0$,
$
(1 - \varepsilon)|B| \leq |p(\bar{a}, \mcT_n) \cap B| \leq (1 + \varepsilon)|B|.
$
Since $\mbbP_n(\mcT_n) = 1$ we get the conclusion of part~(2).
\hfill $\square$ \\

\noindent
{\bf \em
For the rest of 
Section~\ref{Convergence and balance} 
we assume that the height of the underlying DAG of $\mbbG$ is $\rho + 1$ where $\rho \geq -1$.
We also let {\rm
\begin{align*}
&\sigma_\rho = \tau \cup \{ R \in \sigma \setminus \tau : \text{ $R$ is on level $i$ of the underlying graph of $\mbbG$ where $i \leq \rho$} \},\\
&\mbW_n^\rho = \{\mcA \uhrc \sigma_\rho : \mcA \in \mbW_n \},
\end{align*}
}
and we let $\mbbG_\rho$ be the subnetwork of $\mbbG$ which is induced by $\sigma_\rho$.
Moreover, we assume the following:
}

\begin{assump}\label{induction hypothesis} {\bf (Induction hypothesis)} {\rm
 \begin{enumerate}
 	\item For every $R \in \sigma \setminus \tau$, there exists a closure-basic formula
 	$\chi_R \in PLA^*(\sigma_\rho)$, such that $\chi_R$ and $\theta_R$ are asymptotically equivalent with respect to 
 	$\mbbG_\rho$, where $\theta_R$ is the formula of $\mbbG$ associated to $R$.
 	
 	\item If $p(\bar{x})$ is a self-contained closure type over $\sigma_\rho$ and $q(\bar{x})$ is a 
 	self-contained closure type over $\tau$, then $(p, q)$ converges
 	with respect to $\mbbG_\rho$.
 	
 	\item Suppose that $p(\bar{x}, y)$ and $q(\bar{x})$ are complete closure types over $\sigma_\rho$ and that
 	$p_\tau(\bar{x}, y)$ is a self-contained closure type over $\tau$ such that $\rank_y(p_\tau) = 1$. 
 	Also suppose that $p \wedge p_\tau \wedge q$ is cofinally
 	satisfiable.
 	Then there is $\alpha \in [0, 1]$ such that for all $\varepsilon > 0$ there is $c > 0$ such that for all sufficiently
 	large $n$, if $\bar{a} \in (T_n)^{|\bar{x}|}$, $B \subseteq p_\tau(\bar{a}, \mcT_n)$, and $|B| \geq g_3(n)$, then
 	\begin{align*}
 	&\mbbP_n^\rho \big( \big\{ \mcA \in \mbW_n^\rho : \text{ if $\mcA \models q(\bar{a})$ then }
 	(\alpha - \varepsilon)|B| \leq |p(\bar{a}, \mcA) \cap B| \leq (\alpha + \varepsilon)|B| \big\} \big) \ \geq \\
 	&1 - e^{-c g_3(n)}.
 	\end{align*}
 \end{enumerate}
 }
 \end{assump}
 
 \begin{rem}\label{remark about base case}{\rm
 Suppose (in this remark) that $\sigma_\rho = \tau$, or equivalently, that the height of the underlying DAG of $\mbbG$ 
(with vertex set $\sigma$)  is $0$, so all $R \in \sigma$ are on level 0.
 Then part~(1) of Assumption~\ref{induction hypothesis} holds vacuously and
 Lemma~\ref{base case} implies that parts~(2) and~(3) hold.
{\em So if the height of the underlying DAG of $\mbbG$ is 0 then Assumption~\ref{induction hypothesis} holds.}
 }\end{rem}
 
\noindent
The goal of the rest of Section~\ref{Convergence and balance} is to prove, in the following order, that 
 
\begin{itemize}
\item[(A)] part~(2) of Assumption~\ref{induction hypothesis} holds if $\sigma_\rho$ and $\mbbG_\rho$ 
are replaced by $\sigma$ and $\mbbG$, respectively,
\item[(B)] part~(3) of Assumption~\ref{induction hypothesis} holds if $\sigma_\rho$, $\mbbP_n^\rho$,
and $\mbW_n^\rho$ are replaced by $\sigma$, $\mbbP_n$, and $\mbW_n$, respectively, and
\item[(C)] if $\varphi(\bar{x}) \in PLA^*(\sigma)$ satisfies certain conditions
(stated in Proposition~\ref{elimination in the inductive step}) 
then $\varphi$ is asymptotically equivalent (with respect to $\mbbG$) to a closure-basic formula
over~$\sigma$. 
It will follow that if $\sigma \subset \sigma^+$ and $\mbbG$ is an induced subnetwork of a $PLA^*(\sigma^+)$-network
$\mbbG^+$ such that $\mbbG^+$ has height $\rho + 2$, then part~(1)
of Assumption~\ref{induction hypothesis} holds if $\sigma$ and $\mbbG$ are replaced by $\sigma^+$ and $\mbbG^+$,
respectively, and $\sigma_\rho$ and $\mbbG_\rho$ are replaced by $\sigma$ and $\mbbG$, respectively.
\end{itemize}
 
\begin{defin}\label{definition of W-A'}{\rm
For every $n$ and $\mcA' \in \mbW_n^\rho$ let $\mbW^{\mcA'} = \big\{ \mcA \in \mbW_n : \mcA \uhrc \sigma_\rho = \mcA' \big\}$.
}\end{defin}

\subsection{Convergence}\label{Convergence}

In this subsection we prove statement~(A) above. We do it in three steps.

\begin{lem} \label{(p, p-s) convergence} 
Let $p(\bar{x})$ be a self-contained closure type over $\sigma$ and $p_\rho(\bar{x})$ a self-contained complete 
closure type over $\sigma_\rho$. 
Then $(p, p_\rho)$ converges with respect to $\mbbG$.
\end{lem}

\noindent
{\bf Proof.} 
According to Remark~\ref{remark on trivial convergences}
we may assume that $p(\bar{x}) \wedge p_\rho(\bar{x})$ is satisfiable in some $\mcA \in \mbW_n$ for infinitely many $n$.
Let $\varepsilon > 0$. For every $R \in \sigma \setminus \sigma_\rho$ let $\nu_R$ be the arity of $R$ and let 
$\theta_R(x_1, \ldots, x_{\nu_R})$ be
the $PLA^*(\sigma_\rho)$-formula associated to $R$ by $\mbbG$.
By the first part of Assumption~\ref{induction hypothesis}  there is a closure-basic formula 
$\chi_R(x_1, \ldots, x_{\nu_R})$ over $\sigma_\rho$ which is  asymptotically equivalent  to $\theta_R$ with respect to $\mbbG_\rho$.
Let
\begin{align*}
&\mbX_n^{\rho, \varepsilon} = \big\{ \mcA \in \mbW_n^\rho : \text{ for all $R \in \sigma \setminus \sigma_\rho$ 
and all $\bar{a} \in (T_n)^{\nu_R}$, }
\big| \mcA(\theta_R(\bar{a})) - \mcA(\chi_R(\bar{a})) \big| \leq \varepsilon \big\}, \\ 
&\text{ and} \\
&\mbX_n^{\varepsilon} = \big\{ \mcA \in \mbW_n : \text{ for all $R \in \sigma \setminus \sigma_\rho$ 
and all $\bar{a} \in (T_n)^{\nu_R}$, }
\big| \mcA(\theta_R(\bar{a})) - \mcA(\chi_R(\bar{a})) \big| \leq \varepsilon \big\}.
\end{align*}
As $\theta_R$ and $\chi_R$ are asymptotically equivalent with respect to $\mbbG_\rho$
it follows that $\lim_{n\to\infty}$ $\mbbP_n^\rho$ $\big(\mbX_n^{\rho, \varepsilon}\big) = 1$.
By Lemma~\ref{conditioning on A'}~(iii) 
we also get $\lim_{n\to\infty} \mbbP_n\big(\mbX_n^{\varepsilon}\big) = 1$.

Thus, to show that $(p, p_\rho)$ converges it suffices to show that there is $\alpha \in [0, 1]$ such that for every
$\varepsilon' > 0$ there is $\varepsilon > 0$
such that for all sufficiently large $n$ and all $\bar{a} \in T_n^{|\bar{x}|}$, 
\[
\Big|\mathbb{P}_n\Big(\mathbf{E}_n^{p(\bar{a})} \ \big| \ 
\mathbf{E}_n^{p_\rho(\bar{a})}  \cap \mbX_n^\varepsilon \Big) - \alpha \Big| \leq \varepsilon'
\ \ \text{  if  } \ \ \mbbP_n\big(\mbE_n^{p_\rho(\bar{a})}  \cap \mbX_n^\varepsilon \big) > 0.
\]
Let $\bar{a} \in (T_n)^{|\bar{x}|}$ and 
observe that $\mbE_n^{p_\rho(\bar{a})}  \cap \mbX_n^\varepsilon$ is the disjoint union of
sets $\mbW^{\mcA'}$ as $\mcA'$ ranges over structures in $\mbX_n^{\rho, \varepsilon}$ 
such that $\mcA' \models p_\rho(\bar{a})$.
By Lemma~\ref{basic fact about conditional probabilities},
 it suffices to show that there is $\alpha$
such that for all $\varepsilon' > 0$ there is $\varepsilon > 0$ such that for all sufficiently large $n$ 
and all $\mcA' \in \mbX_n^{\rho, \varepsilon}$
such that $\mcA' \models p_\rho(\bar{a})$ and $\mbbP_n\big(\mbW^{\mcA'}\big) > 0$ we have 
\begin{equation}\label{main equation to prove in first convergence lemma}
\Big|\mathbb{P}_n\Big(\mathbf{E}_n^{p(\bar{a})} \ \big| \ \mbW^{\mcA'} \Big) - \alpha \Big| \leq \varepsilon'.
\end{equation}
So suppose that $\mcA' \in \mbX_n^{\rho, \varepsilon}$, $\mcA' \models p_\rho(\bar{a})$ and $\mbbP_n\big(\mbW^{\mcA'}\big) > 0$.
Let $\bar{x} = (x_1, \ldots, x_m)$ and $\bar{a} = (a_1, \ldots, a_m)$.

Since we assume that $p(\bar{x})$ is a {\em self-contained} closure type over $\sigma$
it follows, by the definition of how $\mbbG$ induces $\mbbP_n$, that
\begin{align*}
&\mbbP_n\big( \mbE_n^{p(\bar{a})} \ \big| \ \mbW^{\mcA'}\big) = \\
&\prod_{\substack{R\in\sigma \setminus \sigma_\rho \\ 1 \leq i_1 < \ldots < i_{\nu_R} \leq m\\
p(\bar{x})\models R(x_{i_1},...,x_{i_{\nu_R}})}} \mcA'\big(\theta_R(a_{i_1}, \ldots, a_{i_{\nu_R}})\big)
\prod_{\substack{R\in \sigma\setminus \sigma_\rho\\ 1 \leq i_1 < \ldots < i_{\nu_R} \leq m\\
p(\bar{x})\models \neg R(x_{i_1},...,x_{i_{\nu_R}})}} \Big(1-\mcA'\big(\theta_R(a_{i_1},...,a_{i_{\nu_R}})\big)\Big).
\end{align*}
Let $\Theta_n(\bar{a}) = $
\begin{align*}
\prod_{\substack{R\in\sigma \setminus \sigma_\rho \\ 1 \leq i_1 < \ldots < i_{\nu_R} \leq m\\
p(\bar{x})\models R(x_{i_1},...,x_{i_{\nu_R}})}} \mcA'\big(\chi_R(a_{i_1}, \ldots, a_{i_{\nu_R}})\big)
\prod_{\substack{R\in \sigma\setminus \sigma_\rho\\ 1 \leq i_1 < \ldots < i_{\nu_R} \leq m\\
p(\bar{x})\models \neg R(x_{i_1},...,x_{i_{\nu_R}})}} \Big(1-\mcA'\big(\chi_R(a_{i_1},...,a_{i_{\nu_R}})\big)\Big).
\end{align*}
Let $\varepsilon' > 0$. 
Since $\mcA' \in \mbX_n^{\rho, \varepsilon}$ it follows that if $\varepsilon > 0$ is chosen small enough 
(and the choice depends only on $\varepsilon'$, $p$ and $p_\rho$), then
\begin{equation*}
\big|\mbbP_n\big(\mbE_n^{p(\bar{a})} \ \big| \ \mbW^{\mcA'}\big)  - \Theta_n(\bar{a}) \big| \ \leq \ \varepsilon'.
\end{equation*}
Recall that, for every $R \in \sigma \setminus \sigma_\rho$, $\chi_R(x_1, \ldots, x_{\nu_R})$ 
is a closure-basic formula over $\sigma_\rho$,
so it has the form
\[
\bigwedge_{j=1}^{\eta_R} \big( \psi_{R, j}(x_1, \ldots, x_{\nu_R}) \rightarrow c_{R, j}\big)
\ \ \text{ where $c_{R, j} \in [0, 1]$}
\]
and $\psi_{R, j}$ is a closure type over $\sigma_\rho$.
Since $p_\rho(x_1, \ldots, x_m)$ is a {\em complete} closure type over $\sigma_\rho$ it follows that for all $R \in \sigma \setminus \sigma_\rho$
and all
$1 \leq i_1 < \ldots < i_{\nu_R} \leq m$, either
\[
p_\rho(x_1, \ldots, x_m) \models \psi_{R, j}(x_{i_1}, \ldots, x_{i_{\nu_R}}) \ \text{ or } \
p_\rho(x_1, \ldots, x_m) \models \neg\psi_{R, j}(x_{i_1}, \ldots, x_{i_{\nu_R}}).
\]
So for all $R \in \sigma \setminus \sigma_\rho$
and all $1 \leq i_1 < \ldots < i_{\nu_R} \leq m$,
$\mcA\big(\chi_R(a_{i_1}, \ldots, a_{i_{\nu_R}})\big)$ is determined only by $p_\rho$.
Hence $\Theta_n(\bar{a})$ is determined only by $p_\rho$ and $p$ (and it is a product of numbers of the
form $c_{R, j}$ and $(1 - c_{R, j})$).
Let $\alpha = \Theta_n(\bar{a})$. 
Then, if $\varepsilon > 0$ is small enough,
\begin{equation*}
\big|\mbbP_n\big(\mbE_n^{p(\bar{a})} \ \big| \ \mbW^{\mcA'}\big)  -  \alpha \big| \ \leq \ \varepsilon'
\end{equation*}
 This completes the proof.
\hfill $\square$

\begin{prop}\label{(p, p-tau) convergence}
Suppose that $p(\bar{x})$ is a complete self-contained closure type over $\sigma$ and that
$p_\tau(\bar{x})$ is a closure type over $\tau$.
Then $(p, p_\tau)$ converges with respect to $\mbbG$.
\end{prop}

\noindent
{\bf Proof.}
As said in Remark~\ref{remark on trivial convergences}
we may assume that there are infinitely many $n$ such that 
$p(\bar{x}) \wedge p_\tau(\bar{x})$ is satisfiable in some $\mcA \in \mbW_n$, and it follows that 
$p \uhrc \tau$ is equivalent to $p_\tau$ (as every closure type over $\tau$ is a complete closure type over $\tau$,
by the definition). 
Let $p_\rho = p \uhrc \sigma_\rho$, so $p_\rho$ is a complete closure type over $\sigma_\rho$.
By Assumption~\ref{induction hypothesis}~(2)
there is $\alpha$ such that $(p_\rho, p_\tau)$ converges to $\alpha$ with respect to $\mbbG_\rho$.
By Lemma~\ref{(p, p-s) convergence} there is $\beta$ such that $(p, p_\rho)$ converges to $\beta$ with respect to $\mbbG$.
Hence, for all $\varepsilon > 0$ there is $n_0$ such that for all $n \geq n_0$ and all $\bar{a} \in (T_n)^{|\bar{x}|}$,
\begin{align*}
&\mbbP_n\big( \mbE_n^{p_\rho(\bar{a})} \ \big| \ \mbE_n^{p_\tau(\bar{a})} \big) \ \in \ [\alpha - \varepsilon, \alpha + \varepsilon]
\quad \text{ if $\mbbP_n\big(\mbE_n^{p_\tau(\bar{a})}\big) > 0$ and} \\
&\mbbP_n\big( \mbE_n^{p(\bar{a})} \ \big| \ \mbE_n^{p_\rho(\bar{a})} \big) \ \in \ [\beta - \varepsilon, \beta + \varepsilon]
\quad \text{ if $\mbbP_n\big(\mbE_n^{p_\rho(\bar{a})}\big) > 0$}.
\end{align*}
Now we get
\begin{align*}
&\mbbP_n\big( \mbE_n^{p(\bar{a})} \ \big| \ \mbE_n^{p_\tau(\bar{a})} \big) =
\mbbP_n\big( \mbE_n^{p(\bar{a})} \ \big| \ \mbE_n^{p_\rho(\bar{a})} \big) \cdot 
\mbbP_n\big( \mbE_n^{p_\rho(\bar{a})} \ \big| \ \mbE_n^{p_\tau(\bar{a})} \big) \ \in \ 
[\alpha \beta - 3\varepsilon, \alpha \beta + 3\varepsilon].
\end{align*}
Since $\varepsilon > 0$ can be chosen as small as we like this completes the proof.
\hfill $\square$

\begin{cor}\label{corollary to (p, p-tau) convergence}
Suppose that $p(\bar{x})$ is a (not necessarily complete) self-contained closure type over $\sigma$ and that
$p_\tau(\bar{x})$ is a closure type over $\tau$.
Then $(p, p_\tau)$ converges with respect to $\mbbG$.
\end{cor}

\noindent
{\bf Proof.}
Let $p(\bar{x})$ and $p_\tau(\bar{x})$ be as assumed. Then there are complete self-contained closure types over $\sigma$
$p_1(\bar{x}), \ldots, p_k(\bar{x})$ such that 
$p(\bar{x})$ is equivalent to $\bigvee_{i=1}^k p_i(\bar{x})$ and if $i \neq j$ then $p_i(\bar{x}) \wedge p_j(\bar{x})$ is
inconsistent.
Then for every $n$ and $\bar{a} \in (T_n)^{|\bar{x}|}$ we have
\[
\mbbP_n\big(\mbE_n^{p(\bar{a})} \ \big| \ \mbE_n^{p_\tau(\bar{a})} \big) =
\sum_{i=1}^k \mbbP_n\big( \mbE_n^{p_i(\bar{a})} \ \big| \ \mbE_n^{p_\tau(\bar{a})} \big) 
\]
where, for each $i$, $(p_i, p_\tau)$ converges, by
Proposition~\ref{(p, p-tau) convergence}, to some $\alpha_i$.
It follows that $(p, p_\tau)$ converges to $\alpha = \alpha_1 + \ldots + \alpha_k$.
\hfill $\square$

\begin{rem}\label{induction completed for convergence}{\rm
By Corollary~\ref{corollary to (p, p-tau) convergence},
part~(2) of Assumption~\ref{induction hypothesis} holds if $\sigma_\rho$ 
and $\mbbG_\rho$ are replaced by $\sigma$ and $\mbbG $, respectively.
So the induction step for convergence is completed, that is, claim~(A) above is proved.
}\end{rem}

\begin{rem}\label{remark about basic theta-R}{\rm
Let $p(\bar{x})$, $p_\rho(\bar{x})$, and $p_\tau(\bar{x})$ be complete closure types over $\sigma$, $\sigma_\rho$, and $\tau$,
respectively (and recall that a closure type over $\tau$ is the same as a complete closure type over $\tau$).
Suppose that $p \models p_\rho$ and $p_\rho \models p_\tau$.
Suppose that, for every $R \in \sigma \setminus \sigma_\rho$, 
$\theta_R$ is a closure-basic formula over $\sigma_\rho$.
In the proof of Lemma~\ref{(p, p-s) convergence} we can then let $\chi_R$ be the {\em same} formula as $\theta_R$.
Then, with the notation of that proof, we get 
$\mbbP_n\big(\mbE_n^{p(\bar{a})} \ \big| \ \mbE_n^{p_\rho(\bar{a})} \big) = \Theta_n(\bar{a}) = \alpha$
for all $n$ and $\bar{a} \in (T_n)^{|\bar{x}|}$ such that $\mbbP_n\big(\mbE_n^{p_\rho(\bar{a})}\big) > 0$. 
Hence $(p, p_\rho)$ is eventually constant with value $\alpha$.

Suppose, in addition, that $(p_\rho, p_\tau)$ is eventually constant with value $\beta$. 
It follows straightforwardly from the definition of `eventually constant'
 that $(p, p_\tau)$ is eventually constant with value $\alpha\beta$.
 By induction on the height of the underlying DAG of $\mbbG$, it follows that if, for all $R \in \sigma \setminus \tau$,
 $\theta_R$ is a closure-basic formula, then $(p, p_\tau)$ is eventually constant, for all closure types $p_\tau(\bar{x})$
 over $\tau$ and all complete closure types $p(\bar{x})$ over $\sigma$.
 By the proof of  Corollary~\ref{corollary to (p, p-tau) convergence} it follows that (under the same assumption)
$(p, p_\tau)$ is eventually constant also in the case when $p$ is not complete.
}\end{rem}

\subsection{Balance}\label{Balance}

In this section we prove statement~(B), in the first part of
Lemma~\ref{balance, rank 1}, 
but we also prove other results that will be used to prove statement~(C).

\begin{lem}\label{balance, rank 1 for a subset}
Suppose that $p_\rho(\bar{x}, y)$ is a complete self-contained closure type over $\sigma_\rho$ and that
$p(\bar{x}, y)$ is a self-contained closure type over $\sigma$ such that $p \models p_\rho$ 
(or equivalently, $p_\rho$ is equivalent to $p \uhrc \sigma_\rho$).
Suppose that $\rank_y(p_\rho) \ (= \rank_y(p)) = 1$ and that for every $R \in \sigma \setminus \sigma_\rho$,
if $\bar{z}$ is a subsequence of $\bar{x}y$ and $p \models R(\bar{z})$ or $p \models \neg R(\bar{z})$,
then $\bar{z}$ contains $y$.
Furthermore, suppose that $n \in \mbbN^+$, $\bar{a} \in (T_n)^{|\bar{x}|}$, $\mcA' \in \mbW_n^\rho$,
$P \subseteq p_\rho(\bar{a}, \mcA')$ and $\mbbP_n^\rho(\mcA') > 0$.
There is $\alpha \in [0, 1]$, depending only on $p_\rho$ and $p$, such that for every $\varepsilon > 0$
there is $c > 0$, depending only on $\alpha$ and $\varepsilon$, such that if $n$ and $|P|$ are large enough, then
\[
\mbbP_n\big(\big\{ \mcA \in \mbW_n : (\alpha - \varepsilon)|P| \leq
|P \cap p(\bar{a}, \mcA)| \leq (\alpha + \varepsilon)|P| \big\} \big) \ \big| \  \mbW^{\mcA'} \big) \ 
\geq \ 1 - e^{-c|P|}.
\]
\end{lem}

\noindent
{\bf Proof.}
We adopt all assumptions of the lemma. 
Since $\mbbP_n^\rho(\mcA') > 0$, it follows from 
Lemma~\ref{conditioning on A'}
that $\mbbP_n\big(\mbW^{\mcA'}\big) > 0$.
As $\mcA' \models p_\rho(\bar{a}, b)$ for all $b \in P$, it follows that for all $b \in P$,
$\mbbP_n\big(\mbE_n^{p_\rho(\bar{a}, b)}\big) > 0$.
Lemma~\ref{(p, p-s) convergence} tells that, for some $\alpha$, $(p, p_\rho)$ converges to $\alpha$ with respect to $\mbbG$.
So for every $\varepsilon > 0$, if $n$ is large enough we have
\[
\mbbP_n\big(\mbE_n^{p(\bar{a}, b)} \ \big| \ \mbE_n^{p_\rho(\bar{a}, b)}\big) \ \in \ [\alpha - \varepsilon, \alpha + \varepsilon]
\quad \text{ for all } b \in P.
\]
Since, by assumption, $\mcA' \models p_\rho(\bar{a}, b)$ for all $b \in P$ we have 
$\mbW^{\mcA'} \subseteq \mbE_n^{p_\rho(\bar{a}, b)}$ for all $b \in P$.
Hence
\[
\mbbP_n\big(\mbE_n^{p(\bar{a}, b)} \ \big| \ \mbW^{\mcA'} \big) = 
\mbbP_n\big(\mbE_n^{p(\bar{a}, b)} \ \big| \ \mbE_n^{p_\rho(\bar{a}, b)}\big)
 \ \in \ [\alpha - \varepsilon, \alpha + \varepsilon]
\quad \text{ for all } b \in P.
\]
By the assumption that $p_\rho$ (and hence $p$) is self-contained,
the assumption regarding the literals using a relation symbol from $\sigma \setminus \sigma_\rho$, and 
Lemma~\ref{conditioning on A'}, 
if we condition on $\mbW^{\mcA'}$,
then for every $b \in P$ the event $\mbE_n^{p(\bar{a}, b)}$ is independent
from the events $\mbE_n^{p(\bar{a}, b')}$ as
$b'$ ranges over $P \setminus \{b\}$.
Corollary~\ref{independent bernoulli trials, second version}
now implies that there is $c > 0$, depending only on $\alpha$ and $\varepsilon$, such that if $n$ and $|P|$ are large enough then
\[
\mbbP_n\big(\big\{ \mcA \in \mbW_n : (\alpha - \varepsilon)|P| \leq
|P \cap p(\bar{a}, \mcA)| \leq (\alpha + \varepsilon)|P| \big\} \big) \ \big| \  \mbW^{\mcA'} \big) \ 
\geq \ 1 - e^{-c|P|},
\]
so the proof is complete. 
\hfill $\square$

\begin{lem}\label{balance, rank 1}
Let $p_\tau(\bar{x}, y)$ be a self-contained closure type over $\tau$ and let
$p(\bar{x}, y)$ and $q(\bar{x})$ be complete closure types over $\sigma$.
Suppose that $\rank_y(p_\tau) = 1$.\\
(i) Also suppose that $p \wedge p_\tau \wedge q$ is cofinally satisfiable.
 Then there is $\gamma \in [0, 1]$ such that for all $\varepsilon > 0$ there is $c > 0$ such that for all sufficiently
 large $n$, if $\bar{a} \in (T_n)^{|\bar{x}|}$, $B \subseteq p_\tau(\bar{a}, \mcT_n)$, and $|B| \geq g_3(n)$, then
 \begin{align*}
 &\mbbP_n \big( \big\{ \mcA \in \mbW_n : \text{ if $\mcA \models q(\bar{a})$ then }
 (\gamma - \varepsilon)|B| \leq |p(\bar{a}, \mcA) \cap B| \leq (\gamma + \varepsilon)|B| \big\} \big) \ \geq \\
 &1 - e^{-cg_3(n)}.
 \end{align*} 
(ii) $(p, p_\tau, q)$ is balanced with respect to $\mbbG$.
\end{lem}

\noindent
{\bf Proof.}
(i) The assumption that $p \wedge p_\tau \wedge q$ is cofinally satisfiable implies that $p \wedge p_\tau \wedge q$ is consistent
and this
implies that $q(\bar{x})$ is equivalent to $p \uhrc \bar{x}$ and $p_\tau(\bar{x}, \bar{y})$ is equivalent to $p \uhrc \tau$
(so $\rank_y(p) = \rank_y(p_\tau) = 1$).
Let $p_\rho(\bar{x}, y) = p \uhrc \sigma_\rho$ and $q_\rho(\bar{x}) = q \uhrc \sigma_\rho$, 
so $p_\rho(\bar{x}, y)$ and $q_\rho(\bar{x})$ are complete closure types over $\sigma_\rho$.
By the induction hypothesis, Assumption~\ref{induction hypothesis}, 
there is $\alpha \in [0, 1]$ such that for all $\varepsilon > 0$ there is $d > 0$ such that for all sufficiently
large $n$, if $\bar{a} \in (T_n)^{|\bar{x}|}$, $B \subseteq p_\tau(\bar{a}, \mcT_n)$, $|B| \geq g_3(n)$, and
\begin{align}\label{using the induction hypothesis for balance}
&\mbX_n^{\rho, \varepsilon} = \\
&\big\{ \mcA' \in \mbW_n^\rho : \text{ if $\mcA' \models q_\rho(\bar{a})$ then }
(\alpha - \varepsilon)|B| \leq |p_\rho(\bar{a}, \mcA') \cap B| \leq (\alpha + \varepsilon)|B| \big\}, \nonumber \\
&\text{then } \mbbP_n^\rho \big( \mbX_n^{\rho, \varepsilon} \big) \ \geq \ 1 - e^{-d g_3(n)}. \nonumber
\end{align}
 
First suppose that $\alpha = 0$.
Let $\mbX_n^\varepsilon = \big\{ \mcA \in \mbW_n : \mcA \uhrc \sigma_\rho \in \mbX_n^{\rho, \varepsilon} \big\}$.
By Lemma~\ref{conditioning on A'},
$\mbbP_n\big(\mbX_n^\varepsilon\big)  = \mbbP_n^\rho\big(\mbX_n^{\rho, \varepsilon}\big)$
so $\lim_{n\to\infty} \mbbP_n\big(\mbX_n^\varepsilon\big)  = 1$.
Since $p$ implies $p_\rho$ and $q$ implies $q_\rho$ it follows that if  $\mcA \in \mbX_n^\varepsilon$ and $\mcA \models q(\bar{a})$,
then $|p(\bar{a}, \mcA)| \leq \varepsilon |B|$.
Hence the conclusion of part~(i) of the lemma holds with $\gamma = 0$.
 
Now suppose that $\alpha > 0$.
Let $\varepsilon > 0$, $\bar{a} \in (T_n)^{|\bar{x}|}$, 
$B \subseteq p_\tau(\bar{a}, \mcT_n)$, and $|B| \geq g_3(n)$.
Without loss of generality we may assume that $\alpha > \varepsilon$.
Suppose that $\mcA' \in \mbX_n^{\rho, \varepsilon}$, $\mcA' \models q_\rho(\bar{a})$, and $\mbbP_n^\rho(\mcA') > 0$,
hence $\mbbP_n\big(\mbW^{\mcA'}\big) > 0$ (by Lemma~\ref{conditioning on A'}).
By~(\ref{using the induction hypothesis for balance})
\begin{equation}\label{bounds on A' and p-s}
(\alpha - \varepsilon)|B| \leq |p_\rho(\bar{a}, \mcA') \cap B| \leq (\alpha + \varepsilon)|B|.
\end{equation}
Let $\hat{p}(\bar{x}, y)$ be the conjuction of all $(\sigma \setminus \sigma_\rho)$-literals $\varphi(\bar{z})$
such that $p(\bar{x}, y) \models \varphi(\bar{z})$ and $\bar{z}$ is a subsequence of $\bar{x}y$ such that 
$y$ occurs in $\bar{z}$.

By Lemma~\ref{balance, rank 1 for a subset} (with $P = p_\rho(\bar{a}, \mcA') \cap B$), 
there is $\beta$, depending only on $p_\rho$ and $\hat{p}$, and $d' > 0$, depending only on $\beta$ and $\varepsilon$, 
such that if $n$ is large enough, then
\begin{align}\label{result of using the lemma for the subset of rank 1}
\mbbP_n\Big(\Big\{ \mcA \in \mbW^{\mcA'} : \ &(\beta - \varepsilon)|p_\rho(\bar{a}, \mcA') \cap B| \ \leq \ 
|p_\rho(\bar{a}, \mcA') \cap \hat{p}(\bar{a}, \mcA)  \cap B| \ \leq  \\
\ & (\beta + \varepsilon)|p_\rho(\bar{a}, \mcA') \cap B| \Big\}  \ \Big| \ \mbW^{\mcA'} \Big) \ \geq \nonumber \\
& 1 - e^{-d'|p_\rho(\bar{a}, \mcA') \cap B|} \geq 1 - e^{-d' (\alpha - \varepsilon)|B|} \ \geq \
1 - e^{-d' (\alpha - \varepsilon)g_3(n)}. \nonumber
\end{align}
From~(\ref{result of using the lemma for the subset of rank 1}) and~(\ref{bounds on A' and p-s}) we get
\begin{align}\label{bounds with alpha times beta}
\mbbP_n\Big(\Big\{ \mcA \in \mbW^{\mcA'} :  &(\alpha\beta - 3\varepsilon)|B| \ \leq \ 
|p_\rho(\bar{a}, \mcA') \cap \hat{p}(\bar{a}, \mcA)  \cap B| \ \leq  \\
\ & (\alpha\beta + 3\varepsilon)|B| \Big\}  \ \Big| \ \mbW^{\mcA'} \Big) \ \geq \nonumber \\
&1 - e^{-d' (\alpha - \varepsilon)g_3(n)} = 1 - e^{-d'' g_3(n)} \text{ if } d'' = d' (\alpha - \varepsilon). \nonumber
\end{align}
Since $p_\rho \in PLA^*(\sigma_\rho)$ we have $p_\rho(\bar{a}, \mcA) = p_\rho(\bar{a}, \mcA')$ for all $\mcA \in \mbW^{\mcA'}$.
Also, $q(\bar{x}) \wedge p_\rho(\bar{x}, \bar{y} ) \wedge \hat{p}(\bar{x}, \bar{y}) \models p(\bar{x}, \bar{y})$.
So~(\ref{bounds with alpha times beta}) implies that
\begin{align}\label{alpha times beta and p}
\mbbP_n\Big(\Big\{ \mcA \in \mbW_n : &\text{ if $\mcA \models q(\bar{a})$ then }\\
 &(\alpha\beta - 3\varepsilon)|B| \leq |p(\bar{a}, \mcA) \cap B| \leq (\alpha\beta + 3\varepsilon)|B| \Big\}
\ \Big| \ \mbW^{\mcA'} \Big)  \nonumber \\
& \geq \ 1 - e^{-d'' g_3(n)}. \nonumber
\end{align}
Define $\mbX_n^\varepsilon = \bigcup_{\mcA' \in \mbX_n^{\rho, \varepsilon}} \mbW^{\mcA'}$ and note that the union is disjoint.
From~(\ref{alpha times beta and p}) and Lemma~\ref{basic fact about conditional probabilities} we now get
\begin{align}
\mbbP_n\Big(\Big\{ \mcA \in \mbW_n : &\text{ if $\mcA \models q(\bar{a})$ then }\\
 &(\alpha\beta - 3\varepsilon)|B| \leq |p(\bar{a}, \mcA) \cap B| \leq (\alpha\beta + 3\varepsilon)|B| \Big\}
\ \Big| \ \mbX_n^\varepsilon \Big)  \nonumber \\
& \geq \ 1 - e^{-d'' g_3(n)}. \nonumber
\end{align}
Let 
\begin{align*}
\mbY_n^\varepsilon = 
\big\{ \mcA \in \mbW_n : &\text{ if $\mcA \models q(\bar{a})$ then }\\
 &(\alpha\beta - 3\varepsilon)|B| \leq |p(\bar{a}, \mcA) \cap B| \leq (\alpha\beta + 3\varepsilon)|B| \big\},
\end{align*}
so $\mbbP_n\big(\mbY_n^\varepsilon \big| \mbX_n^\varepsilon \big) \geq 1 - e^{-d'' g_3(n)}$.
By Lemma~\ref{conditioning on A'} 
we have $\mbbP_n\big(\mbW^{\mcA'}\big) = \mbbP_n^\rho(\mcA')$ for all $\mcA' \in \mbW_n^\rho$
and therefore $\mbbP_n\big(\mbX_n^\varepsilon\big) = \mbbP_n^\rho\big(\mbX_n^{\rho, \varepsilon}\big)$.
By~(\ref{using the induction hypothesis for balance}) we get
$\mbbP_n\big(\mbX_n^\varepsilon\big) \geq 1 - e^{-d g_3(n)}$ for all sufficiently large $n$.
Hence, for all large enough $n$,
\begin{align*}
&\mbbP_n\big(\mbY_n^\varepsilon\big) \geq \mbbP_n\big(\mbY_n^\varepsilon \ \big| \ \mbX_n^\varepsilon\big)
\cdot \mbbP_n\big(\mbX_n^\varepsilon) \ \geq \ \big(1 - e^{-d'' g_3(n)}\big)\big(1 - e^{-d g_3(n)}\big) \ \geq \\
&1 - e^{-c g_3(n)} \quad \text{ for an appropriate choice of $c > 0$ which depends only on $d$ and $d''$.}
\end{align*}
Now the conclusion of part~(i) follows if $\gamma = \alpha\beta$ because $\varepsilon > 0$ can be chosen as small as we like.

(ii) By Remark~\ref{balance follows from inconsistency}
we may assume that $p \wedge p_\tau \wedge q$ is cofinally satisfiable.
It follows that that $q(\bar{x})$ is equivalent to $p \uhrc \bar{x}$ and $p_\tau(\bar{x}, y)$ is equivalent to $p \uhrc \tau$.
Let $\bar{x} = (x_1, \ldots, x_k)$.
Since $\rank_y(p_\tau) = 1$ it follows that for some $i \in \{1, \ldots, k\}$
\[
p_\tau(\bar{x}, y) \models_{tree} \text{ ``$y$ is a child of $x_i$ and $y \neq x_j$ for all $j = 1, \ldots, k$''.}
\]
For notational simplicity (and without loss of generality) let us assume that $i = 1$.

Suppose that (for some $n$) $\bar{a} = (a_1, \ldots, a_k) \in (T_n)^{k}$, $p_\tau(\bar{a}, \mcA) \neq \es$ 
(so $\bar{a}$ satisfies the restriction of $p_\tau$ to $\bar{x}$).
Then $p_\tau(\bar{a}, \mcA)$ is the set of all children of $a_1$ that do not belong to $\rng(\bar{a})$.
By Assumption~\ref{properties of the trees},
$|p_\tau(\bar{a}, \mcA)| \geq g_1(n) - |\bar{x}|$.
By the same assumption again, we have $\lim_{n\to\infty} (g_1(n) - g_3(n)) = \infty$, 
so if $n$ is large enough then $|p_\tau(\bar{a}, \mcA)| \geq g_3(n)$.

It now follows from part~(i), with $B = p_\tau(\bar{a}, \mcA)$, that 
there is $\gamma$, depending only on $p_\tau$ and $p$, such that for every $\varepsilon > 0$
there is $c > 0$, depending only on $\varepsilon$ and $\gamma$, such that if $n$ is sufficiently large 
 \begin{align*}
 \mbbP_n \big( \big\{ \mcA \in \mbW_n : &\text{ if $\mcA \models q(\bar{a})$ then }\\
& (\gamma - \varepsilon)|p_\tau(\bar{a}, \mcA)| \leq |p(\bar{a}, \mcA)| \leq 
 (\gamma + \varepsilon)|p_\tau(\bar{a}, \mcA)| \big\} \big) \ \geq \\
 & 1 - e^{-cg_3(n)}.
 \end{align*} 
The above holds for all $\bar{a} \in (T_n)^{|\bar{x}|}$. 
(If $p_\tau(\bar{a}, \mcA) = \es$ then the inequalities that define the event above
are trivial since all cardinalities involved are 0 in this case.)
The number of $\bar{a} \in (T_n)^{|\bar{x}|}$ is 
(by Assumption~\ref{properties of the trees}) 
bounded from above by
$(g_4(n)^\Delta)^{|\bar{x}|}$ which is bounded by $f(n)$ for some polynomial $f$.
It follows that 
\begin{align*}
\mbbP_n\big( \big\{ \mcA \in \mbW_n : (p, p_\tau, q) \text{ is {\bf not} $(\gamma, \varepsilon)$-balanced in $\mcA$} \big\} \big) \ 
\leq \ f(n) e^{-c g_3(n)}.
\end{align*}
By  Lemma~\ref{consequences of the first assumption about the trees} ,
$\lim_{n\to\infty} f(n) e^{-c g_3(n)} = 0$.
As $\varepsilon > 0$ was arbitrary it follows that $(p, p_\tau, q)$ is $\gamma$-balanced with respect to $\mbbG$.
\hfill $\square$

\begin{rem}\label{remark about completed induction for balance}{\rm
By part~(i) of Lemma~\ref{balance, rank 1}
we have proved statement~(B) above (at the end of Section~\ref{Proofs of results about convergence and balance}). 
}\end{rem}

\noindent
Although we have completed the inductive step for part~(3) of Assumption~\ref{induction hypothesis}
we will continue to prove results about balanced triples because we need them for proving statement~(C)
and the main results of this article.

\begin{prop}\label{balance, self-contained}
Let $p_\tau(\bar{x}, \bar{y})$ be a self-contained and $\bar{y}$-independent closure type over $\tau$ and let
$p(\bar{x}, \bar{y})$ and $q(\bar{x})$ be complete closure types over $\sigma$.
Then $(p, p_\tau, q)$ is balanced with respect to $\mbbG$.
\end{prop}

\noindent
{\bf Proof.}
By Remark~\ref{balance follows from inconsistency}
we may assume that $p \wedge p_\tau \wedge q$ is cofinally satisfiable, hence consistent.
Since $p$ and $q$ are complete closure types over $\sigma$, it follows that
$q(\bar{x})$ is equivalent to $p \uhrc \bar{x}$ and $p_\tau(\bar{x}, \bar{y})$ is equivalent to $p \uhrc \tau$.
We use induction on $\rank_{\bar{y}}(p_\tau)$ (which equals $\rank_{\bar{y}}(p)$).
If $\rank_{\bar{y}}(p_\tau) = 0$ then $p$ is not $\bar{y}$-independent, contradicting the assumption, 
so $\rank_{\bar{y}}(p_\tau) \geq 1$.
If $\rank_{\bar{y}}(p_\tau) = 1$ then $(p, p_\tau, q)$ is balanced with respect to $\mbbG$, by
Lemma~\ref{balance, rank 1}. 

So suppose that $\rank_{\bar{y}}(p_\tau) = \kappa +1$ where $\kappa \geq 1$.
The induction hypothesis is that 
the proposition holds if we add the assumption that $\rank_{\bar{y}}(p_\tau) \leq \kappa$.
By Remark~\ref{decomposing a type of higher rank},
we may assume that $\bar{y} = z\bar{w}$, 
$\rank_{\bar{w}}(p_\tau) = \kappa$, 
$\rank_z(r_\tau) = 1$, where $r(\bar{x}, z) = p \uhrc \bar{x}z$ and $r_\tau(\bar{x}, z) = p_\tau \uhrc \bar{x}z$.
By the same remark we may also assume that $r_\tau$ is self-contained.
As $\rank_z(r_\tau) = 1$ it follows that $r$ is $z$-independent.
Since $\bar{w}$ is a subsequence of $\bar{y}$ and $p_\tau$ is $\bar{y}$-independent it follows thay $p_\tau$
is also $\bar{w}$-independent.
Note that $r(\bar{x}, z)$ is a complete closure type over $\sigma$.
Since $\rank_z(r_\tau) = 1$ it follows that 
\[
p(\bar{x}, z, \bar{w}) \models_{tree} \text{ ``$z$ is a child of a member of $\bar{x}$''.}
\]

By reordering $\bar{w}$ if necessary we may also assume that $\bar{w} = \bar{u}\bar{v}$ and
\begin{align*}
p(\bar{x}, z, \bar{u}, \bar{v}) \models_{tree} & \text{ ``all members of $\bar{u}$ are successors of $z$, and } \\
& \text{ no member of $\bar{v}$ is a successor of $z$''.}
\end{align*}
It follows that
\[
p(\bar{x}, z, \bar{u}, \bar{v}) \models_{tree} \text{ ``every member of $\bar{v}$ is a successor of some member of $\bar{x}$''.}
\]
Let  $s(\bar{x}, z, \bar{u}) = p \uhrc \bar{x}z\bar{u}$ and $t(\bar{x}, \bar{v}) = p \uhrc \bar{x}\bar{v}$
and note that $s$ and $t$ are complete closure types over $\sigma$.
Let $s_\tau = s \uhrc \tau$ and $t_\tau = t \uhrc \tau$.
Then $p_\tau \uhrc \bar{x}z\bar{u}$ is equivalent to $s_\tau(\bar{x}, z, \bar{u})$ 
and $p_\tau \uhrc \bar{x}\bar{v}$ is equivalent to $t_\tau(\bar{x}, \bar{v})$.
By the choices of $z$, $\bar{u}$, and $\bar{v}$, 
$s(\bar{x}, z, \bar{u})$ is self-contained and $z\bar{u}$-independent,
and $t(\bar{x}, \bar{v})$ is self-contained and $\bar{v}$-independent.
Moreover, in every $\sigma$-structure $\mcA$ that expands a tree,
\[
\text{$p_\tau(\bar{x}, z, \bar{u}, \bar{v})$ is equivalent to $s_\tau(\bar{x}, z, \bar{u}) \wedge t_\tau(\bar{x}, \bar{v})$.}
\]
It follows that
\begin{equation}\label{decomposition into factors}
\text{for all $n$, all  $\bar{a} \in (T_n)^{|\bar{x}|}$ and all $\mcA \in \mbW_n$}, 
|p_\tau(\bar{a}, \mcA)| = |s_\tau(\bar{a}, \mcA)| \cdot |t_\tau(\bar{a}, \mcA)|.
\end{equation}

\medskip
\noindent
{\bf Case 1.} Suppose that $\bar{v}$ is {\em not} empty.

\medskip
\noindent
Then $1 \leq \rank_{\bar{v}}(p) \leq \kappa$ and $1 \leq \rank_{z\bar{u}}(s) \leq \kappa$.
The induction hypothesis implies that $(s, s_\tau, q)$ is $\alpha$-balanced and $(p, p_\tau, s)$ is $\beta$-balanced
(with respect to $\mbbG$) for some $\alpha$ and $\beta$.

Let $\varepsilon > 0$ and define
\begin{align*}
&\mbX_n^\varepsilon = \big\{ \mcA \in \mbW_n : (s, s_\tau, q) \text{ is $(\alpha, \varepsilon)$-balanced in $\mcA$} \big\}
\ \text{ and} \\
&\mbY_n^\varepsilon = \big\{ \mcA \in \mbW_n : (p, p_\tau, s) \text{ is $(\beta, \varepsilon)$-balanced in $\mcA$} \big\}.
\end{align*}
Then $\lim_{n\to\infty} \mbbP_n\big(\mbX_n^\varepsilon \cap \mbY_n^\varepsilon \big) = 1$.

Let $\mcA \in \mbX_n^\varepsilon \cap \mbY_n^\varepsilon$ and $\bar{a} \in (T_n)^{|\bar{x}|}$.
Then
\begin{align*}
&|p(\bar{a}, \mcA)| \ = \ \sum_{b\bar{c} \in s(\bar{a}, \mcA)} |p(\bar{a}, b, \bar{c}, \mcA)| \ \leq \ 
(\beta + \varepsilon) \sum_{b\bar{c} \in s(\bar{a}, \mcA)} |p_\tau(\bar{a}, b, \bar{c}, \mcA)| \\
&= \ (\beta + \varepsilon) \sum_{b\bar{c} \in s(\bar{a}, \mcA)}  |t_\tau(\bar{a}, \mcA)|   \quad \quad
\text{ by (\ref{decomposition into factors})} \\
&= \ (\beta + \varepsilon) |s(\bar{a}, \mcA)| \cdot |t_\tau(\bar{a}, \mcA)| \ \leq \ 
(\beta + \varepsilon)(\alpha + \varepsilon) |s_\tau(\bar{a}, \mcA)| \cdot |t_\tau(\bar{a}, \mcA)| \\
&= \ (\beta + \varepsilon)(\alpha + \varepsilon) |p_\tau(\bar{a}, \mcA)| \quad \quad \text{ by (\ref{decomposition into factors})} \\
&\leq \ (\alpha\beta + 3\varepsilon) |p_\tau(\bar{a}, \mcA)|.
\end{align*}
In a similar way we can derive that $(\alpha\beta - 3\varepsilon) |p_\tau(\bar{a}, \mcA)| \leq |p(\bar{a}, \mcA)|$.
Hence $(p, p_\tau, q)$ is $(\alpha\beta, 3\varepsilon)$-balanced in every $\mcA \in \mbX_n^\varepsilon \cap \mbY_n^\varepsilon$.
As $\varepsilon > 0$ can be taken as small as we like it follows that $(p, p_\tau, q)$ is $\alpha\beta$-balanced 
with respect to $\mbbG$.

\medskip
\noindent
{\bf Case 2.} Suppose that $\bar{v}$ is empty.

\medskip
\noindent
Then $\bar{y} = z\bar{u}$ so $s(\bar{x}, z, \bar{u})$ is equivalent to $p(\bar{x}, z, \bar{u})$
(and $p_\tau$ is equivalent to $s_\tau$). 
Since $p_\tau(\bar{x}, z, \bar{u})$ is self-contained it follows that if $\mcT_n \models p_\tau(\bar{a}, b, \bar{c})$
then $\mcT' := \mcT_n\uhrc \rng(b\bar{c})$ is a subtree of $\mcT_n$ which is rooted in $b$ and the isomorphism type
of $\mcT'$ is determined by $p_\tau$ alone.
It is easy to see that if $\mcT_n \models r_\tau(\bar{a}, b)$ and $\mcT''$ is a subtree of $\mcT_n$ which is rooted in $b$
and isomorphic to $\mcT'$, then the elements of $T'' \setminus \{b\}$ can be ordered as $\bar{c}$ so that
$\mcT_n \models p_\tau(\bar{a}, b, \bar{c})$.
Also, note that if $\mcT^*$ is a different (from $\mcT''$) subtree of $\mcT_n$ that is rooted in $b$ and isomorphic to $\mcT'$,
then $T'' \setminus \{b\} \neq T^* \setminus \{b\}$.
Therefore, $|p_\tau(\bar{a}, b, \mcT_n)|$ is at least as large as the number of subtrees of $\mcT_n$ that are 
rooted in $b$ and isomorphic to $\mcT'$.
Recall that $\msfN_{\mcT_n}(b, \mcT')$ is the number of subtrees of $\mcT_n$ that are rooted in $b$
and isomorphic to $\mcT'$.
Now we have $|p_\tau(\bar{a}, b, \mcT_n)| \geq \msfN_{\mcT_n}(b, \mcT')$ if $p_\tau(\bar{a}, b, \mcT_n) \neq \es$.
Recall that, by Remark~\ref{remark on the properties of the trees},
we have $\msfN_{\mcT_n}(b, \mcT') \geq g_1(n)$ whenever $n$ is large enough and $\msfN_{\mcT_n}(b, \mcT') > 0$.

Suppose that $\bar{x} = (x_1, \ldots, x_k)$.
Without loss of generality we can assume that
$p_\tau(\bar{x}, z, \bar{u}) \models$ ``$z$ is a child of $x_1$''.
It follows that for all $n$ and all $\bar{a} = (a_1, \ldots, a_k) \in (T_n)^k$,
if $r_\tau(\bar{a}, \mcT_n) \neq \es$ then
$r_\tau(\bar{a}, \mcT_n)$
is the set of all children of $a_1$ that do not belong to $\bar{a}$.

We now assume that $p_\tau(\bar{a}, \mcT_n) \neq \es$; otherwise there is nothing to prove.
Then $r_\tau(\bar{a}, \mcT_n) \neq \es$ and for every $b \in r_\tau(\bar{a}, \mcA)$ we have
$|p_\tau(\bar{a}, b, \mcT_n)| \geq \msfN_{\mcT_n}(b, \mcT') \geq g_1(n)$ if $n$ is large enough.
We will now define a partition $P_1, \ldots, P_\mu$ of $r_\tau(\bar{a}, \mcT_n)$.
Recall the functions $g_1, g_2, g_3 : \mbbN \to \mbbR^+$ from Assumption~\ref{properties of the trees}.
\begin{enumerate}
\item Let $\lambda \geq 1$ be the minimal integer such that there is $b \in r_\tau(\bar{a}, \mcA)$ such that
$g_1(n) + (\lambda - 1)g_2(n) \leq \msfN_{\mcT_n}(b, \mcT') < g_1(n) + \lambda g_2(n)$.
\begin{enumerate}
\item If $\lambda = 1$ then let $P_1$ contain all $b \in  r_\tau(\bar{a}, \mcT_n)$ such that
$m_{1, 1} \leq \msfN_{\mcT_n}(b, \mcT') < m_{1, 2}$ where
$m_{1, 1} := g_1(n)$ and $m_{1, 2} := m_{1, 1} + (\lambda + 3)g_2(n)$.

\item If $\lambda > 1$ then let $P_1$ contain all $b \in  r_\tau(\bar{a}, \mcT_n)$ such that
$m_{1, 1} \leq \msfN_{\mcT_n}(b, \mcT') < m_{1, 2}$ where
$m_{1, 1} := g_1(n) + (\lambda - 1)g_2(n)$ and $m_{1, 2} := m_{1, 1} + (\lambda + 3)g_2(n)$.
\end{enumerate}
Assumption~\ref{properties of the trees} implies that, in both cases, $|P_1| \geq g_3(n)$.

\item Now suppose that the parts $P_1, \ldots, P_l$ have been defined such that $|P_i| \geq g_3(n)$ for all $i$
and that numbers $m_{i, j}$ (with $j \in \{1, 2\}$) have been defined such that
$P_i$ contains all $b \in r_\tau(\bar{a}, \mcA)$ such that 
$m_{i, 1} \leq \msfN_{\mcT_n}(b, \mcT') < m_{i, 2}$,
and $m_{i+1, 1} = m_{i, 2}$ if $i < l$.

Let $\lambda \geq 1$ be the minimal integer (if it exists) such that there is $b \in r_\tau(\bar{a},  \mcA)$ such that
$m_{l, 2} + \lambda g_2(n) \leq \msfN_{\mcT_n}(b, \mcT') < m_{l, 2} + (\lambda + 1) g_2(n)$.
\begin{enumerate}
\item If $\lambda = 1$ then let $P_{l+1}$ contain all $b \in r_\tau(\bar{a}, \mcA)$ such that
$m_{l+1, 1} \leq \msfN_{\mcT_n}(b, \mcT') < m_{l+1, 2}$ where
$m_{l+1, 1} := m_{l, 2}$ and $m_{l+1, 2} := m_{l, 2} + 3g_2(n)$.

\item If $\lambda > 1$ then  
\begin{enumerate} 
\item let $P_{l+1}$ contain all $b \in r_\tau(\bar{a}, \mcT_n)$ such that
$m_{l+1, 1} \leq \msfN_{\mcT_n}(b, \mcT') < m_{l+1, 2}$ where
$m_{l+1, 1} := m_{l, 2} + (\lambda - 1) g_2(n)$ and $m_{l+1, 2} := m_{l, 2} + (\lambda + 2) g_2(n)$, and

\item {\em redefine} $P_l$ so that it contains all $b \in  r_\tau(\bar{a}, \mcT_n)$ such that
$m_{l, 1} \leq \msfN_{\mcT_n}(b, \mcT') < m_{l, 2} + g_2(n)$ 
and then {\em redefine} $m_{l, 2}$ by
$m_{l, 2} := m_{l, 2} + g_2(n)$.
\end{enumerate}
\end{enumerate}
In both cases we have $|P_{l+1}| \geq g_3(n)$ by Assumption~\ref{properties of the trees}.
If no $\lambda$ as above exists then $r_\tau(\bar{a}, \mcT_n) = P_1 \cup \ldots \cup P_l$ 
and we are done and let $\mu = l$.
\end{enumerate}
Observe that when a part $P_i$ was first defined then, for some real number $m_i$, 
all $b \in P_i$ satisfy $m_i \leq \msfN_{\mcT_n}(b, \mcT') < m_i + 3g_2(n)$.
Also, each part $P_i$ was redefined at most once and in such a way that (with the same $m_i$)
all $b \in P_i$ satisfy $m_i \leq \msfN_{\mcT_n}(b, \mcT') < m_i + 4g_2(n)$.
By the construction we have $m_i \geq g_1(n)$ for all $i$.
It follows that
\begin{align}\label{the span of a part}
&\text{for each part $P_i$, $|P_i| \geq g_3(n)$ and there is $m_i \geq g_1(n)$ such that} \\ 
&\text{for all $b \in P_i$, $m_i \leq \msfN_{\mcT_n}(b, \mcT') \leq m_i + 4g_2(n)$.} \nonumber
\end{align}

It follows from part~(i) of Lemma~\ref{balance, rank 1}
that there is $\alpha$, depending only on $r$, such that for every 
$\varepsilon > 0$ there is $C > 0$, depending only on $\alpha$ and $\varepsilon$, such that if $n$ is sufficiently large,
then for every part $P_i \subseteq r_\tau(\bar{a}, \mcT_n)$,
\begin{align}\label{balance for one part}
\mbbP_n\big(\mbX_n^{\varepsilon, i}(\bar{a})\big) \ \geq \ 1 - &e^{-Cg_3(n)}, \text{ where}   \\
\mbX_n^{\varepsilon, i}(\bar{a}) = \big\{ \mcA \in \mbW_n : 
&\text{ if $\mcA \models q(\bar{a})$ then } \nonumber \\
&(\alpha - \varepsilon)|P_i| \leq |r(\bar{a}, \mcA) \cap P_i| \leq (\alpha + \varepsilon)|P_i| \big\}. \nonumber
\end{align}
Recall that (by Assumption~\ref{properties of the trees}) every element of $\mcT_n$ has at most $g_4(n)$
children, where $g_4$ is a polynomial.
Since all members of $r_\tau(\bar{a}, \mcT_n)$ are children of $a_1$
it follows that $|r_\tau(\bar{a}, \mcT_n)| \leq g_4(n)$.
As $P_1, \ldots, P_\mu$ is a partition of $r_\tau(\bar{a}, \mcT_n)$ it follows that $\mu \leq g_4(n)$.
Let
\[
\mbX_n^\varepsilon(\bar{a}) = \bigcap_{i = 1}^\mu \mbX_n^{\varepsilon, i}(\bar{a}).
\]
Then
$\mbbP_n\big(\mbX_n^\varepsilon(\bar{a})\big) \ \geq \ 1 - f_4(n) e^{-C g_3(n)}$ and as 
$f_4$ is a polynomial it follows from
Lemma~\ref{consequences of the first assumption about the trees}  
that there is $C' > 0$ such that if $n$ is large enough, then
$\mbbP_n\big(\mbX_n^\varepsilon(\bar{a})\big) \ \geq \ 1 - e^{-C' g_3(n)}$.
So far we have considered a fixed $\bar{a} \in (T_n)^{|\bar{x}|}$.
Now let 
\[
\mbX_n^\varepsilon = \bigcap_{\bar{a} \in (T_n)^{|\bar{x}|}} \mbX_n^\varepsilon(\bar{a}).
\]
Since all elements of $\mcT_n$ have at most $f_4(n)$ children and $\mcT_n$ has height at most $\Delta$ 
(by Assumption~\ref{properties of the trees}) it follows that
$\big|(T_n)^{|\bar{x}|}\big| \leq \big(1 + f_4(n)^\Delta\big)^{|\bar{x}|} = f(n)$ where $f$ is a polynomial.
Thus $\mbbP_n\big( \mbX_n^\varepsilon \big) \geq 1 - f(n) e^{-C' g_3(n)}$ and by 
Lemma~\ref{consequences of the first assumption about the trees}  
there is $C'' > 0$
such that  $\mbbP_n\big( \mbX_n^\varepsilon \big) \geq 1 - e^{-C'' g_3(n)}$ for all sufficiently large $n$.

\medskip
\noindent
{\bf Claim.}  Suppose that $\mcA \in \mbX_n^\varepsilon$. If $n$ is sufficiently large then,
for all $\bar{a} \in (T_n)^{|\bar{x}|}$,
\begin{align*}
(\alpha - 3\varepsilon)\sum_{b \in r_\tau(\bar{a}, \mcA)} |p_\tau(\bar{a}, b, \mcA)|   \leq  
\sum_{b \in r(\bar{a}, \mcA)} |p_\tau(\bar{a}, b, \mcA)| \leq 
(\alpha + 3\varepsilon)\sum_{b \in r_\tau(\bar{a}, \mcA)} |p_\tau(\bar{a}, b, \mcA)|.
\end{align*}

 \noindent
{\bf Proof.}
Let $\mcA \in \mbX_n^\varepsilon$,
so $\mcA \in \mbX_n^{\varepsilon, i}$ for all $i = 1, \ldots, \mu$, and let $\bar{a} \in (T_n)^{|\bar{x}|}$.
Suppose that $r_\tau(\bar{a}, \mcA) \neq \es$, since otherwise the inequalities are trivial.
Fix any $i \in \{1, \ldots, \mu\}$.
By~(\ref{the span of a part}) there is a number $m_i \geq g_1(n)$ such that if $b \in P_i$
then $m_i \leq \msfN_{\mcT_n}(b, \mcT') < m_i + 4g_2(n)$ and hence
$m_i \leq |p_\tau(\bar{a}, b, \mcA)| < m_i + 4g_2(n)$.

Since $\mcA \in \mbX_n^{\varepsilon, i}(\bar{a})$ we get, by the use of~(\ref{balance for one part}),
\begin{align*}
&\frac{\sum_{b \in r(\bar{a}, \mcA) \cap P_i} |p_\tau(\bar{a}, b, \mcA)|}
{\sum_{b \in P_i} |p_\tau(\bar{a}, b, \mcA)|} \leq 
\frac{\sum_{b \in r(\bar{a}, \mcA) \cap P_i} (m_i + 4g_2(n))}
{\sum_{b \in P_i} m_i} \leq \\
&\frac{|r(\bar{a}, \mcA) \cap P_i|(m_i + 4g_2(n))}{|P_i|m_i} \leq
(\alpha + \varepsilon)\frac{(m_i + 4g_2(n))}{m_i} = \\
&(\alpha + \varepsilon)\bigg(1 + \frac{4g_2(n)}{m_i}\bigg) \leq  \\
&(\alpha + \varepsilon)(1 + \varepsilon) \leq (\alpha + 3\varepsilon) \ \text{ if $n$ is large enough }\\
&\text{because $m_i \geq g_1(n)$ and by Assumption~\ref{properties of the trees}
$g_2(n)/g_1(n) \to 0$ as $n\to\infty$.}
\end{align*}
In a similar way we get
\[
(\alpha - 3\varepsilon) \leq \frac{\sum_{b \in r(\bar{a}, \mcA) \cap P_i} |p_\tau(\bar{a}, b, \mcA)|}
{\sum_{b \in P_i} |p_\tau(\bar{a}, b, \mcA)|}
\]
and hence
\begin{align*}
(\alpha - 3\varepsilon)\sum_{b \in P_i} |p_\tau(\bar{a}, b, \mcA)|  \ \leq  \
\sum_{b \in r(\bar{a}, \mcA) \cap P_i} |p_\tau(\bar{a}, b, \mcA)| \ \leq \
(\alpha + 3\varepsilon)\sum_{b \in P_i} |p_\tau(\bar{a}, b, \mcA)|.
\end{align*}
Since
\begin{align*}
&\sum_{b \in r(\bar{a}, \mcA)} |p_\tau(\bar{a}, b, \mcA)| = 
\sum_{i=1}^\mu \ \sum_{b \in r(\bar{a}, \mcA) \cap P_i} |p_\tau(\bar{a}, b, \mcA)| \quad \text{ and} \\
&\sum_{b \in r_\tau(\bar{a}, \mcA)} |p_\tau(\bar{a}, b, \mcA)| = 
\sum_{i=1}^\mu \ \sum_{b \in P_i} |p_\tau(\bar{a}, b, \mcA)|
\end{align*}
the claim follows.
\hfill $\square$

\medskip
\noindent
Since $\rank_z(r) = 1$ it follows that $\rank_{\bar{u}}(p) = \kappa$.
By the induction hypothesis, $(p, p_\tau, r)$ is $\beta$-balanced with respect to $\mbbG$, for some $\beta$.
Let 
\[
\mbY_n^\varepsilon = \big\{ \mcA \in \mbW_n : (p, p_\tau, r) \text{ is $(\beta, \varepsilon)$-balanced in $\mcA$} \big\},
\]
so $\lim_{n\to\infty} \mbbP_n\big( \mbY_n^\varepsilon \big) = 1$.
Suppose that $\mcA \in \mbX_n^\varepsilon \cap \mbY_n^\varepsilon$ and $\bar{a} \in (T_n)^{|\bar{x}|}$.
Then
\begin{align*}
&|p(\bar{a}, \mcA)| \ = \ \sum_{b \in r(\bar{a}, \mcA)} |p(\bar{a}, b, \mcA)| \ \leq \ 
(\beta + \varepsilon) \sum_{b \in r(\bar{a}, \mcA)} |p_\tau(\bar{a}, b, \mcA)| \\
&\leq \ (\beta + \varepsilon) (\alpha + \varepsilon) \sum_{b \in r_\tau(\bar{a}, \mcA)} |p_\tau(\bar{a}, b, \mcA)|
\quad \quad \text{ by the claim} \\
&= \ (\beta + \varepsilon) (\alpha + \varepsilon) |p_\tau(\bar{a}, \mcA)| \ \leq \ 
(\alpha\beta + 3\varepsilon) |p_\tau(\bar{a}, \mcA)|.
\end{align*}
In a similar way we get
$(\alpha\beta - 3\varepsilon) |p_\tau(\bar{a}, \mcA)| \leq |p(\bar{a}, \mcA)|$.
Hence, for every
$\mcA \in \mbX_n^\varepsilon \cap \mbY_n^\varepsilon$, $(p, p_\tau, q)$ is
is $(\alpha\beta, 3\varepsilon)$-balanced in $\mcA$.
As $\varepsilon > 0$ can be chosen as small as we like it and 
$\lim_{n\to\infty}\mbbP_n\big( \mbX_n^\varepsilon \cap \mbY_n^\varepsilon \big) = 1$
it follows that $(p, p_\tau, q)$ is $\alpha\beta$-balanced with respect to $\mbbG$.
This completes the proof of Proposition~\ref{balance, self-contained}.
\hfill $\square$

\begin{cor}\label{balanced, complete types}
Let $p_\tau(\bar{x}, \bar{y})$ be a closure type over $\tau$ and let
$p(\bar{x}, \bar{y})$ and $q(\bar{x})$ be complete closure types over $\sigma$.
Then $(p, p_\tau, q)$ is balanced with respect to $\mbbG$.
\end{cor}

\noindent
{\bf Proof.}
Let $p_\tau$, $p$ and $q$ be as assumed. 
By Remark~\ref{balance follows from inconsistency}
we may assume that $p \wedge p_\tau \wedge q$ is cofinally satisfiable, hence consistent.
By Remark~\ref{0-dimensional types are balanced}
we may also assume that $\rank_{\bar{y}}(p_\tau) > 0$.
Since $p$ is a complete closure type over $\sigma$, $p(\bar{x}, \bar{y}) \models_{tree} p_\tau(\bar{x}, \bar{y}) \wedge q(\bar{x})$.
By Remark~\ref{remark on getting a y-independent type},
there are sequences of variables $\bar{u}$ and $\bar{v}$ such that
$\bar{y}$ is a subsequence of $\bar{u}\bar{v}$, and 
a complete closure type over $\tau$, say $p^*_\tau(\bar{x}, \bar{u}, \bar{v})$, such that
$p^*_\tau$ is self-contained, $\bar{v}$-independent,
and for all $n$, all $\mcA \in \mbW_n$ and all $\bar{a} \in (T_n)^{|\bar{x}|}$, 
if $p_\tau(\bar{a}, \mcA) \neq \es$ then there is a unique $\bar{b} \in (T_n)^{|\bar{u}|}$ such that 
$|p_\tau(\bar{a}, \mcA)| = |p^*_\tau(\bar{a}, \bar{b}, \mcA)|$ (and we have $\bar{b} \in \cl_{\mcT_n}(\bar{a})$).
By Remark~\ref{remark on getting a y-independent type} again there is a self-contained and $\bar{v}$-independent
closure type $p^*(\bar{x}, \bar{u}, \bar{v})$ over $\sigma$ such that 
$p^* \models_{tree} p^*_\tau$ and 
if $p(\bar{a}, \mcA) \neq \es$ then there is a unique $\bar{b} \in (T_n)^{|\bar{u}|}$ such that 
$|p(\bar{a}, \mcA)| = |p^*(\bar{a}, \bar{b}, \mcA)|$.
Since $p^* \models_{tree} p^*_\tau$ this $\bar{b}$ must be the same tuple for which
$|p_\tau(\bar{a}, \mcA)| = |p^*_\tau(\bar{a}, \bar{b}, \mcA)|$.
By Proposition~\ref{balance, self-contained},
$(p^*, p^*_\tau, q^*)$ is $\alpha$-balanced with respect to $\mbbG$ for some $\alpha$.
Because of the identities of cardinalities above it follows that also $(p, p_\tau, q)$ is $\alpha$-balanced with respect to $\mbbG$.
\hfill $\square$

\begin{cor}\label{balance, incomplete types}
Let $p_\tau(\bar{x}, \bar{y})$ be a closure type over $\tau$, let
$p(\bar{x}, \bar{y})$ be a (possibly not complete)
closure type over $\sigma$, and $q(\bar{x})$ be a {\rm complete} closure type over $\sigma$.
Then $(p, p_\tau, q)$ is balanced with respect to $\mbbG$.
\end{cor}

\noindent
{\bf Proof.}
By Remark~\ref{balance follows from inconsistency}
we may assume that $p \wedge p_\tau \wedge q$ is cofinally satisfiable.
Let $p_1(\bar{x}, \bar{y}), \ldots p_m(\bar{x}, \bar{y})$ be an enumeration of all,
up to equivalence, {\em complete} closure types over $\sigma$
such that, for each $i$, $p_i(\bar{x}, \bar{y}) \models_{tree} p(\bar{x}, \bar{y}) \wedge q(\bar{x})$.
So $p(\bar{x}, \bar{y}) \wedge q(\bar{x})$ is equivalent,
in every $\sigma$-structure that expands a tree,
to $\bigvee_{i=1}^m p_i(\bar{x}, \bar{y})$ and 
we may assume that if $i \neq j$ then $p_i \wedge p_j$ is inconsistent.
It follows that if $\mcA \in \mbW_n$, $\bar{a} \in (T_n)^{|\bar{x}|}$ and $\mcA \models q(\bar{a})$, then
$
|p(\bar{a}, \mcA)| = \sum_{i=1}^m |p_i(\bar{a}, \mcA)|
$.
By Corollary~\ref{balanced, complete types}, 
for each $i$, $(p_i, p_\tau, q)$ is $\alpha_i$-balanced with respect to $\mbbG$ for some $\alpha_i$.
It is now straightforward to verify that $(p, p_\tau, q)$ is $(\alpha_1 + \ldots + \alpha_m)$-balanced with respect to $\mbbG$.
\hfill $\square$

\begin{defin}\label{definition of y-positive}{\rm 
Let $p(\bar{x}, \bar{y})$ be a closure type over $\sigma$.
We call $p$ {\bf \em $\bar{y}$-positive (with respect to $\mbbG$)} if the following holds:
if $q(\bar{x})$ is a complete closure type over $\sigma$, $p \wedge q$ is cofinally satisfiable,
and $p_\tau(\bar{x}, \bar{y}) = p \uhrc \tau$, then there is $\alpha > 0$ such that $(p, p_\tau, q)$ is $\alpha$-balanced
with respect to $\mbbG$.
}\end{defin}

\begin{rem}\label{remark about y-positive types}{\rm
(i) Let $p(\bar{x}, \bar{y})$ be a closure type over $\tau$, so in particular $p$ is an incomplete  closure type over $\sigma$
(if $\tau$ is a proper subset of $\sigma$). 
Then $p \uhrc \tau$ is the same as (or at least equivalent to) $p$ and therefore it is evident from the definitions that 
$(p, p\uhrc \tau, q)$ is 1-balanced with respect to $\mbbG$ and hence $p$ is $\bar{y}$-positive.\\
(ii) Let $p(\bar{x}, \bar{y})$ be a closure type over $\tau$ and suppose that $\rank_{\bar{y}}(p) = 0$.
The argument in Remark~\ref{0-dimensional types are balanced}
shows that if $q(\bar{x})$ is a complete closure type over $\sigma$
and $p \wedge q$ is cofinally satisfiable (hence $q(\bar{x}) \models \exists \bar{y} p(\bar{x}, \bar{y})$), 
then  $(p, p\uhrc \tau, q)$ is 
1-balanced with respect to $\mbbG$. Hence $p$ is $\bar{y}$-positive.
}\end{rem}

\begin{prop}\label{balance for positive types}
Let $p(\bar{x}, \bar{y})$ and $r(\bar{x}, \bar{y})$ be (possibly incomplete) closure types over $\sigma$,
where $r$ is $\bar{y}$-positive with respect to $\mbbG$,  and let 
$q(\bar{x})$ be a complete closure type over $\sigma$, 
Then  $(p, r, q)$ is balanced with respect to $\mbbG$.
\end{prop}

\noindent
{\bf Proof.}
By Remark~\ref{balance follows from inconsistency}
we may assume that $p \wedge r \wedge q$ is cofinally satisfiable. 
Let $p_\tau(\bar{x}, \bar{y}) = p \uhrc \tau$. 
Then $p_\tau$ is equivalent to $r \uhrc \tau$, $p \models p_\tau$ and $r \models p_\tau$.
Since $r$ is $\bar{y}$-positive it follows that $(r, p_\tau, q)$ is $\alpha$-balanced for some $\alpha > 0$.
By Corollary~\ref{balance, incomplete types},
$(p \wedge r, p_\tau, q)$ is $\beta$-balanced for some $\beta$.
Now it is straightforward to verify that 
$(p, r, q)$ is $\beta/\alpha$-balanced.
\hfill $\square$

\begin{rem}\label{remark on balance and eventual constancy}{\rm
Let $p(\bar{x}, \bar{y})$, $r(\bar{x}, \bar{y})$, and $q(\bar{x})$ be as in 
Proposition~\ref{balance for positive types}
and suppose that $p \wedge r \wedge q$ is cofinally satisfiable.
By analysing the proofs of Section~\ref{Balance} (this section) we see that if $(p, r, q)$ is $0$-balanced, 
then $(p, r)$ converges to $0$.
If $(p, r)$ is also eventually constant, then it must be eventually constant with value $0$.
This implies that for all sufficiently large $n$, all $\bar{a} \in (T_n)^{|\bar{x}|}$, and all $\mcA \in \mbW_n$, we have
$p(\bar{a}, \mcA) \cap r(\bar{a}, \mcA) = \es$.

By involving the conclusions of Remark~\ref{remark about basic theta-R}
we can now conclude the following, by induction on the height of $\mbbG$:
{\em Suppose that for all $R \in \sigma \setminus \tau$, $\theta_R$ is a closure basic formula.
If $p(\bar{x}, \bar{y})$, $r(\bar{x}, \bar{y})$, and $q(\bar{x})$ are as in 
Proposition~\ref{balance for positive types}
and $(p, r, q)$ is $0$-balanced, then for all sufficiently large $n$ and all $\bar{a} \in (T_n)^{|\bar{x}|}$, 
$p(\bar{a}, \mcA) \cap r(\bar{a}, \mcA) = \es$.}
}\end{rem}

\subsection{Asymptotic elimination of aggregation functions}\label{Asymptotic elimination of aggregation functions}

\noindent
In this subsection we prove statement (C) above, that is, we prove that given
assumptions~\ref{properties of the trees} and~\ref{induction hypothesis} (the induction hypothesis)
and the results proved earlier in Section~\ref{Convergence and balance}
we can asymptotically eliminate aggregation functions, provided that some conditions are satisfied.
Some motivation for these conditions are given in the beginning of Section~\ref{The main results}.

\begin{prop}\label{elimination in the inductive step}
Suppose that assumptions~\ref{properties of the trees} and~\ref{induction hypothesis} hold.\\
(i) Let $\varphi(\bar{x})\in PLA^*(\sigma)$
and suppose that for every subformula of $\varphi$ of the form
\[
F\big(\varphi_1(\bar{y},\bar{z}), \ldots, \varphi_m(\bar{y},\bar{z}) : \bar{z} : 
\chi_1(\bar{y},\bar{z}), \ldots, \chi_m(\bar{y},\bar{z})\big),
\]
it holds that, for all $i = 1, \ldots, m$,
$\chi_i(\bar{y},\bar{z})$ is a $\bar{z}$-positive closure type over $\sigma$,
and either $F$ is continuous, or
$\rank_{\bar{z}}(\chi_i) = 0$ for all $i = 1, \ldots, m$ and $F$ is admissible.
Then $\varphi(\bar{x})$ is asymptotically equivalent to a closure-basic formula over $\sigma$
with respect to the sequence of probability distributions $\mbbP = (\mbbP_n : n\in \mbbN^+)$
induced by $\mbbG$.

(ii) Suppose that for every $R\in \sigma\setminus\sigma_\rho$
the formula $\theta_R$ associated to $R$ (in $\mbbG$) is a closure-basic formula.
Then, in part~(i), we can replace the occurence of ``continuous'' by the weaker condition ``admissible''.
\end{prop}

\noindent
{\bf Proof.} 
We prove part~(i) by induction on the complexity of $\varphi(\bar{x})$.
In the base case we assume that $\varphi(\bar{x})$ is aggregation-free.
Then it follows from 
Lemma~\ref{connectives and basic formulas}
that $\varphi(\bar{x})$ is equivalent, hence asymptotically equivalent, to a closure-basic formula over $\sigma$.

Next, suppose that $\varphi(\bar{x})$ has the form $\msfC\big(\psi_1(\bar{x}),\ldots, \psi_k(\bar{x})\big)$
where $\msfC : [0,1]^k \rightarrow [0,1]$ is continuous.
By the induction hypothesis, each $\psi_i(\bar{x})$ is asymptotically equivalent to a closure-basic formula $\psi'_i(\bar{x})$.
By Lemma~\ref{connectives and basic formulas},
$\msfC\big(\psi'_1(\bar{x}),\ldots, \psi'_k(\bar{x})\big)$ is asymptotically equivalent to a closure-basic formula $\chi(\bar{x})$.
Since $\msfC$ is continuous it follows that $\msfC\big(\psi_1(\bar{x}),\ldots, \psi_k(\bar{x})\big)$
is asymptotically equivalent to $\msfC\big(\psi'_1(\bar{x}),\ldots, \psi'_k(\bar{x})\big)$.
By transitivity of asymptotic equivalence, $\msfC\big(\psi_1(\bar{x}),\ldots, \psi_k(\bar{x})\big)$
is asymptotically equivalent to $\chi(\bar{x})$.

Finally, suppose that $\varphi(\bar{x})$ is of the form
\[
F\big(\varphi_1(\bar{x},\bar{y}), \ldots, \varphi_m(\bar{x},\bar{y}):\bar{y}:
\chi_1(\bar{x},\bar{y}), \ldots, \chi_m(\bar{x},\bar{y})\big)
\]
where $F : \big([0,1]^{<\omega}\big)^m \rightarrow [0,1]$ is continuous and, for all $i = 1, \ldots, m$,
$\chi_i(\bar{x}, \bar{y})$ is a $\bar{y}$-positive closure type over $\sigma$.
By the induction hypothesis, each $\varphi_i(\bar{x},\bar{y})$ is asymptotically equivalent to a closure-basic formula 
$\varphi'_i(\bar{x},\bar{y})$ over $\sigma$.
Then each $\varphi'_i(\bar{x},\bar{y})$ has the form
\[
\bigwedge_{j=1}^{s_i} \big(\varphi_{i,j}(\bar{x},\bar{y}) \rightarrow c_{i,j}\big)
\]
where each $\varphi_{i, j}(\bar{x}, \bar{y})$ is a complete closure type over $\sigma$ and $c_{i, j} \in [0, 1]$.

We will use Theorem~\ref{general asymptotic elimination} 
to show that $\varphi(\bar{x})$ is asymptotically equivalent to a closure-basic formula.
We first assume that $F$ is continuous and show that $\varphi(\bar{x})$ is asymptotically equivalent to a closure basic formula.
Then we make some observations from which we can make the other conclusions of part~(i) of the proposition.
In order to use 
 Theorem~\ref{general asymptotic elimination} 
we need to define appropriate subsets $L_0, L_1 \subseteq PLA^*(\sigma)$ and show that 
Assumption~\ref{assumptions on the basic logic} 
holds with this choice of $L_0$ and $L_1$.
Let $L_0$ be the the set of all complete closure types over $\sigma$ and let $L_1$ be the set of all
(not necessarily complete) closure types over $\sigma$.
(Note that $L_0$ also contains closure types without free variables and recall that these express what relations the root satisfies.
Thus the argument that follows makes sense also if $\bar{x}$ is empty.)
Note that with this choice of $L_0$, each $\varphi'_i$ is an $L_0$-basic formula.
We now show that Assumption~\ref{assumptions on the basic logic} holds for $L_0$ and $L_1$.
Part~(1) of Assumption~\ref{assumptions on the basic logic} follows from
Lemma~\ref{connectives and basic formulas}.
It remains to verify that part~(2) of Assumption~\ref{assumptions on the basic logic} holds.

For each $p(\bar{x}, \bar{y}) \in L_0$ define $L_{p(\bar{x}, \bar{y})}$ to be the set of all 
$\bar{y}$-positive $\lambda(\bar{x}, \bar{y}) \in L_1$. (So $L_{p(\bar{x}, \bar{y})}$ depends only on 
the free variables $\bar{x}$ and $\bar{y}$ but not on other properties of $p$.)
Suppose that $p_i(\bar{x}, \bar{y}) \in L_0$ for $i = 1, \ldots, k$
and let $\lambda_i(\bar{x}, \bar{y}) \in L_{p_i(\bar{x}, \bar{y})}$ for $i = 1, \ldots, k$, so 
$\lambda_i(\bar{x}, \bar{y})$ is a $\bar{y}$-positive closure type over $\sigma$.
Let $q_i(\bar{x}) \in L_0$, $i = 1, \ldots, s$, enumerate all, up to equivalence, complete closure types over $\sigma$ in the variables 
$\bar{x}$.
We may assume that $q_i(\bar{x})$ is not equivalent to $q_j(\bar{x})$ if $i \neq j$. 

Observe that by Assumption~\ref{properties of the trees}, Remark~\ref{remark on the properties of the trees}
and the definition of $\mbW_n$, it follows that if $p(\bar{x}, \bar{y})$ and $q(\bar{x})$ are complete closure types over $\sigma$
and $p(\bar{x}, \bar{y}) \wedge q(\bar{x})$ is satisfied in some $\mcA \in \mbW_n$ for some $n$, then for all 
sufficiently large $n$
the same formula is satisfied in some $\mcA \in \mbW_n$.
Now let $\lambda'_1(\bar{x}), \ldots, \lambda'_t(\bar{x})$ be an enumeration of all
$q_i(\bar{x})$ such that, for some $j$, $q_i(\bar{x}) \wedge \lambda_j(\bar{x}, \bar{y})$ is not satisfied
in any $\mcA \in \mbW_n$ for any $n$.
With these choices we have, for every $n$ and every $\mcA \in \mbW_n$, that
\begin{align}\label{a, b, c essentially verified}
&\mcA \models \forall \bar{x} \big(\bigvee_{i=1}^s q_i(\bar{x})\big), \\
&\mcA \models \forall \bar{x} \neg \big(q_i(\bar{x}) \wedge q_j(\bar{x})\big) \text{ if $i \neq j$, and } \nonumber \\
&\mcA \models 
\Big(\bigvee_{i=1}^k \neg \exists \bar{y} \lambda_i(\bar{x},\bar{y})\Big) \leftrightarrow 
\Big(\bigvee _{i=1}^t \lambda'_i(\bar{x})\Big). \nonumber
\end{align}
We now verify that condition~(d) of part~(2) of
Assumption~\ref{assumptions on the basic logic}
holds. 
It will then be immediate from~(\ref{a, b, c essentially verified}) that also conditions (a), (b) and (c) of 
part~(2) of Assumption~\ref{assumptions on the basic logic} hold.
From Proposition~\ref{balance for positive types}
it follows that, for all $i = 1, \ldots, k$ and all $j = 1, \ldots, s$, the triple $(p_i, \lambda_i, q_j)$ is $\alpha_{i, j}$-balanced for 
some $\alpha_{i, j}$. This means that for all $\varepsilon > 0$ and all $n$, if
\[
\mbY_{n, \varepsilon}^{i, j} = \big\{\mcA \in \mbW_n : (p_i, \lambda_i, q_j) \text{ is $(\alpha_{i, j}, \varepsilon)$-balanced in }
\mcA \big\}
\]
then $\lim_{n\to\infty}\mbbP_n(\mbY_{n, \varepsilon}^{i, j}) = 1$.
Let $\mbY_n^\varepsilon = \bigcap_{i=1}^k \bigcap_{j=1}^s \mbY_{n, \varepsilon}^{i, j}$.
Then $\lim_{n\to\infty}\mbbP_n(\mbY_{n, \varepsilon}^{i, j}) = 1$ and, for every choice of $i$ and $j$,
if $\bar{a} \in (T_n)^{|\bar{x}|}$, $\mcA \in \mbY_n^\varepsilon$ and $\mcA \models q_j(\bar{a})$, then
\[
(\alpha_{i, j} - \varepsilon)|\lambda_i(\bar{a}, \mcA)| \leq
|p_i(\bar{a}, \mcA) \cap \lambda_i(\bar{a}, \mcA)| \leq (\alpha_{i, j} + \varepsilon)|\lambda_i(\bar{a}, \mcA)|.
\]
Hence condition~(d) of part~(2) of
Assumption~\ref{assumptions on the basic logic}
holds.
As~(\ref{a, b, c essentially verified}) holds for all $\mcA \in \mbW_n$ it also holds for all $\mcA \in \mbY_n^\varepsilon$.
Hence conditions~(a), (b) and~(c) of part~(2) of
Assumption~\ref{assumptions on the basic logic}
hold. 
Now part (i) of Theorem~\ref{general asymptotic elimination} 
implies that $\varphi(\bar{x})$ is asymptotically equivalent to an $L_0$-basic formula, or equivalently, to
a closure-basic formula over $\sigma$.

So far we assumed that $F$ is continuous.
Now suppose that $F$ is admissible and that $\rank_{\bar{y}}(\lambda_i) = 0$ for all $i = 1, \ldots, k$.
Then for all $i = 1, \ldots, k$ and $j = 1, \ldots, s$, 
either $q_j(\bar{x}) \wedge \lambda_i(\bar{x}, \bar{y}) \models_{tree} p_i(\bar{x}, \bar{y})$, or
$q_j(\bar{x}) \wedge \lambda_i(\bar{x}, \bar{y}) \models_{tree} \neg p_i(\bar{x}, \bar{y})$.
So if $\alpha_{i, j} = 0$ then, for all $n$, all $\bar{a} \in (T_n)^{|\bar{x}|}$, and all $\mcA \in \mbW_n$,
if $\mcA \models q_j(\bar{a})$ then $p_i(\bar{a}, \mcA) \cap \lambda_i(\bar{a}, \mcA) = \es$.
This means that the extra condition in part~(ii) of Theorem~\ref{general asymptotic elimination}
is satisfied and therefore $\varphi(\bar{x})$ is asymptotically equivalent to a closure-basic formula.

Part~(ii) follows from the conclusion of
Remark~\ref{remark on balance and eventual constancy}
and the above argument (where we use part~(ii) of Theorem~\ref{general asymptotic elimination}).
\hfill $\square$

\begin{rem}\label{completing the induction hypothesis for part 1}{\rm
Recall that we have assumed that $\mbbG$ is a $PLA^*(\sigma)$-network of height $\rho + 1$.
Suppose that $\sigma \subset \sigma^+$ and that $\mbbG^+$ is a $PLA^*(\sigma^+)$-network of height $\rho + 2$
such that $\mbbG$ is the subnetwork of $\mbbG^+$ which is induced by $\sigma$.
For every $R \in \sigma^+ \setminus \sigma$, let $\theta_R$ be the $PLA^*(\sigma)$-formula which $\mbbG^+$
associates to $R$. Suppose that for each $R \in \sigma^+ \setminus \sigma$, $\theta_R$ satisfies
the assumptions on $\varphi$ in Proposition~\ref{elimination in the inductive step}.
Then Proposition~\ref{elimination in the inductive step} implies that 
part~(1) of Assumption~\ref{induction hypothesis} holds if 
$\sigma_\rho$, $\sigma$ and $\mbbG_\rho$ are replaced by $\sigma$, $\sigma^+$ and $\mbbG$, respectively.
Hence statement (C) is proved.
}\end{rem}

\section{The main results}\label{The main results}

\noindent
In this section we prove the main results.
The reader may note that we only ``asymptotically eliminate'' aggregations that are conditioned on
closure types. From the perspective of first-order logic this may seem like a strong constraint since 
first-order quantifications range over the whole domain.
However, in applications one is often {\em not} interested in searching through the whole domain (of a database for example),
but instead one may be interested in some part of the domain (defined by some constraints). 
In the current context, when conditioning aggregations on closure types we restrict the range of aggregations to 
vertices, or tuples of vertices, on specified levels and, possibly,
with specified ancestors, in the underlying tree.
Also note that, with the assumptions that we have adopted on the sequence $\mbT = (\mcT_n : n \in \mbbN^+)$
of underlying trees, the number of vertices on level $l+1$ divided by the number of vertices on level $l$ tends to infinity
(unless level $l+1$ is empty). So if we use a continuous aggregation function in an aggregation, 
then the highest level
(in the tree) that the aggregation ranges over will, up to a small discrepancy, determine the output.

\begin{theor}\label{main result}
Let $\tau \subset \sigma$,
let $\mbT = (\mcT_n : n \in \mbbN^+)$ be a sequence of trees such that  Assumption~\ref{properties of the trees} holds,
and let $\mbbG$ be a $PLA^*(\sigma)$-network based on $\tau$ (and recall that $\mbbG$ associates a formula
$\theta_R \in PLA^*(\mr{par}(R))$ to every $R \in \sigma \setminus \tau$).
Furthermore let $\mbbP_n$ be the probability distribution that $\mbbG$ induces on $\mbW_n$, 
the set of expansions of $\mcT_n$ to $\sigma$.

Suppose that for every $R \in \sigma \setminus \tau$
and every subformula of $\theta_R$ of the form
\begin{equation}\label{a subformula in the main results}
F\big(\varphi_1(\bar{y}, \bar{z}), \ldots, \varphi_m(\bar{y}, \bar{z}) : \bar{z} : 
\chi_1(\bar{y}, \bar{z}), \ldots, \chi_m(\bar{y}, \bar{z})\big),
\end{equation}
it holds that, for all $i = 1, \ldots, m$,
$\chi_i(\bar{y},\bar{z})$ is a (not necessary complete) $\bar{z}$-positive closure type over $\sigma$,
and either $F$ is continuous, or
$\rank_{\bar{z}}(\chi_i) = 0$ for all $i = 1, \ldots, m$ and $F$ is admissible.

Let $\varphi(\bar{x}) \in PLA^*(\sigma)$ and suppose that for every subformula of $\varphi(\bar{x})$ of the
form~(\ref{a subformula in the main results}) the same condition holds.
Then:\\
(i) $\varphi(\bar{x})$ is asymptotically equivalent to a closure-basic formula over $\sigma$
with respect to $(\mbbP_n : n \in \mbbN^+)$.\\
(ii) For every closure type $p(\bar{x})$ over $\tau$ there are $k \in \mbbN^+$, 
$c_1, \ldots, c_k \in [0, 1]$, and $\alpha_1, \ldots, \alpha_k \in [0, 1]$ such that 
for every $\varepsilon > 0$ there is $n_0$ such that if $n \geq n_0$, $\bar{a} \in (T_n)^{|\bar{x}|}$
and $\mcT_n \models p(\bar{a})$, then 
\begin{align*}
&\mbbP_n \Big( \big\{ \mcA \in \mbW_n : \mcA(\varphi(\bar{a})) \in 
\bigcup_{i = 1}^k [c_i - \varepsilon, c_i + \varepsilon] \big\}\Big)
\geq 1 - \varepsilon  \text{ and, for all $i = 1, \ldots, k$,} \\
&\mbbP_n\big(\{\mcA \in \mbW_n : \mcA(\varphi(\bar{a})) \in [c_i - \varepsilon, c_i + \varepsilon] \}\big) \in 
[\alpha_i - \varepsilon, \alpha_i + \varepsilon].
\end{align*}
\end{theor}

\noindent
{\bf Proof.}
(i) Suppose that $\tau$ is a proper subset of $\sigma$.
$\mbbG$ be a $PLA^*(\sigma)$-network $\mbbG$ based on $\tau$ with height $\rho +1$ (where $\rho \geq -1$).
Let $\sigma_\rho = \{R \in \sigma \setminus \tau : \text{ $R$ is on level $l$ for some $l \leq \rho$}\}$.
We use induction on the height $\rho + 1$.
The base case $\rho + 1 = 0$ (i.e. $\rho = -1$) is equivalent to assuming that $\sigma_\rho = \tau$.
It was pointed out in Remark~\ref{remark about base case}
that, by Lemma~\ref{base case}, Assumption~\ref{induction hypothesis} holds if $\sigma_\rho = \tau$.

By the observations in 
remarks~\ref{induction completed for convergence},
\ref{remark about completed induction for balance}
and~\ref{completing the induction hypothesis for part 1}
it follows that if Assumption~\ref{induction hypothesis}
holds for all $\sigma \supseteq \tau$ and all $PLA^*(\sigma)$-networks based on $\tau$ with height $\rho + 1$ such that, 
for all $R \in \sigma \setminus \tau$,
$\theta_R$ satisfies the conditions of
Theorem~\ref{main result}, then 
Assumption~\ref{induction hypothesis}
also holds for all $\sigma \supseteq \tau$ and all $PLA^*(\sigma)$-networks based on $\tau$ with height $\rho + 2$ such that,
 for all $R \in \sigma \setminus \tau$, $\theta_R$ satisfies the conditions of
Theorem~\ref{main result}.
Thus Assumption~\ref{induction hypothesis} holds for all $\sigma \supseteq \tau$ and all
$PLA^*(\sigma)$-networks based on $\tau$ such that, for all $R \in \sigma \setminus \tau$,
$\theta_R$ satisfies the conditions of
Theorem~\ref{main result}.
Proposition~\ref{elimination in the inductive step} was derived by using only parts~(2) and~(3) 
of Assumption~\ref{induction hypothesis}, so it follows that 
if $\varphi(\bar{x})$ is as assumed in the theorem, then it is asymptotically equivalent to a closure-basic formula.

Since all results in Section~\ref{Convergence and balance} were derived from Assumption~\ref{induction hypothesis}
we can, by induction, conclude that all results in that section hold for every $\sigma \supseteq \tau$
and every $PLA^*(\sigma)$-network such that, for all $R \in \sigma \setminus \tau$, $\theta_R$ satisfies the conditions of
Theorem~\ref{main result}.

(ii) Let $p(\bar{x})$ be a closure type over $\tau$.
By part~(i), $\varphi(\bar{x})$ is asymptotically equivalent to a closure-basic formula
\[
\bigwedge_{i = 1}^m (\varphi_i(\bar{x}) \to c_i),
\]
 where each $\varphi_i(\bar{x})$ is a 
complete closure type over $\sigma$. 
Without loss of generality we may assume that $\varphi_i(\bar{x})$, $i = 1, \ldots, m$, enumerate
all, up to equivalence, complete closure types over $\sigma$ that are cofinally satisfiable and
that $\varphi_i \wedge \varphi_j$ is inconsistent if $i \neq j$.
Let $p(\bar{x})$ be a closure type over $\tau$.
By Proposition~\ref{(p, p-tau) convergence} (which as noted above holds under the assumptions
of the theorem), for all $i$,
$(\varphi_i, p)$ converges to some $\alpha_i$.
So for every $\varepsilon > 0$, 
if $n$ is large enough, $\bar{a} \in (T_n)^{|\bar{x}|}$ and $\mcT_n \models p(\bar{a})$, then 
\[
\mbbP_n\big(\mbE_n^{\varphi_i(\bar{a})} \ | \ \mbE_n^{p(\bar{a})}\big) \in [\alpha_i - \varepsilon, \alpha_i + \varepsilon].
\]
and (as $\varphi(\bar{x})$ and $\bigwedge_{i = 1}^m (\varphi_i(\bar{x}) \to c_i)$ are asymptotically equivalent)
\[
\mbbP_n\Big(\Big\{\mcA \in \mbW_n : 
\Big|\mcA(\varphi(\bar{a})) - \mcA\Big(\bigwedge_{i = 1}^m \big(\varphi_i(\bar{a}) \to c_i\big)\Big)\Big| 
\leq \varepsilon \Big\}\Big) \geq 1 - \varepsilon.
\]
Note that if $\mcA \models \varphi_i(\bar{a})$ then $\mcA(\bigwedge_{i = 1}^m (\varphi_i(\bar{a}) \to c_i)) = c_i$.
Since we assume that $\varphi_i(\bar{x})$, $i = 1, \ldots, m$, enumerate all, up to equivalence,
complete closure types over $\sigma$ that are cofinally satisfiable it follows that for all sufficiently large $n$
\[
\mbbP_n \Big( \big\{ \mcA \in \mbW_n : \mcA(\varphi(\bar{a})) \in 
\bigcup_{i = 1}^k [c_i - \varepsilon, c_i + \varepsilon] \big\}\Big)
\geq 1 - \varepsilon.
\]
Let $c \in \{c_1, \ldots, c_m\}$ and for simplicity of notation suppose that, for some $1 \leq s \leq m$,
$c = c_i$ if $i \leq s$ and $c \neq c_i$ if $i > s$.
Let $\beta_c = \alpha_1 + \ldots + \alpha_s$.
It now follows that if $n$ is large enough and $\mcT_n \models p(\bar{a})$ then 
\[
\mbbP_n\big(\{\mcA \in \mbW_n : \mcA(\varphi(\bar{a})) \in [c - \varepsilon, c + \varepsilon]\}\big)
\in  [\beta_c - s\varepsilon, \beta_c + s\varepsilon].
\]
The claim now follows since $\varepsilon > 0$ can be chosen as small as we like.
\hfill $\square$

\medskip

\noindent
Recall that the aggregation functions min and max are admissible, but not continuous.
Therefore Theorem~\ref{main result} only applies to min and max if they are used
together with a conditioning closure type $\chi(\bar{y}, z)$ with $z$-rank 0.
The next corollary states that if we strengthen the assumptions on the $PLA^*(\sigma)$-network $\mbbG$
(thus limiting the range of distributions that such $\mbbG$ can induce), then 
the conclusions~(i) and~(ii) of Theorem~\ref{main result} hold if the formula $\varphi(\bar{x})$ uses 
(only) admissible aggregation functions. In particular the conclusions of the theorem hold for first-order formulas.

\begin{cor}\label{corollary to main result}
Let $\tau$, $\sigma$, $\mbT = (\mcT_n : n \in \mbbN^+)$, $\mbW_n$, $\mbbG$ and $\mbbP_n$ be as in
Theorem~\ref{main result}.
In particular Assumption~\ref{properties of the trees} is adopted.
Suppose that, for every $R \in \sigma \setminus \tau$, the formula 
$\theta_R$ that $\mbbG$ associates to $R$ is a closure-basic formula
over $\mr{par}(R)$.
Suppose that $\varphi(\bar{x}) \in PLA^*(\sigma)$ 
and that if
\[
F\big(\varphi_1(\bar{y}, \bar{z}), \ldots, \varphi_m(\bar{y}, \bar{z}) : \bar{z} : 
\chi_1(\bar{y}, \bar{z}), \ldots, \chi_m(\bar{y}, \bar{z})\big)
\]
is a subformula of $\varphi(\bar{x})$, then
$F$ is {\rm admissible}, and 
for all $i = 1, \ldots, m$, $\chi_i(\bar{y}, \bar{z})$ is a
$\bar{z}$-positive closure type over $\sigma$.
Then the conclusions~(i) and~(ii) of 
Theorem~\ref{main result}
hold.
\end{cor}

\noindent
{\bf Proof.}
Part~(ii) follows from part~(i) in exctly the same way as 
part~(ii) of 
Theorem~\ref{main result}
follows from part~(i) of that theorem.

Part~(i) is proved in essentially the same way as part~(i) of Theorem~\ref{main result}.
But now we have a stronger assumption on all $\theta_R$, $R \in \sigma \setminus \tau$.
As concluded in Remark~\ref{remark about basic theta-R}
the stronger assumption implies the following:
{\em If $p(\bar{x}, \bar{y})$ and $r(\bar{x}, \bar{y})$ are closure types over $\sigma$,
where $r$ is $\bar{y}$-positive, and $q(\bar{x})$ is a complete closure type over $\sigma$, and
$(p, r, q)$ is $0$-balanced, then for all sufficiently large $n$ and all $\bar{a} \in (T_n)^{|\bar{x}|}$, 
$p(\bar{a}, \mcA) \cap r(\bar{a}, \mcA) = \es$.}
This means that the extra condition in part~(ii) of
Theorem~\ref{general asymptotic elimination} 
is satisfied and therefore every {\em admissible} aggregation function can be asymptotically eliminated
as in the proof of Proposition~\ref{elimination in the inductive step}.
It follows that every $\varphi(\bar{x})$ subject to the assumptions of the corollary is asymptotically equivalent
to a closure-basic formula over $\sigma$.
\hfill $\square$

\medskip

\noindent
Now we consider a lighter version of Assumption~\ref{properties of the trees},
where the functions $g_2$ and $g_3$ (in Assumption~\ref{properties of the trees}) have been removed.
It turns out that ``aggregations of dimension 1'' can still be asymptotically eliminated if we only
assume the following:

\begin{assump}\label{less properties of the trees} {\rm $\text{   }$
\begin{enumerate}
\item $\Delta \in \mbbN^+$,
\item $g_1$ and $g_4$ are functions from  $\mbbN$ to the positive reals such that
\begin{enumerate}
\item $\lim_{n\to\infty} g_1(n) = \infty$, 
\item $g_4$ is a polynomial,
\end{enumerate}
\item $\mbT = (\mcT_n : n \in \mbbN^+)$ and each $\mcT_n$ is a tree such that
\begin{enumerate}
\item the height of $\mcT_n$ is $\Delta$ and all leaves of $\mcT$ are on level $\Delta$, and
\item every nonleaf has at least $g_1(n)$ children and at most $g_4(n)$ children.
\end{enumerate}
\end{enumerate}
}\end{assump}

\noindent
Corollaries~\ref{second corollary to main result} 
and~\ref{third corollary to main result}
below are analogoues of 
Theorem~\ref{main result} and Corollary~\ref{corollary to main result}
that are adapted to Assumption~\ref{less properties of the trees}.

\begin{cor}\label{second corollary to main result}
Let $\tau \subset \sigma$,
let $\mbT = (\mcT_n : n \in \mbbN^+)$ be a sequence of trees such that  Assumption~\ref{less properties of the trees} holds,
and let $\mbbG$ be a $PLA^*(\sigma)$-network based on $\tau$ .
Furthermore let $\mbbP_n$ be the probability distribution that $\mbbG$ induces
on $\mbW_n$, the set of expansions of $\mcT_n$ to $\sigma$.

Suppose that for every $R \in \sigma \setminus \tau$
and every subformula of $\theta_R$ of the form
\begin{equation}\label{a subformula in the main result with weaker assumption}
F\big(\varphi_1(\bar{y}, \bar{z}), \ldots, \varphi_m(\bar{y}, \bar{z}) : \bar{z} : 
\chi_1(\bar{y}, \bar{z}), \ldots, \chi_m(\bar{y}, \bar{z})\big),
\end{equation}
it holds that, for all $i = 1, \ldots, m$,
$\chi_i(\bar{y},\bar{z})$ is a (not necessary complete) $\bar{z}$-positive closure type over $\sigma$
and, either $F$ is continuous
and $\rank_{\bar{z}}(\chi_i) = 1$ for all $i$, or
$F$ is admissible and $\rank_{\bar{z}}(\chi_i) = 0$ for all $i$.

Let $\varphi(\bar{x}) \in PLA^*(\sigma)$ and suppose that for every subformula of $\varphi(\bar{x})$ of the
form~(\ref{a subformula in the main result with weaker assumption}) the same condition holds.
Then conclusions~(i) and~(ii) of 
Theorem~\ref{main result} hold.
\end{cor}

\noindent
{\bf Proof.}
The conditions of Assumption~\ref{properties of the trees} that have been removed
in Assumption~\ref{less properties of the trees}
are {\em only} used in the proof of Proposition~\ref{balance, self-contained}
when proving the inductive step, and the inductive step
is {\em only} necessary if we only consider $p_\tau(\bar{x}, \bar{y})$ in that proposition with $\rank_{\bar{y}}(p_\tau) \geq 2$.
As a consequence, all results in Section~\ref{Balance}
after  Proposition~\ref{balance, self-contained} restricted to closure types $p(\bar{x}, \bar{y})$ with $\rank_{\bar{y}}(p) \leq 1$
follow from results
before Proposition~\ref{balance, self-contained}.
Hence, Proposition~\ref{elimination in the inductive step} restricted to
$\chi_i(\bar{y}, \bar{z})$ with $\bar{z}$-rank at most 1 does not need Proposition~\ref{balance, self-contained}.
Consequently, the same holds for Corollary~\ref{second corollary to main result}.
\hfill $\square$

\begin{exam}\label{example with only rank-1 aggregations} {\rm
We illustrate Corollary~\ref{second corollary to main result} and a contrast to 
Example~\ref{example with few vertices with many more children}.
Let $\sigma = \tau \cup \{R\}$, where $R$ is unary, and let $\mbT = (\mcT_n : n \in \mbbN^+)$ be as in
Example~\ref{example with few vertices with many more children}.
Then $\mbT$ does not satisfy Assumption~\ref{properties of the trees}, but it does
satisfy Assumption~\ref{less properties of the trees}.
We also let, as in Example~\ref{example with few vertices with many more children},
$\mbW_n$ be the set of all $\sigma$-structures that expand $\mcT_n$ and we let
$\mbbP_n$ be he uniform probability distribution on $\mbW_n$.
($\mbbP_n$ is induced by a $PLA^*(\sigma)$-network based on $\tau$.)
As in that example let $q(x)$ be a closure type over $\tau$ which expresses ``$x$ is a child of the root'',
and let $p(x, y)$ be a closure type over $\tau$ which expresses ``$q(x)$ and $y$ is a child of $x$''.
Then $\rank_x(q) = 1$, $\rank_{(x, y)}(p) = 2$, and $\exists x p(x, y)$ is a closure type over $\tau$
with $y$-rank 2.
Let $\varphi$ be the sentence
\[
\mr{am}\big( \exists x (p(x, y) \wedge R(x) \wedge R(y)) : y : \exists x p(x, y) \big).
\]
We saw in Example~\ref{example with few vertices with many more children}
that for all sufficiently large $n$, the distribution of the values $\mcA(\varphi)$ for random $\mcA \in \mbW_n$
was quite different for odd $n$ compared to even $n$ (so the conclusions of Theorem~\ref{main result} fail,
showing that Assumption~\ref{properties of the trees} is necessary for that theorem).

Let us now replace the aggregation that conditions on $\exists x p(x, y)$ with two aggregations that 
condition on closure types with $x$-rank, respectively $y$-rank, equal to 1, by letting $\psi$ be the sentence
\[
\mr{am}\big( R(x) \wedge \mr{am}\big( p(x, y) \wedge R(y) : y : p(x, y) \big) : x : q(x) \big).
\]
Suppose that $n$ is large.
For every $a \in (T_n)$ such that $\mcT_n \models q(a)$ (that is, such that $a$ is a child of the root)
the probability, for a random $\mcA \in \mbW_n$, that roughly half of the children of $a$ satisfy $R(y)$ is close to 1.
Hence the probability that
\[
\mcA\big(\mr{am}\big( p(a, y) \wedge R(y) : y : p(a, y) \big)\big) \approx 1/2
\]
is close to 1.
With probability close to 1, roughly half of all $a \in T_n$ such that $\mcT_n \models q(a)$
will satisfy $R(x)$, so for roughly half of such $a$, 
the value of $\mcA\big(\mr{am}\big( p(a, y) \wedge R(y) : y : p(a, y) \big)\big)$ is close to  $1/2$,
and for the other such $a$ the value is 0.
It follows that, with probability tending to 1 as $n\to\infty$, for a random $\mcA \in \mbW_n$, $\mcA(\psi) \approx 1/4$.
Or more precisely, for all $\varepsilon > 0$, if $n$ is large enough then
$\mbbP_n\big(\big\{ \mcA \in \mbW_n : \big|\mcA(\psi) - 1/4\big| \leq \varepsilon \big\} \big) \geq 1 - \varepsilon$,
no matter if $n$ is even or odd (and Corollary~\ref{second corollary to main result} tells that we must have
such kind of convergence).
Hence $\psi$ and $\varphi$ are {\em not} asymptotically equivalent.
This illustrates that it can matter if we
condition an aggregation on a closure type of higher rank than 1, or if we instead
``break up'' such an aggregation and aggregate several times, each time conditioning the
aggregation on a closure type of rank 1 (with respect to the variable that the aggregation binds).
}\end{exam}

\begin{cor}\label{third corollary to main result}
Let $\tau$, $\sigma$, $\mbT = (\mcT_n : n \in \mbbN^+)$, $\mbW_n$, $\mbbG$ and $\mbbP_n$ be as in
Corollary~\ref{second corollary to main result}.
Suppose that, for every $R \in \sigma \setminus \tau$, the formula 
$\theta_R$ that $\mbbG$ associates to $R$ is a closure-basic formula
over $\mr{par}(R)$.
Suppose that $\varphi(\bar{x}) \in PLA^*(\sigma)$ 
and that if
\[
F\big(\varphi_1(\bar{y}, \bar{z}), \ldots, \varphi_m(\bar{y}, \bar{z}) : \bar{z} : 
\chi_1(\bar{y}, \bar{z}), \ldots, \chi_m(\bar{y}, \bar{z})\big)
\]
is a subformula of $\varphi(\bar{x})$ then,
$F$ is {\rm admissible}, for all $i = 1, \ldots, m$, 
$\chi_i(\bar{y}, \bar{z})$ is a $\bar{z}$-positive closure type over $\sigma$
and the $\bar{z}$-rank of $\chi_i$ is 0 or 1.
Then the conclusions~(i) and~(ii) of 
Theorem~\ref{main result}
hold.
\end{cor}

\noindent
{\bf Proof.}
This corollary follows by the observations made in the proofs of 
corollaries~\ref{corollary to main result} and~\ref{second corollary to main result}.
\hfill $\square$

\begin{rem}\label{not possible to generalize to admissibility}{\rm
Theorem~\ref{main result} and Corollary~\ref{second corollary to main result}
can not be generalized by replacing `continuous' with `admissible'.
The reason is as follows.
Suppose that $\mcT_n$ is a tree of height 1 such that the root has exactly $n$ children.
Let $\sigma = \tau \cup \{R\}$ where $R$ is a binary relation symbol.
Let $\alpha \in (0, 1)$ be rational.
With the kind of $PLA^*(\sigma)$-network that is allowed in 
Theorem~\ref{main result} and Corollary~\ref{second corollary to main result}
we can induce a probability distribution $\mbbP_n$ on $\mbW_n$ such that for all different $a, b \in T_n$,
the probability that $R(a, b)$ holds is $n^{-\alpha}$ independently of what the case is for other pairs.
(To get such $\mbbP_n$ we can use the aggregation function $\mr{length}^{-\alpha}$ which is continuous.)
It follows from a result by Shelah and Spencer \cite{SS} that there is a first-order sentence $\varphi$
using no other relation symbol than $R$
such that $\mbbP_n\big(\{\mcA \in \mbW_n : \mcA  \models \varphi\}\big)$ does not converge as $n \to \infty$.
Without loss of generality we can assume that all quantifications in $\varphi$ are relativized to the set of children of the root
(thus excluding the root from quantifications). 
Recall that the aggregation functions min and max are admissible.
By letting $q(x)$ be a closure type over $\tau$ which expresses ``$x$ is a child of the root'' and
replacing, in $\varphi$, every quantification `$\forall x \ldots$', respectively `$\exists x \dots$', by
`$\min(\ldots : x : q(x))$', respectively by `$\max(\ldots : x : q(x))$', we get a 
0/1-valued $PLA^*(\sigma)$-sentence, say $\varphi'$, such that
$\mbbP_n\big(\{\mcA \in \mbW_n : \mcA  \models \varphi'\}\big)$ does not converge.
}\end{rem}

\noindent
{\bf Acknowledgements.}
The authors thank the anonymous referees for their insightful remarks and suggestions which have improved this article.
Vera Koponen was partially supported by the Swedish Research Council, grant 2023-05238\_VR.
Vera Koponen and Yasmin Tousinejad were partially supported by 
the Wallenberg AI, Autonomous Systems and Software Program (WASP) funded by the Knut and Alice Wallenberg Foundation.


\begin{thebibliography}{99}\label{References}

\bibitem{ADGP} F. Abu Zaid, A. Dawar, E. Grädel, W. Pakusa,
Definability of summation problems for Abelian groups and semigroups,
in Proceedings of 32th Annual ACM/IEEE Symposium on Logic in Computer Science (LICS) (2017).

\bibitem{AK17} O. Ahlman, V. Koponen, Random $l$-colourable structures with a pregeometry,
{\em Mathematical Logic Quarterly}, Vol. 63 (2017) 32--58.

\bibitem{AlonSpencer} N. Alon, J. H. Spencer, {\em The Probabilistic Method}, Second Edition, 
John Wiley \& Sons (2000).

\bibitem{Bal} J. T. Baldwin, Expansions of geometries, 
{\em The Journal of Symbolic Logic}, Vol. 68 (2003) 803--827.

\bibitem{Ber} M. Bergmann, {\em An Introduction to Many-Valued and Fuzzy Logic: Semantics, Algebras,
and Derivation Systems}, Cambridge University Press (2008).

\bibitem{BP} S. Brin, L. Page, The anatomy of a large-scale hypertextual Web search engine,
{\em Computer Networks and ISDN Systems}, Vol. 30 (1998) 107--117.

\bibitem{BKNP} G. Van den Broeck, K. Kersting, S. Natarajan, D. Poole, (Editors)
{\em An Introduction to Lifted Probabilistic Inference}, The MIT Press (2021).

\bibitem{Chernoff} H. Chernoff, A measure of the asymptotic efficiency for tests of a hypothesis based on
the sum of observations, {\em Annals of Mathematical Statistics}, Vol. 23 (1952) 493--509.

\bibitem{CM} F. G. Cozman, D. D. Maua, 
The finite model theory of Bayesian network specifications: Descriptive complexity and zero/one laws,
{\em International Journal of Approximate Reasoning}, Vol. 110 (2019) 107--126.

\bibitem{DGH} A. Dawar, E. Grädel, M. Hoelzel, 
Convergence and nonconvergence laws for random expansions of product structures,
in A. Blass et al (Editors),
{\em Gurevich Festschrift, Lecture Notes in Computer Science 12180}, 118--132, Springer (2020).

\bibitem{DKNP} L. De Raedt, K. Kersting, S. Natarajan, D. Poole,
{\em Statistical Relational Artificial Intelligence: Logic, Probability, and Computation},
Synthesis Lectures on Artificial Intelligence and Machine Learning \#32,
Morgan \& Claypool Publishers (2016).

\bibitem{Fag} R. Fagin, Probabilities on finite models, {\em The Journal of Symbolic Logic}, 
Vol. 41 (1976) 50-58.

\bibitem{GKLT} Y. Glebskii, D. Kogan, M. Liogon'kii, V. Talanov, Range and degree of realizability of formulas in the restricted predicate calculus. {\em Cybernetics}, Vol. 5 (1969), 142--154.  

\bibitem{Jae98a} M. Jaeger, Convergence results for relational Bayesian networks, 
{\em Proceedings of the 13th Annual IEEE Symposium on Logic in Computer Science (LICS 98)} (1998).

\bibitem{Jae98b} Reasoning about infinite random structures with relational bayesian networks,
{\em Proceedings of KR-98}, Morgan Kaufman, San Francisco, CA (1998).

\bibitem{JW} G. Jeh, J. Widom, SimRank: A Measure of Structural-Context Similarity, in 
D. Hand, D. A. Keim, R. NG (Ed.),
{\em KDD'02: Proceedings of the eighth ACM SIGKDD international conference on Knowledge discovery and data mining}, ACM Press (2002) 538-543.

\bibitem{KMG} A. Kimmig, L. Mihalkova, L. Getoor, Lifted graphical models: a survey,
{\em Machine Learning}, Vol. 99 (2015) 1--45.

\bibitem{Koller} D. Koller, N. Friedman,
{\em Probabilistic Graphical Models: Principles and Techniques},
MIT Press (2009).

\bibitem{Kop20} V. Koponen, Conditional probability logic, lifted Bayesian networks, and almost sure quantifier elimination,
{\em Theoretical Computer Science}, Vol. 848 (2020) 1--27.

\bibitem{Kop24} V. Koponen, Random expansions of finite structures with bounded degree,
submitted, \url{https://arxiv.org/abs/2401.04802}.

\bibitem{KW1} V. Koponen, F. Weitkämper, Asymptotic elimination of partially continuous aggregation functions
in directed graphical models, 
{\em Information and Computation}, Vol. 293 (2023)  105061,
\url{https://doi.org/10.1016/j.ic.2023.105061}.

\bibitem{KW2} V. Koponen, F. Weitkämper, On the relative asymptotic expressivity of inference frameworks,
{\em Logical Methods in Computer Science}, Vol. 20 (2024) 13:1--13:52.

\bibitem{KW3} V. Koponen, F. Weitkämper, A general approach to asymptotic elimination of aggregation functions
and generalized quantifiers,
submitted, \url{https://arxiv.org/abs/2304.07865}.

\bibitem{LT} J. Lukasiewicz, A. Tarski, Untersuchungen  \"{u}ber den Aussagenkalk\"{u}l, 
 Comptes Rendus des Séances de la Société des Sciences
et des Lettres de Varsovie, Class III, vol. 23 (1930) 30--50.

\bibitem{Lyn} J. F. Lynch, Almost sure theories,
{\em Annals of Mathematical Logic}, Vol. 18 (1980) 91--135.

\bibitem{SS} S. Shelah, J. Spencer, Zero-one laws for sparse random graphs,
{\em Journal of the American Mathematical Society}, Vol. 1 (1988) 97--115.

\bibitem{She} S. Shelah, Zero-one laws for graphs with edge
probabilities decaying with distance. Part I,
{\em Fundamenta Mathematicae} Vol. 175 (2002) 195--239.

\bibitem{WeiPLP} F. Weitkämper, An asymptotic analysis of probabilistic logic programming,
{\em Theory and Practice of Logic Programming}, Vol. 21 (2021) 802--817.

\bibitem{WeiFLBN}  F. Weitkämper, Functional lifted Bayesian networks,
in S. Muggleton and A. Tamaddoni-Nezhad (eds.), 
{\em Proceedings of the 31st International Conference on Inductive Logic Programming 2022}, Springer 2024, 142--156.
\end{thebibliography}
\end{document}